\documentclass[pdflatex,sn-vancouver-num]{uact-arxiv}
\usepackage{setspace}
\usepackage[T1]{fontenc}
\usepackage[utf8]{inputenc}
\usepackage{graphicx}
\usepackage{amsmath,amssymb}
\usepackage{xcolor}
\DeclareRobustCommand{\SupplementHighlight}[1]{#1}
\usepackage{booktabs,array}
\usepackage{placeins}
\usepackage{tikz}
\usetikzlibrary{arrows.meta}
\usepackage[title]{appendix}
\usepackage{microtype}
\usepackage{xurl}
\newlength{\TableWidth}
\numbered
\makeatletter
\renewcommand{\@seccntformat}[1]{}
\renewcommand{\Authorfont}{\reset@font\fontsize{10bp}{12bp}\selectfont\boldmath\titraggedcenter}
\renewcommand{\addressfont}{\reset@font\fontsize{9bp}{11bp}\selectfont\titraggedcenter}
\makeatother
\renewcommand{\figurename}{Figure}
\hypersetup{
  unicode=true,
  colorlinks=true,
  citecolor=blue,
  linkcolor=blue,
  urlcolor=black
}
\DeclareRobustCommand{\SectionRef}[1]{\hyperref[#1]{Section~\ref*{#1}}}
\DeclareRobustCommand{\TableRef}[1]{\hyperref[#1]{Table~\ref*{#1}}}
\DeclareRobustCommand{\FigureRef}[1]{\hyperref[#1]{Figure~\ref*{#1}}}
\begin{document}

\title{Uncertainty aware prediction from incomplete clinical records: a conformal prediction study of Parkinsonian classification using C-OPN data}

\author[1]{\fnm{Jiaqi} \sur{Bi}}
\author[2]{\fnm{Jialu} \sur{Sun}}
\author[1]{\fnm{Hanrui} \sur{Dou}}
\author[1]{\fnm{Jiaqi} \sur{Yuan}}
\author[2,3]{\fnm{Chengqian} \sur{Xian}}
\author[1,4,5]{\fnm{Osvaldo} \sur{Espin-Garcia}}

\makeatletter
\g@addto@macro\artauthors{\par\smallskip{\Authorfont for the Canadian Open Parkinson Network\textsuperscript{\hyperref[copn-attribution]{*}}}}
\makeatother

\affil[1]{\orgdiv{Department of Epidemiology and Biostatistics}, \orgname{Schulich School of Medicine \& Dentistry, Western University}, \orgaddress{\city{London}, \state{ON}, \country{Canada}}}
\affil[2]{\orgdiv{Department of Statistical and Actuarial Sciences}, \orgname{Western University}, \orgaddress{\city{London}, \state{ON}, \country{Canada}}}
\affil[3]{\orgdiv{School of Science}, \orgname{Harbin Institute of Technology}, \orgaddress{\city{Shenzhen}, \country{China}}}
\affil[4]{\orgdiv{Department of Biostatistics}, \orgname{University Health Network}, \orgaddress{\city{Toronto}, \state{ON}, \country{Canada}}}
\affil[5]{\orgdiv{Dalla Lana School of Public Health and Department of Statistical Sciences}, \orgname{University of Toronto}, \orgaddress{\city{Toronto}, \state{ON}, \country{Canada}}}

\abstract{
\singlespacing\abstractfont
\textbf{Background:} Incomplete clinical records and imbalanced diagnostic groups complicate Parkinsonian classification. We examined whether incomplete records could support useful lists of diagnostic possibilities through conformal prediction, assessing how often the lists retained the recorded diagnosis, their length and the uncertainty remaining.

\textbf{Methods:} We analyzed records from 3,434 Canadian Open Parkinson Network participants: 2,853 with Parkinson's disease, 171 in the atypical Parkinsonian group (AP) and 410 healthy controls. In three repeats of five-fold cross-validation, we compared single-category predictions, conformal prediction sets and lists combined across 21 analyses using alternative models and assumptions about missing entries. An information deficit score summarized unavailable information.

\textbf{Results:} The median proportion of eligible predictor entries unrecorded per participant was 75.6\%. Single-category prediction correctly classified 89.29\% of test records (balanced accuracy 55.40\%), but only 10.33\% of AP records. Conformal prediction sets included the recorded category in 93.35\% of evaluation records and averaged 1.846 categories; combined lists increased inclusion to approximately 99\% and averaged 2.289 categories. AP inclusion increased from 98.05\% to 100\%. Combined lists averaged 1.010 more categories in the group with the greatest information deficit than in the group with the least (95\% interval 0.996--1.023).

\textbf{Conclusions:} Even with extensive missing information and imbalanced diagnostic groups, clinical records can support informative prediction when uncertainty is quantified and expressed through sets of possible diagnoses rather than a single label. Joint assessment of list length and information deficit showed both the narrowing achieved and the records for which all three categories remained open.
}


\newgeometry{top=26mm,text={31pc,245mm},bindingoffset=6mm,headheight=5.5pt,headsep=5.6mm,footskip=10mm}
\maketitle
\restoregeometry

\hypertarget{background}{%
\section{Background}\label{background}}

Prediction from incomplete health records involves more than choosing a classifier. A record can be incomplete because an item was not administered, a questionnaire branch did not apply, a value was unknown, or a data artifact obscured the value. These circumstances need not have the same statistical meaning \citep{Rubin1976}. Missingness may also be associated with the recorded outcome, so apparently good classification can reflect both underlying clinical characteristics and patterns in how information was collected \citep{Agniel2018}. 
How predictions behave may therefore depend on assumptions about the process that generated the missing data \citep{Rubin1976}.

The Canadian Open Parkinson Network (C-OPN) provides a setting in which these issues are consequential. The C-OPN is a Canadian research network for Parkinson's disease and related conditions \citep{Cressatti2024}. In the analytic cohort of 3,434 participants, 2,853 (83.08\%) had Parkinson's disease (PD), 171 (4.98\%) belonged to the study's atypical Parkinsonian category (AP), and 410 (11.94\%) were healthy controls (HC), so aggregate performance could be dominated by the much larger PD group \citep{Brodersen2010}. 
Participant records were assembled from multiple types of record sources, including self-reported questionnaires and professionally collected clinical assessments. 
 
Because participants did not necessarily have the same predictor variables recorded, the resulting records differed in information availability and patterns of missingness.

Set-valued prediction has been developed for situations in which forcing every observation into a single class can hide genuine ambiguity. 
A prediction set contains the diagnostic labels considered plausible for a given participant.
Sadinle et al. proposed least ambiguous and set-valued classifiers that return one or more plausible labels while controlling classification error \citep{Sadinle2016}. 
Romano et al. and Angelopoulos et al. subsequently developed conformal methods for adaptive multiclass prediction sets with formal coverage guarantees \citep{Romano2020,Angelopoulos2020}, and Vazquez and Facelli reviewed the growing use of conformal prediction in clinical applications \citep{Vazquez2022}. 
More recent work has addressed several challenges that are directly relevant to health data. Zaffran et al. studied conformal prediction when predictor values are missing \citep{Zaffran2023}; Sreenivasan et al. used conformal prediction to express individualized uncertainty in classification of disease course in multiple sclerosis \citep{Sreenivasan2025}. 
In Parkinson's disease, Diaz-Rincon et al. used conformal prediction intervals to quantify uncertainty in future medication requirements rather than diagnostic class membership \citep{DiazRincon2025}. 
These studies address important aspects of the present problem, but generally examine missing predictors, diagnostic ambiguity and class imbalance in separate analyses. To our knowledge, prior work has not combined multiclass Parkinsonian diagnostic classification with a rare atypical Parkinsonian group, extensive structured missingness, and predefined implementation levels/budgets within one conformal uncertainty evaluation framework. 
This combination motivates the present study.

We examined whether extensively incomplete C-OPN records could support informative diagnostic possibilities despite imbalance between the parkinsonian groups when uncertainty was represented explicitly through conformal prediction sets. 
We addressed this aim through three questions. 
First, how accurately do incomplete C-OPN records distinguish PD, AP and HC overall and within each diagnostic group given the imbalance between the parkinsonian groups?
Second, how much diagnostic inclusion is gained, particularly for the less represented parkinsonian group, and at what cost in set size, by conformalizing the primary model and combining scenarios for missing data?
Third, how do information deficit, disagreement across scenarios and estimated prediction error describe different aspects of uncertainty, and how do these patterns differ between the imbalanced parkinsonian groups?
We assessed these constructs using the information deficit score developed for this study, denoted by \(Q\), the Scenario Sensitivity Index (SSI) and reliability scores, respectively.

\hypertarget{methods}{%
\section{Methods}\label{methods}}

\begin{figure}[htbp]
\centering
\includegraphics[width=\linewidth]{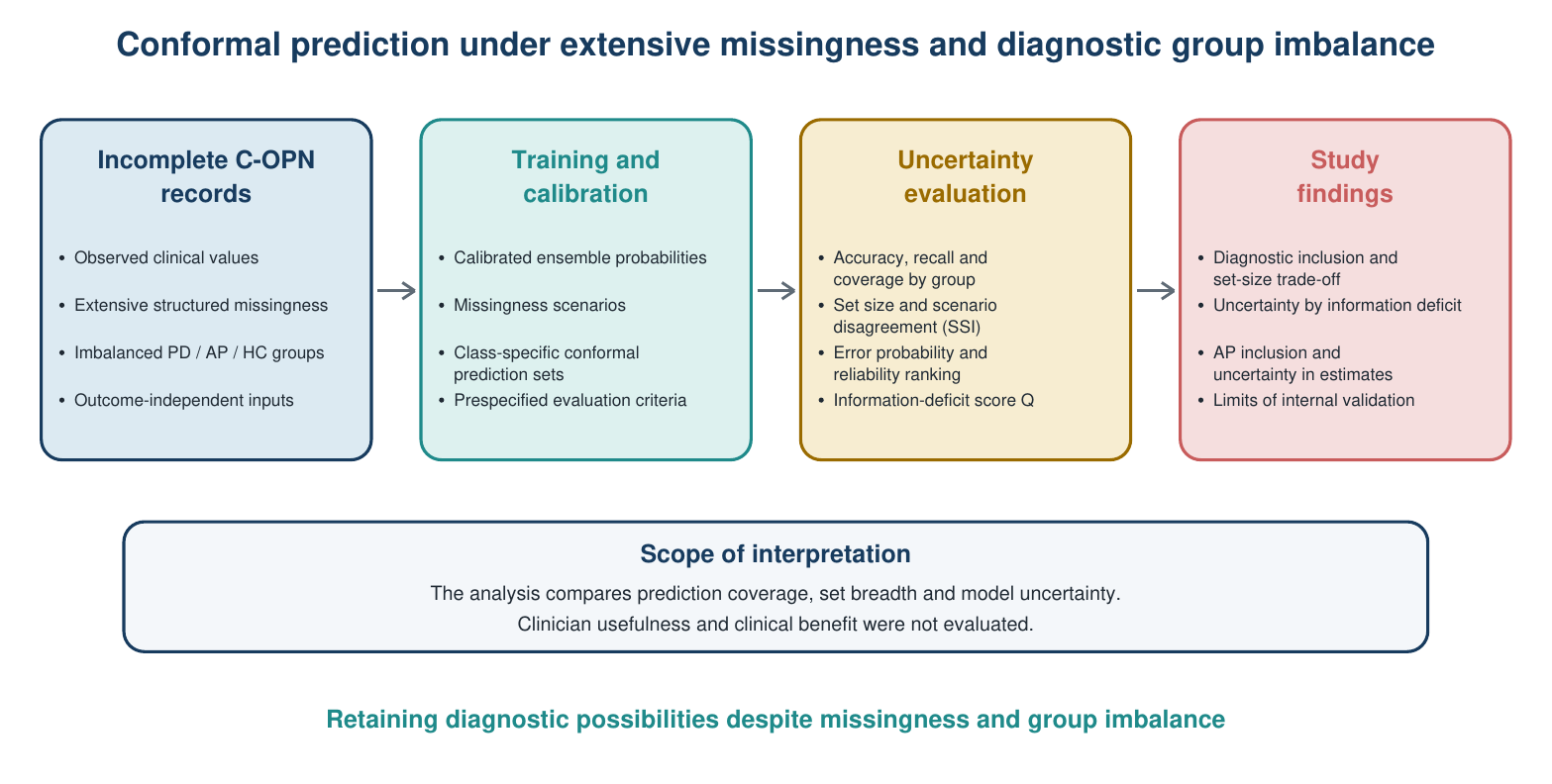}
\caption{\textbf{Overview of prediction and uncertainty evaluation.} The schematic links incomplete participant records, model development and calibration, uncertainty measures and test evaluation. Arrows represent analysis dependencies. The comparisons assess single-label prediction (A0), conformal prediction sets from the same ensemble (A1), the union of sets across missing-data scenarios (A2), and sets constructed from combined scenario scores (A3). Diagnostic inclusion and set size are evaluated across diagnostic groups. \(Q\), SSI and reliability scores describe information deficit, scenario disagreement and estimated prediction error, respectively; \(Q\) is used for evaluation rather than set construction. Coverage, set breadth, ranking of error risk and precision for rare classes address different aspects of performance. C-OPN, Canadian Open Parkinson Network; PD, Parkinson's disease; AP, atypical Parkinsonian category; HC, healthy control; SSI, Scenario Sensitivity Index.}
\label{fig:1}
\end{figure}

The analysis compared what could be learned from incomplete records when reporting one diagnostic category or a set of categories. Models supplied diagnostic probabilities; separate calibration records determined which categories entered prediction sets; and test records assessed diagnostic inclusion and list length. Additional analyses examined sensitivity to missing-data assumptions, information deficit and prediction reliability.
\FigureRef{fig:1} summarizes how incomplete records, prediction models, calibration and uncertainty assessment were connected.

\hypertarget{study-design-setting-and-analytic-cohort}{%
\subsection{Study design, setting and analytic cohort}\label{study-design-setting-and-analytic-cohort}}

This retrospective cohort study used existing, de-identified participant records from C-OPN, which was launched in 2020 with support from Parkinson Canada and Brain Canada \citep{Cressatti2024}. Its records combine participant questionnaires with clinical assessments. Questionnaires were completed in person or remotely, including by telephone, with help from research staff where needed. Trained staff performed motor assessments in person \citep{Cressatti2024}.
C-OPN collected and managed these data using REDCap, an electronic data collection system \citep{Harris2009,Harris2019}, hosted at the University of Calgary.
Our team did not recruit participants or collect new data for this analysis.

Our research team received the data under signed C-OPN Data Use Agreements and authorization from C-OPN. The dataset was accompanied by a data dictionary and documentation describing how to interpret the records. No direct identifiers or key linking the records to participants' identities were provided. The records were collected from 11 May 2016 to 9 February 2023.

The released dataset contained 3,541 unique participants.
Of these, 107 participants could not be assigned a PD/AP/HC outcome, leaving 3,434 participants for the analysis of the three diagnostic groups: Parkinson's disease (PD), atypical Parkinsonian (AP) and healthy controls (HC).

PD diagnoses were made by movement disorder specialists using established diagnostic criteria \citep{Cressatti2024}.
Participants were classified as PD, AP or HC using recorded C-OPN diagnostic information.
The AP group retained the dataset's grouping of atypical parkinsonian disorders and other neurological conditions, as detailed in Supplement S1.2.

The sample size was determined by the eligible records available in the C-OPN release. 
We summarized participant characteristics by diagnostic group using counts and percentages or medians and interquartile ranges, with counts of available observations and no imputation for these descriptive summaries. For reporting, nonzero counts below five are suppressed, together with percentages and additional quantities that would permit their reconstruction. Prediction record counts are distinguished from participant counts. Sparse repeated record events are also withheld when fewer than five distinct contributors cannot be ruled out.
We assessed prediction performance separately within each diagnostic group, with particular attention to the smaller AP group. We distinguished uncertainty about the diagnostic category from uncertainty in the estimated performance of the prediction rules.

\hypertarget{keyed-preparation-source-variables-and-implementation-levels}{%
\subsection{Predictor variables and assessment budgets}\label{keyed-preparation-source-variables-and-implementation-levels}}

The dataset included information on participants' backgrounds, health histories, environmental exposures and clinical characteristics, together with questionnaire responses and standardized assessments \citep{Cressatti2024}.
The Movement Disorder Society sponsored revision of the Unified Parkinson's Disease Rating Scale (MDS-UPDRS) assessed movement related and other symptoms and their effects on daily life \citep{Goetz2008}.
The 39 item Parkinson's Disease Questionnaire (PDQ-39) assessed quality of life \citep{Peto1995}, while the Scales for Outcomes in Parkinson's Disease-Autonomic Dysfunction (SCOPA-AUT) assessed symptoms such as bowel and bladder difficulties \citep{Visser2004}.
The Montreal Cognitive Assessment (MoCA), a brief test of cognitive function, and more detailed neuropsychological tests assessed abilities such as memory and attention \citep{Nasreddine2005,Litvan2012}.
We linked questionnaire and assessment records using each participant's study identifier.
Where different versions of the same form contained comparable information, we combined them using documented rules for matching fields and choosing between conflicting entries. Forms that could not be reliably matched remained separate. Supplement S1.2 describes these rules.

Before analysis, variables were reviewed to distinguish potential predictors from fields used only for administrative purposes.
Administrative fields and other variables not eligible as predictors were excluded from the final analysis.
Variables derived from the recorded outcome were excluded.
The information deficit analysis, which assessed how much relevant information was missing from each record, also excluded variables derived from diagnosis.
Missing entries remained marked as missing in the prepared dataset. We summarized each participant's data availability before imputation as the proportion of eligible predictor entries that were unrecorded, without distinguishing the reasons for their absence. The information deficit score \(Q\), described below, additionally accounted for model contribution and verified recording status.
Model settings were chosen using training data alone to avoid using evaluation records during model development \citep{Varma2006}.

Predictors were grouped into the four implementation levels predefined by C-OPN according to the resources required to obtain them: level 1, self-report; level 2, questionnaire composite scores; level 3, routine professional clinical assessment; and level 4, professional assessment requiring additional resources.
Lower implementation levels were prioritized with the aim of limiting assessment costs and the need for additional professional resources.
We labeled the four cumulative assessment budgets J1--J4: J1 included eligible predictors from level 1, J2 from levels 1--2, J3 from levels 1--3 and J4 from levels 1--4.
The reported uncertainty analyses retained J3 as the primary budget and J4 as a sensitivity budget that added assessments requiring additional resources.
All four budgets used the same participants and differed in the predictors available to the models.

\hypertarget{prediction-model-approaches}{%
\subsection{Prediction model approaches}\label{prediction-model-approaches}}

We compared five approaches, labeled M1--M5, to estimate probabilities of PD, AP and HC (\TableRef{tab:1}). Grouped elastic net regression (M1) provided a regularized regression comparator \citep{Zou2005,Yuan2006}; CatBoost (M2) captured nonlinear predictor effects through boosted trees \citep{Prokhorenkova2018}; and the additive model (M3) combined effects defined within intervals with selected pairwise interactions \citep{Lou2013}. The two-stage hierarchical model (M4) estimated PD/AP versus HC and then divided the disease probability between PD and AP. M2 and M4 supplied the primary ensemble. M2 handled missing numeric values directly \citep{CatBoostDocs} and represented missing categorical values using designated tokens. M4 represented predictor values and missingness features separately, allowing their contributions to differ.

We developed an adaptive elastic net regression model (M5) for this study, incorporating adaptive selection penalties informed by an initial fit and adapting ideas from Wong et al.'s work on missing data \citep{Wong2023}.
M5 supplied sensitivity scenarios for missing data and predictor weights for \(Q\). Supplement S1 provides technical specifications and the mapping between manuscript labels and original code identifiers.

\begin{table}[htbp]
\caption{\textbf{Prediction model prefixes and approaches.}}\label{tab:1}
\centering
\tablebodyfont
\renewcommand{\arraystretch}{1.2}
\begin{minipage}{0.82\linewidth}
\centering
\begin{tabular}{@{}l@{\hspace{1.5em}}l@{}}
\toprule
\textbf{Model prefix} & \textbf{Model approach} \\
\midrule
M1 & Grouped elastic net regression \\
M2 & Boosted decision trees (CatBoost) \\
M3 & Additive model with pairwise interactions \\
M4 & Two-stage hierarchical classification \\
M5 & Adaptive elastic net regression \\
\bottomrule
\end{tabular}

\par\vspace{0.4em}
\begingroup\singlespacing\tablefootnotefont
\raggedright
Only the five retained model approaches are listed. Supplement S1 describes their specifications, procedures for missing data and evaluation methods.
\par\endgroup
\end{minipage}
\end{table}

The primary ensemble combined calibrated probabilities from CatBoost (M2; 60\%) and the hierarchical model (M4; 40\%). The 60:40 weights were fixed throughout the uncertainty evaluation and across assessment budgets. Probability calibration used development predictions (\SectionRef{sec:probability-calibration}).

M5 assessed how predictions changed under different assumptions about missing information.
Missing predictor values were denoted \textbf{MissC}, while missing predictor values together with missing diagnostic labels were denoted \textbf{MissCO}.
Analyses that also considered missing diagnostic labels used inverse-probability weighting (IPW): records with observed labels received more weight when their labels were estimated to be less likely to have been recorded \citep{Robins1994}.
The diagnosis-observation model used the permitted labeled training records together with the supplied outcome-missing records; held-out labeled records did not enter this fit. Outcome-missing records informed observation probabilities but were not used to calculate diagnostic classification performance.
Under a missing-at-random (MAR) assumption, missingness can depend on recorded information, with no additional dependence on unrecorded values after accounting for that information.
A missing-not-at-random (MNAR) assumption allows additional dependence on unrecorded values \citep{Rubin1976}.
Predefined changes to the MNAR assumptions assessed sensitivity to unrecorded information.

Within the M5 sensitivity calculations, the reference procedure drew values for missing predictors from the distribution in the training data for each assessment and tracked which entries had originally been missing.
These procedures were used within the sensitivity model; missing entries in the prepared analysis dataset remained marked as missing.
Supplement S1 details the sensitivity procedure for missing data.

To examine sensitivity to incomplete information, we prespecified 21 prediction scenarios in five families: the primary 60\% M2/40\% M4 ensemble (one scenario), M5 MAR MissC (one), M5 MAR MissCO-IPW (one), M5 MNAR MissC (three) and M5 MNAR MissCO-IPW (fifteen). Supplement S1.4 specifies the MNAR predictor adjustments and weighting grid.
Each scenario supplied predicted probabilities for PD, AP and HC under a specified model and set of missingness assumptions.
These scenarios represented alternative assumptions; none was established as the true missingness mechanism \citep{Molenberghs2008}.
The scenario analysis was distinct from the four methods for probability calibration described in \SectionRef{sec:probability-calibration}.

\hypertarget{cross-validation-design}{%
\subsection{Cross-validation design}\label{cross-validation-design}}

We used three repeats of five-fold outer cross-validation, with each participant tested once per repeat. In each outer split, 20\% of participants formed the test sample and 80\% formed the training sample. We divided the training sample into a subset for model development (75\% of training records, also termed the proper fitting subset) and a separate subset for conformal calibration (25\%). Thus approximately 60\% of the full cohort supported model fitting, tuning and probability calibration, 20\% supported conformal calibration and 20\% supported testing. Within the subset for model development, four-fold inner cross-validation supported tuning and out-of-fold predictions: each model was fitted to three inner folds and predicted the fourth. The outer folds evaluated the fitted procedures on unseen participants; the inner folds supported model development without using the outer test records \citep{Varma2006}. The subset for conformal calibration supplied the prediction set cutoffs for each class and was separate from both model development and testing \citep{Sadinle2016}. \hyperlink{box:validation}{Box~1} shows these sample roles.

\begin{table}[htbp]
\centering
\hypertarget{box:validation}{}
{\tablecaptionfont \textbf{Box 1. Distinct samples in each outer validation split.}
The solid arrows indicate sample allocation; the dashed arrows indicate that a fitted procedure is applied to a different sample, without reusing its records for fitting.\par}
\medskip
\definecolor{flowNavy}{HTML}{193B59}
\definecolor{flowBlue}{HTML}{315C86}
\definecolor{flowBluePale}{HTML}{ECF3F9}
\definecolor{flowTeal}{HTML}{1C7978}
\definecolor{flowTealPale}{HTML}{E9F5F3}
\definecolor{flowOchre}{HTML}{96611F}
\definecolor{flowOchrePale}{HTML}{FAF1E5}
\begin{tikzpicture}[
  >=Latex,
  role/.style 2 args={draw=#1,fill=#2,line width=0.85pt,rounded corners=4pt,
    text width=3.23cm,align=left,inner sep=6pt,
    font=\sffamily\footnotesize},
  allocation/.style={draw=flowNavy!75,line width=0.8pt,-{Latex[length=2.0mm]}},
  transfer/.style={draw=flowNavy!70,line width=1.05pt,densely dashed,
    -{Latex[length=1.8mm]}}
]
\node[fill=flowNavy,rounded corners=4pt,text=white,
  minimum width=11.75cm,minimum height=0.80cm,
  font=\sffamily\small,align=center] (cohort) at (6,0)
  {\textbf{3,434 participants}\quad $\cdot$ \quad three repeats of five-fold validation};

\node[role={flowBlue}{flowBluePale}] (fit) at (1.86,-2.98)
  {\begin{minipage}[t][3.10cm][t]{\linewidth}
   \raggedright
   \textbf{\color{flowBlue}Model development}\\
   \textbf{60\% of the cohort}\\[5pt]
   Fit and tune\\prediction models\\[3pt]
   Four inner folds for\\development predictions\\from excluded folds\\[3pt]
   Calibrate probabilities
   \end{minipage}};
\node[role={flowTeal}{flowTealPale}] (cal) at (6.00,-2.98)
  {\begin{minipage}[t][3.10cm][t]{\linewidth}
   \raggedright
   \textbf{\color{flowTeal}Conformal calibration}\\
   \textbf{20\% of the cohort}\\[5pt]
   Apply fitted predictions\\[3pt]
   Estimate thresholds for\\labels in each class\\[3pt]
   No model retuning
   \end{minipage}};
\node[role={flowOchre}{flowOchrePale}] (test) at (10.14,-2.98)
  {\begin{minipage}[t][3.10cm][t]{\linewidth}
   \raggedright
   \textbf{\color{flowOchre}Test evaluation}\\
   \textbf{20\% of the cohort}\\[5pt]
   Apply fitted model and\\set thresholds\\[3pt]
   Assess predictions\\and label sets\\[3pt]
   No changes based\\on test outcomes
   \end{minipage}};

\draw[draw=flowNavy!65,line width=0.8pt] (cohort.south) -- (6,-0.77);
\draw[draw=flowNavy!65,line width=0.8pt] (1.86,-0.77) -- (10.14,-0.77);
\draw[allocation] (1.86,-0.77) -- (fit.north);
\draw[allocation] (6.00,-0.77) -- (cal.north);
\draw[allocation] (10.14,-0.77) -- (test.north);
\draw[transfer] ([yshift=-0.14cm]fit.east) -- ([yshift=-0.14cm]cal.west);
\draw[transfer] ([yshift=-0.14cm]cal.east) -- ([yshift=-0.14cm]test.west);

\node[fill=flowBluePale,draw=flowBlue!40,rounded corners=2pt,
  minimum width=7.92cm,minimum height=0.43cm,
  font=\sffamily\footnotesize,text=flowNavy] at (3.93,-5.20)
  {\textbf{Outer training: 80\%} (60\% development + 20\% set calibration)};
\node[fill=flowOchrePale,draw=flowOchre!40,rounded corners=2pt,
  minimum width=3.73cm,minimum height=0.43cm,
  font=\sffamily\footnotesize,text=flowNavy] at (10.14,-5.20)
  {\textbf{Outer test: 20\%}};
\node[text=flowNavy!85,font=\sffamily\footnotesize,align=center] at (6,-5.79)
  {Per assessment budget: 10,302 test predictions from 3,434 people; inference resamples people.};
\end{tikzpicture}
\end{table}

Preprocessing and tuning of the base models were restricted to their training roles.

\subsubsection{Probability calibration}\label{sec:probability-calibration}
Four prespecified approaches to probability calibration were compared using out of fold development predictions and multiclass log loss \citep{Guo2017,Kull2019}. 
Dirichlet calibration was selected for both primary ensemble components (M2 and M4) in all 15 outer splits at each assessment budget (J3 and J4). 
The fitted mappings were then applied to records for conformal calibration and testing. Separate conformal calibration determined thresholds for prediction sets within each class \citep{Kull2019,Sadinle2016}. 
Supplement S1 describes the four approaches, fitting and selection procedures.

\hypertarget{conformal-prediction-sets-and-comparison-approaches}{%
\subsection{Conformal prediction sets and comparison approaches (A0--A5)}\label{conformal-prediction-sets-and-comparison-approaches}}

We labeled the six reporting rules A0--A5. A0 reports the diagnosis with the highest probability; A1--A5 return sets of possible diagnoses using the procedures summarized in \TableRef{tab:2}.
Prediction sets could contain one, two or all three study categories (PD, AP and HC). A set containing PD and AP means that both labels were retained by the statistical rule; it does not mean that a participant has both conditions or that the labels are equally likely. Returning all three labels expresses greater uncertainty about which category to select.
We call a set containing one label a singleton, a set containing two labels a pair, and a set containing all three labels a full set.

Each prediction scenario supplies three probabilities for the same participant, one for each diagnostic category. For policies A1--A3, these probabilities are converted to least ambiguous classification (LAC) scores: the score for a label is one minus its predicted probability, \(1-\widehat p_k\), where \(\widehat p_k\) is the probability assigned to diagnostic category \(k\). Smaller scores therefore mean stronger model support \citep{Sadinle2016,Angelopoulos2021}. The separate records for conformal calibration described in \SectionRef{cross-validation-design} determine a cutoff for each category: for example, the AP cutoff is calculated from the AP scores of calibration participants whose recorded diagnosis is AP \citep{Sadinle2016}. A label enters the prediction set when its score is no larger than its cutoff. A4 and A5 begin with the A3 set and can add labels.

The prespecified coverage targets for each class were 90\% for PD, 95\% for AP and 90\% for HC, corresponding to target miscoverage rates of 10\%, 5\% and 10\%, respectively. These targets specify intended inclusion rates within each recorded diagnostic group and are distinct from class prevalence or an individual set's probability of correctness \citep{Angelopoulos2021}. Given these targets, Supplement S1 describes how calibration scores within each class determine the finite-sample cutoffs and states the assumptions required for the coverage guarantee \citep{Sadinle2016,Angelopoulos2021}.

\begin{table}[htbp]
\caption{\textbf{Six rules for reporting possible diagnostic categories.} The rules use the model probabilities described in the text. For A1--A5, least ambiguous classification (LAC) scores and calibrated cutoffs determine the initial sets.}\label{tab:2}
\centering
\tablebodyfont
\setlength{\tabcolsep}{4pt}
\renewcommand{\arraystretch}{1.2}
\setlength{\TableWidth}{\dimexpr\linewidth-4\tabcolsep\relax}
\begin{tabular}{@{}>{\raggedright\arraybackslash}p{0.11\TableWidth}>{\raggedright\arraybackslash}p{0.54\TableWidth}>{\raggedright\arraybackslash}p{0.35\TableWidth}@{}}
\toprule
\textbf{Rule} & \textbf{What the rule does} & \textbf{Why we compared it} \\
\midrule
A0 & Report the label with the highest probability from the primary 60\% M2/40\% M4 ensemble. & Familiar comparison returning one label. \\
A1 & Apply LAC cutoffs to that same primary ensemble; retain every qualifying label. & Show the effect of allowing a set using the primary model alone. \\
A2 & Apply LAC cutoffs separately to all 21 scenarios, then combine all labels in their sets. & Preserve alternatives across assumptions about missing data. \\
A3 & For each label, combine its 21 LAC scores using a weighted 90th percentile, then calibrate new cutoffs. & Explore the balance between shorter sets and coverage. \\
A4 & Check A3 singletons against a fitted reliability cutoff; enlarge failing sets using costs specified for this study. & Examine a safeguard against potentially unreliable results containing one label. \\
A5 & Compare A4 with its allowed larger sets using the same costs; enlarge every set containing only HC. & Examine the balance between set size and potentially harmful omissions. \\
\bottomrule
\end{tabular}

\begingroup\singlespacing\tablefootnotefont
HC, healthy control. A0 and A1 are reference approaches that use the primary ensemble alone. A2--A5 also use 20 additional analyses from M5 under different assumptions about missing data. A2 is the comparator based on the union of scenario sets; A3--A5 are exploratory alternatives. A4 and A5 preserve all starting labels. 
\par\endgroup
\end{table}

The development protocol for the revised analysis designated the policy based on combined scores (A3) as its main challenger to the broader scenario union (A2). A later analysis of the information deficit score \(Q\) focused on A2, with A5 as an exploratory sensitivity comparison. Those choices were fixed before linking \(Q\) to outcomes, following earlier policy development in the same cohort (Supplement S5). Policy performance was compared through class coverage, number of labels returned and cost as defined for this study.

A0 and A1 use the primary M2/M4 ensemble. A0 reports its most probable label. A1 applies the LAC rule to the same probabilities, using cutoffs calibrated specifically for that ensemble. For an illustration using hypothetical values throughout, suppose the ensemble gives PD, AP and HC probabilities of 0.60, 0.30 and 0.10. A0 reports PD. The corresponding LAC scores are 0.40, 0.70 and 0.90. If the PD, AP and HC cutoffs were 0.60, 0.80 and 0.75, A1 would return \(\{\mathrm{PD},\mathrm{AP}\}\): PD and AP pass their respective cutoffs, while HC does not. AP can therefore remain a possibility even though PD has the largest probability.

A2 combines the sets from all 21 scenarios. We first calculate LAC scores and calibrate category specific cutoffs separately for each scenario, using that scenario's predictions on the records for conformal calibration. 
For a new participant, each scenario returns a set; A2 includes every label present in at least one of those sets. 
Continuing the illustration, suppose the M5 MAR MissC scenario gives probabilities of 0.50, 0.20 and 0.30. Its LAC scores are 0.50, 0.80 and 0.70; with its own hypothetical cutoffs of 0.60, 0.75 and 0.75, it returns \(\{\mathrm{PD},\mathrm{HC}\}\). Combining this with the primary ensemble's \(\{\mathrm{PD},\mathrm{AP}\}\) already gives A2 all three labels. The other 19 scenarios cannot remove them. This rule retains alternatives arising under different assumptions, at the cost of potentially broader sets.

A3 combines the LAC scores before calibrating a set. For each participant and each label, we order the 21 scenario scores from smallest to largest and select their weighted 90th percentile: the first score at which the accumulated weight reaches 90\%. Each of the five scenario families receives one fifth of the total weight, divided equally among that family's settings. This prevents the family with fifteen settings from dominating simply because it contains more scenarios. The selected value becomes one combined LAC score for that label. We apply this same calculation to the records for conformal calibration to obtain new PD, AP and HC cutoffs, then retain labels whose combined scores meet those cutoffs. The 90th percentile controls how scenario scores are combined; the separate 90\%/95\%/90\% coverage targets control their calibration.

In the hypothetical example, if the remaining 19 scenarios give the same probabilities as the primary ensemble, A3's combined PD, AP and HC scores are 0.50, 0.80 and 0.90. With hypothetical A3 cutoffs of 0.60, 0.85 and 0.85, it returns \(\{\mathrm{PD},\mathrm{AP}\}\), because HC's combined score exceeds its cutoff. A3 can therefore be smaller than A2, although the separately calibrated rules are not required to produce nested sets. 
If no label passes a cutoff, A1 and each separate A2 scenario set add that scenario's highest probability label; A3 instead adds the label with the smallest combined score relative to its own cutoff. 
Supplement S1 specifies this fallback exactly.

A4 checks whether an A3 singleton should be enlarged. It keeps A3 pairs and sets containing all three labels. 
For a singleton, it uses the Support Reliability Score (SRS), equal to 100 minus the estimated error percentage for the primary ensemble's most probable label; higher scores indicate lower estimated risk. The model and cutoff were fitted using out-of-fold development predictions (\SectionRef{endpoints-and-benchmark-interpretation}). 
If the score meets its fitted cutoff, A4 keeps the singleton. If it falls below the cutoff, or no valid cutoff is available, A4 compares the larger sets containing the original label. For example, starting from \(\{\mathrm{PD}\}\), the choices are \(\{\mathrm{PD},\mathrm{AP}\}\), \(\{\mathrm{PD},\mathrm{HC}\}\), and all three labels. 

Each candidate set receives a penalty for excluding the recorded diagnosis, plus a penalty for set size. 
The exclusion penalty is zero when the recorded diagnosis is included; otherwise, it is the lowest prespecified misclassification cost among the labels in the set. 
The set size penalty is 0 for one label, 0.5 for two and 1 for all three, reflecting the loss of informativeness as sets grow. 
Supplement S1.5 gives the misclassification cost matrix. Within each scenario, the expected penalty is the average of a candidate set's penalties for PD, AP and HC, weighted by their predicted probabilities. A4 selects the set with the smallest maximum expected penalty across the 21 scenarios. 
If the reliability score cannot be calculated, it returns all three labels. 
The error target was the primary ensemble's most probable label, which differed from the A3 singleton label in 238/2,882 J3 outputs (8.3\%) and 262/2,528 J4 outputs (10.4\%). 

A5 compares A4 with its possible enlargements. It also uses the same misclassification and set size penalties as A4. 
It evaluates the current A4 set and every larger set containing all its labels, using the same worst scenario expected penalty. 
It may keep A4 or add labels when a larger set has a lower penalty; ties favor the smaller set. 
The penalties assign greater weight to some errors, including reporting HC for someone with PD or AP, while also charging for less informative, larger sets. 
A5 enlarged every set containing only HC. 

\hypertarget{endpoints-and-benchmark-interpretation}{%
\subsection{Endpoints and benchmark interpretation}\label{endpoints-and-benchmark-interpretation}}

Coverage is the proportion of evaluated records whose returned set contains the recorded diagnosis\citep{Sadinle2016}. For A0, which returns one label, overall inclusion is accuracy and class inclusion is recall. We assessed informativeness jointly through diagnostic inclusion, mean list length and the frequency of returning all three categories. Mean list length ranged from one to three. Returning all three categories necessarily includes the recorded category but provides no diagnostic narrowing; shorter lists were therefore interpreted alongside how often they retained the recorded category. Supplement S2.1 defines the secondary Set Information Gain (SIG) transformation of set size.

The Scenario Sensitivity Index (SSI) measures variation in the possible diagnoses returned across scenarios. For each test prediction record, we compared the sets returned by all 21 scenarios for models and missing data. For each pair of sets, we divided the number of shared labels by the number of distinct labels returned by either set, subtracted this fraction from 1 and multiplied by 100 \citep{Jaccard1912}. For example, \(\{\mathrm{PD}\}\) and \(\{\mathrm{PD},\mathrm{AP}\}\) share one of two labels, giving \(100\times(1-1/2)=50\). SSI is the average across all 210 scenario pairs, on a scale from 0 to 100. A score of 0 means every scenario returns the same set; larger values mean greater variation. Identical sets can still omit the recorded diagnosis.

For the comparison across the whole cohort, we pooled test records across the three cross-validation repeats and ranked them by SSI, separately for J3 and J4. Cutoffs at the 20th, 40th, 60th and 80th percentiles formed five groups: lowest, lower, middle, higher and highest variation, each containing roughly one fifth of the records. Records with equal scores stayed together, so group sizes could differ. Highest variation means scores above the 80th percentile within that budget. All six policies were compared within the same groups, which are separate from the groups based on the information deficit score \(Q\). Supplement S2.1 gives the formal definition.

The prespecified misclassification cost matrix assigns relative penalties to mislabeling, including errors that classify PD or AP as HC. 
For a returned set, a size penalty is added to the smallest class misclassification penalty within the set. 
Robust expected cost takes the worst expected set loss over the prespecified model and missingness scenarios, then averages across records. 
Realized set harm instead evaluates the returned set against the recorded outcome. 

For probabilities, unweighted log loss evaluates the negative logarithm of the probability assigned to the recorded class, and the multiclass Brier score sums squared differences from the one-hot class indicator. Smaller values are better \citep{Gneiting2007}. 
Overall accuracy is the proportion of predictions of one label that match the recorded diagnosis. Balanced accuracy averages the fractions correctly classified within each of the three diagnostic groups \citep{Brodersen2010}. For accuracy comparisons, calibration adjusted each model's diagnostic probability estimates using training records. We tested M2, M4 and their combination on the same records, which had not been used for fitting or calibration. Each chose the diagnosis with the highest adjusted probability. Accuracy and class sensitivities were calculated across all test predictions. The balanced accuracy results in \FigureRef{fig:budget-performance} describe the earlier development stage and average the 15 values from the outer folds.
AP precision is the fraction of predicted AP labels that are truly AP. AP recall is the fraction of AP records detected. AP F1 combines the two. 
AP precision recall area summarizes ranking of the rare class and must be reported with its scoring convention and cohort. 
Accuracy of point predictions and coverage of prediction sets assess different outputs. Prediction sets described diagnostic ambiguity. \(Q\) summarized weighted information deficit. SSI described disagreement across scenarios, and reliability scores estimated the chance of an incorrect label. Confidence intervals separately described the precision of estimated performance, conditional on the fitted predictions.

\subsubsection{Reliability of predictions of one label}\label{sec:reliability-methods}

An estimated error probability is the model's estimated chance that its predicted label disagrees with the recorded diagnosis. The multicomponent score combined six signals: spread of diagnostic probabilities (entropy), closeness of the top two probabilities, model disagreement, missing information, similarity to predictions based on missingness patterns alone, and sensitivity to assumptions about missing data. We fitted a logistic regression model to training predictions and their observed errors, using nonnegative weights to combine the six signals into one estimated error probability. The simple confidence measures were the highest diagnostic probability \citep{Geifman2017}, the gap between the top two probabilities, and entropy; training data alone selected the comparator. Within each testing split, predictions were ranked from lowest to highest estimated error risk; keeping only the most confidently ranked predictions is called selective retention \citep{ElYaniv2010}.

The area under the resulting risk--coverage curve (AURC) summarizes error rates as more predictions are kept; smaller values are better, and coverage here means the fraction kept \citep{Ding2020}. Error Brier score and error log loss assess how well the numerical error probabilities match observed errors, with smaller values preferred \citep{Gneiting2007}. We assessed A3 outputs containing one label separately from predictions selecting the most probable diagnosis for every record, and only an AURC improvement meeting the primary J3 rule could support adopting the combined score (Supplements S2 and S5).

\hypertarget{the-information-deficit-score-q}{%
\subsection{The information deficit score (\texorpdfstring{\(Q\)}{Q})}\label{the-information-deficit-score-q}}

We defined an information deficit score, \(Q\), for this study. It combines predictor weights from the selected M5 model with penalties for each predictor's original recording status. Higher \(Q\) indicates a greater deficit in available information, weighted by model contribution.

We derived predictor weights for \(Q\) from the selected adaptive elastic net regression model (M5), using coefficient magnitude to measure model contribution in this study.
M5 assessed sensitivity to missing data, while primary predictions came from the combination of CatBoost (M2) and the two-stage hierarchical classifier (M4).
For \(Q\), we used M5's MAR MissC branch, in which predictors may be missing but diagnostic categories are recorded.
\(Q\) was used to compare fixed prediction policies across groups with different levels of information deficit; it did not enter the primary probability ensemble, reliability fitting or construction of prediction sets.
Supplement S2.2 specifies the separate input for missing information used in the evaluated multicomponent reliability score.

We combined these weights with each variable's original missingness status, before any missing values were filled in for model fitting:

\begin{equation}
\label{eq:information-deficit}
Q_i = \sum_{j\in\mathcal{J}} \nu_j\,d\!\left(M_{ij}\right),
\qquad
\nu_j =
\frac{\lVert B_j\rVert_F/\sqrt{q_j}}
     {\displaystyle\sum_{h\in\mathcal{J}}\lVert B_h\rVert_F/\sqrt{q_h}}.
\end{equation}
In this equation, \(i\) identifies a participant and \(\mathcal{J}\) the eligible predictors represented in the selected M5 fit. For predictor \(j\), \(B_j\) contains coefficients for its \(q_j\) encoded columns across the contrasts of PD versus HC and AP versus HC; HC is the zero reference. The Frobenius norm \(\lVert B_j\rVert_F\) summarizes coefficient magnitude. Dividing by \(\sqrt{q_j}\) adjusts for encoded block size; normalizing across predictors gives nonnegative weights summing to one. Excluded predictors and zero coefficient blocks carry zero weight (Supplement S3). 

\(M_{ij}\) denotes the original recording status of predictor \(j\) for participant \(i\). The deficit penalty used in this study \(d(M_{ij})\) is 0 for observed values including valid zeros, 0.5 only when response codes or verified skip rules in the C-OPN data dictionary\footnote{\label{fn:copn-dictionary-access}The C-OPN data dictionary provides response definitions and questionnaire skip rules; it does not explain every blank. The authors cannot redistribute it publicly. Researchers must request it directly from C-OPN; access information is provided in the Availability of data and materials section.} identify an entry as not applicable or structurally absent, and 1 for other missing entries. The half penalty assigns intermediate deficit to verified structural absence; these numerical penalties are study choices, not empirically established clinical utilities.
The weighted sum \(Q_i\) ranges from 0 to 1: 0 means all positively weighted variables were observed, and 1 means all carried a full missingness penalty. 
For a hypothetical example, weights 0.50, 0.30 and 0.20 and respective penalties 1, 0 and 0.5 give \(Q_i=0.50(1)+0.30(0)+0.20(0.5)=0.60\).
We checked whether \(Q\) depended on these choices by repeating the calculation in three ways: removing the adjustment for the number of numerical columns used to represent each predictor, assigning no penalty (0) for missing information to entries verified as not applicable or structurally absent, or assigning those entries a full penalty (1) instead of the main rule's 0.5. Only one choice was changed at a time.

We divided the model development records into four groups and calculated each participant's \(Q\) using a model selected and fitted without that participant's group. Calibration and test records were scored using a final model selected and fitted on all model development records.

We checked the stability of \(Q\) scores and participants' ordering by \(Q\) under six ways of dividing the training records between model development and calibration of prediction sets. Calibration of prediction sets uses separate records with known diagnoses to determine the cutoffs for retaining diagnostic labels (\SectionRef{conformal-prediction-sets-and-comparison-approaches}). Three divisions reserved 25\%, 30\% or 35\% of the training records for this step; the other three used alternative selections of calibration records while keeping the test records fixed. Reserving 25\% was the reference. These choices changed the records available to fit M5 and derive weights for \(Q\).

Within each division of the training records and each outer validation split, the model development \(Q\) scores provided cutoffs at the 25th, 50th and 75th percentiles. These defined four groups with increasing information deficit: lowest (Q1), lower middle (Q2), upper middle (Q3) and highest (Q4). Q1 includes scores at or below the first cutoff; Q2 lies above the first and at or below the second; Q3 lies above the second and at or below the third; and Q4 lies above the third. The same cutoffs were applied to calibration and test records, with equal scores kept together, so group sizes could differ. Prediction performance was assessed after grouping. Supplement S3 gives the detailed training and grouping rules.

\hypertarget{q-stratified-outcome-analysis-and-inference}{%
\subsection{Statistical inference and exploratory chronology}\label{q-stratified-outcome-analysis-and-inference}}

The policy protocol designated A3 as the main challenger to A2. After model and policy development in this cohort, but before linking \(Q\) to outcomes, the exploratory \(Q\) analysis designated J3 A2 mean set size in Q4 minus Q1 as its primary contrast, with Q4 class coverage checks and A5 as an exploratory comparison. Each budget supplied 10,302 test records from 3,434 participants across three repeats.

We used 2,000 bootstrap replicates, sampling participant identifiers with replacement and retaining each sampled participant's repeated records together \citep{Field2007}. Paired policy and group contrasts used the same participant multiplicities on both sides. Intervals required at least 1,900 usable replicates; undefined denominators made a replicate unusable for that metric. Two-sided 95\% intervals used the 2.5th and 97.5th percentiles and one-sided 95\% bounds the relevant 5th or 95th percentile \citep{Efron1986}. The \(Q\) analysis used linear percentile interpolation; the A3 reliability comparison used the ``higher'' convention (Supplement S5). Coverage was undefined for strata containing no records from the relevant diagnostic group.

These intervals hold fitted predictions fixed and describe uncertainty from resampling participants conditional on the fitted procedures and analysis design. They exclude the full variability from model training, selection and calibration, and do not automatically resolve dependence arising from shared cross-validation fits \citep{Bates2024}. Intervals are pointwise, without simultaneous control across comparisons. Separate one-sided exact binomial sensitivities assessed AP coverage within each repeat and the distinct event that all relevant Q4 records for a participant were covered \citep{Clopper1934}. Those calculations assume independent Bernoulli trials with a common event probability.

\hypertarget{results}{%
\section{Results}\label{results}}

\hypertarget{cohort-and-analysis-sample-sizes}{%
\subsection{Participant characteristics and data availability}\label{cohort-and-analysis-sample-sizes}}

Of the 3,541 participants in the dataset, 107 could not be assigned a PD/AP/HC outcome under the specified diagnosis hierarchy, leaving 3,434 for analysis.
The analytic sample comprised 2,853 participants with Parkinson's disease (PD; 83.08\%), 171 in the atypical Parkinsonian category (AP; 4.98\%) and 410 healthy controls (HC; 11.94\%).
\TableRef{tab:3} summarizes their characteristics and recorded assessments before imputation.
Among participants with age available, the median age was 68 years (interquartile range 61--73).
Among participants with the relevant assessment recorded, the AP group had the lowest median cognitive assessment score and the highest median motor examination score.

The four cumulative assessment budgets contained 638 eligible predictor variables at J1, 695 at J2, 833 at J3 and 980 at J4.
Within participants, the median proportions of unrecorded predictor entries were 80.9\% at J1, 80.6\% at J2, 75.6\% at J3 and 77.9\% at J4.
Cognitive assessment and motor examination scores were available for 60.2\% and 52.1\% of participants, respectively, compared with 16.3\% for the quality of life index.
HC participants had a higher median proportion of unrecorded predictor entries at every budget: 90.4--92.4\% across J1--J4, compared with 72.1--79.9\% for PD. At the primary J3 budget, the corresponding medians were 92.4\% and 72.1\% (\TableRef{tab:3}). 

\begin{table}[htbp]
\caption{\textbf{Participant characteristics and data availability in the analytic cohort.}}\label{tab:3}
\centering
\tablebodyfont
\setlength{\tabcolsep}{3pt}
\renewcommand{\arraystretch}{1.05}
\setlength{\TableWidth}{\dimexpr\linewidth-8\tabcolsep\relax}
\begin{tabular}{@{}>{\raggedright\arraybackslash}p{0.28\TableWidth}*{4}{>{\raggedleft\arraybackslash}p{0.18\TableWidth}}@{}}
\toprule
\textbf{Characteristic} & \textbf{Overall} & \textbf{PD} & \textbf{AP} & \textbf{HC} \\
\midrule
\multicolumn{5}{@{}l}{\textbf{Participants and demographic characteristics}} \\
Participants, n & 3434 & 2853 & 171 & 410 \\
Age at study visit, years & 68.0 (61.0--73.0) & 68.0 (61.0--73.0) & 68.0 (60.0--74.0) & 66.0 (55.0--71.0) \\
Age available, n (\%) & 2371 (69.0) & 2034 (71.3) & 111 (64.9) & 226 (55.1) \\
Recorded gender, n (\%) & & & & \\
\quad Female & 1294 (37.7) & 990 (34.7) & 75 (43.9) & 229 (55.9) \\
\quad Male & 1797 (52.3) & 1611 (56.5) & 79 (46.2) & 107 (26.1) \\
\quad Other recorded categories & $<5$ & $<5$ & 0 (0.0) & 0 (0.0) \\
\quad Not recorded & NR & NR & 17 (9.9) & 74 (18.0) \\
\midrule
\multicolumn{5}{@{}l}{\textbf{Clinical characteristics}} \\
Cognitive assessment score (MoCA), points & 26.0 (24.0--28.0) & 26.0 (24.0--28.0) & 24.0 (21.0--27.0) & 27.5 (26.0--29.0) \\
Motor examination score (MDS-UPDRS III), points & 26.0 (17.0--38.0) & 27.0 (18.0--38.0) & 41.0 (31.0--51.0) & 4.0 (1.0--7.0) \\
\midrule
\multicolumn{5}{@{}l}{\textbf{Data availability}} \\
Cognitive assessment score available, n (\%) & 2066 (60.2) & 1742 (61.1) & 88 (51.5) & 236 (57.6) \\
Motor examination score available, n (\%) & 1789 (52.1) & 1625 (57.0) & 78 (45.6) & 86 (21.0) \\
Quality of life index available, n (\%) & 560 (16.3) & 522 (18.3) & 31 (18.1) & 7 (1.7) \\
Autonomic symptom score available, n (\%) & NR & 178 (6.2) & $<5$ & $<5$ \\
Unrecorded predictor entries (J1), \% & 80.9 (73.5--98.7) & 79.9 (72.6--94.2) & 79.8 (73.0--99.1) & 92.3 (85.7--99.8) \\
Unrecorded predictor entries (J2), \% & 80.6 (73.4--97.4) & 79.9 (72.4--93.2) & 79.7 (73.0--97.7) & 91.4 (85.3--98.4) \\
Unrecorded predictor entries (J3), \% & 75.6 (65.8--96.6) & 72.1 (65.1--92.0) & 80.3 (65.8--97.7) & 92.4 (82.8--98.7) \\
Unrecorded predictor entries (J4), \% & 77.9 (69.8--93.6) & 76.1 (68.5--92.2) & 81.3 (69.0--95.4) & 90.4 (81.8--98.9) \\
\bottomrule
\end{tabular}

\begingroup\singlespacing\tablefootnotefont
PD, Parkinson's disease; AP, atypical Parkinsonian category; HC, healthy control. Continuous summaries are medians (25th--75th percentiles) of available values; counts are n (\%) of all participants in the column. Each participant contributes once. $<5$: one to four participants; NR: additional withholding to prevent recovery. Associated percentages are withheld.
Age uses the calculated age at the study visit, excluding an invalid value of zero.
Cognitive assessment is the Montreal Cognitive Assessment (MoCA) total score; motor examination is Part III of the Movement Disorder Society-sponsored revision of the Unified Parkinson's Disease Rating Scale (MDS-UPDRS); quality of life is the 39-item Parkinson's Disease Questionnaire (PDQ-39) index; autonomic symptoms are the Scales for Outcomes in Parkinson's Disease-Autonomic Dysfunction (SCOPA-AUT) total score.
Assessment availability refers to the specified total or index, rather than any recorded item from that assessment, and gives the number of participants represented in each clinical score summary.
Percentages of unrecorded predictor entries are calculated within each participant before imputation, using all eligible predictors in the relevant budget as the denominator (J1: 638; J2: 695; J3: 833; J4: 980). They count empty entries without distinguishing their causes and are separate from the information deficit score \(Q\), which uses model weights.
\par\endgroup
\end{table}

\FloatBarrier

\hypertarget{probability-model-performance}{%
\subsection{Performance of point predictions overall and by diagnostic group}\label{probability-model-performance}}

At J3, the final calibrated ensemble achieved 89.29\% overall accuracy and 55.40\% balanced accuracy, with PD, AP and HC recalls of 98.63\%, 10.33\% and 57.24\%, respectively. At J4, the corresponding values were 89.75\%, 55.98\%, and 99.02\%, 10.72\% and 58.21\%. Balanced accuracy is the unweighted mean of the three class recalls. Always predicting PD would yield 83.08\% overall accuracy in this cohort. The final point predictions therefore identified PD much more often than AP or HC.

\begin{figure}[htbp]
\centering
\includegraphics[width=\linewidth]{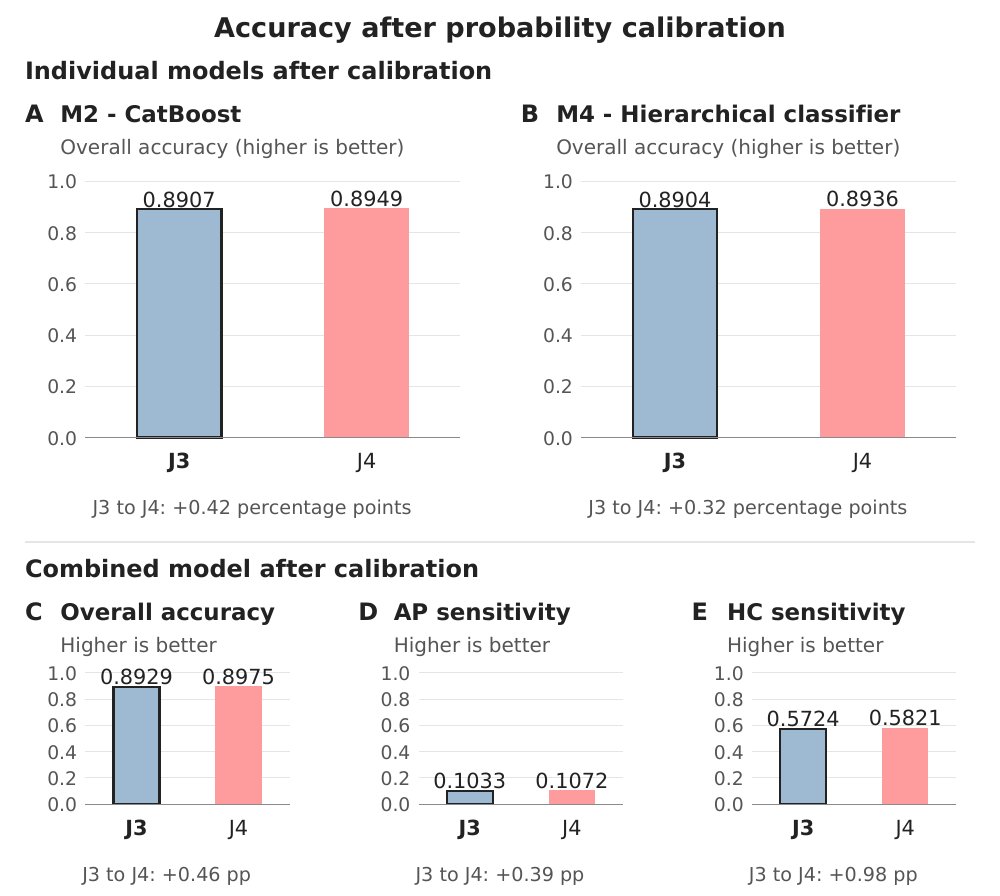}
\caption{\textbf{Accuracy after probability calibration for the individual models and their combination.} Panels A and B show accuracy for CatBoost (M2) and the hierarchical classifier (M4). Panel C shows accuracy for their fixed 60\%/40\% combination; panels D and E show its sensitivities for the atypical Parkinsonian category (AP) and healthy controls (HC). Calibration adjusts each model's diagnostic probability estimates using training records. All three models are tested on the same test records, choosing the diagnosis with the highest adjusted probability. Each evaluation of a model at a given budget uses 10,302 predictions from 3,434 participants across 15 outer splits. J3 includes assessment levels 1--3; J4 also includes level 4. Outlined bars mark the primary J3 budget. All axes run from 0 to 1; higher values indicate better performance. Differences use unrounded values; pp denotes percentage points. }
\label{fig:accuracy-budget}
\end{figure}

After probability calibration, M2 and M4 accuracies were 89.07\% and 89.04\% at J3 and 89.49\% and 89.36\% at J4 (\FigureRef{fig:accuracy-budget}). Relative to the more accurate component, the ensemble's descriptive accuracy increase was 0.22 percentage points at J3 and 0.26 at J4. Ensemble log loss, Brier score and AP average precision were 0.3054, 0.1670 and 0.3426 at J3, versus 0.3025, 0.1639 and 0.3443 at J4.

\begin{figure}[htbp]
\centering
\includegraphics[width=\linewidth]{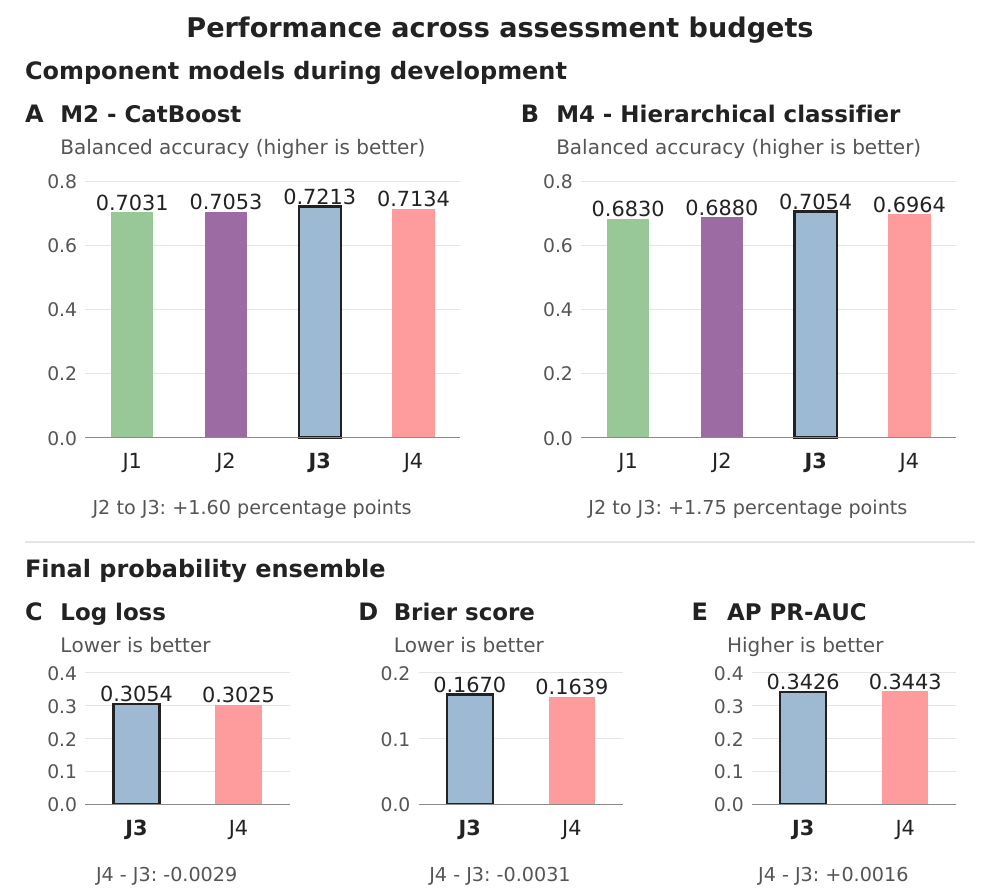}
\caption{\textbf{Balanced accuracy of the component models and ensemble probability performance across assessment budgets.} Panels A and B show mean balanced accuracy across the outer folds during model development for CatBoost (M2) and the hierarchical classifier (M4). Panels C--E show the final calibrated ensemble's test log loss, multiclass Brier score and area under the precision--recall curve (PR-AUC, reported as average precision) for the atypical Parkinsonian category (AP). Each evaluation of a model at a given budget uses 10,302 repeated prediction rows from 3,434 participants. J1--J4 include cumulative implementation levels 1, 1--2, 1--3 and 1--4; outlined bars mark the primary J3 budget. Higher balanced accuracy and AP average precision, and lower log loss and Brier score, indicate better performance. 
Annotated differences use unrounded values. 
}
\label{fig:budget-performance}
\end{figure}

During earlier model development, mean balanced accuracies across the outer folds for M2 were 0.7053, 0.7213 and 0.7134 at J2, J3 and J4; M4 values were 0.6880, 0.7054 and 0.6964 (\FigureRef{fig:budget-performance}). These component results supported J3 as the primary assessment budget, which includes routine clinical assessment without level 4 resource requirements. 

\FloatBarrier

\hypertarget{efficiency-cannot-be-separated-from-class-coverage}{%
\subsection{Coverage and efficiency across budgets and policies}\label{efficiency-cannot-be-separated-from-class-coverage}}

\begin{figure}[htbp]
\centering
\includegraphics[width=\linewidth]{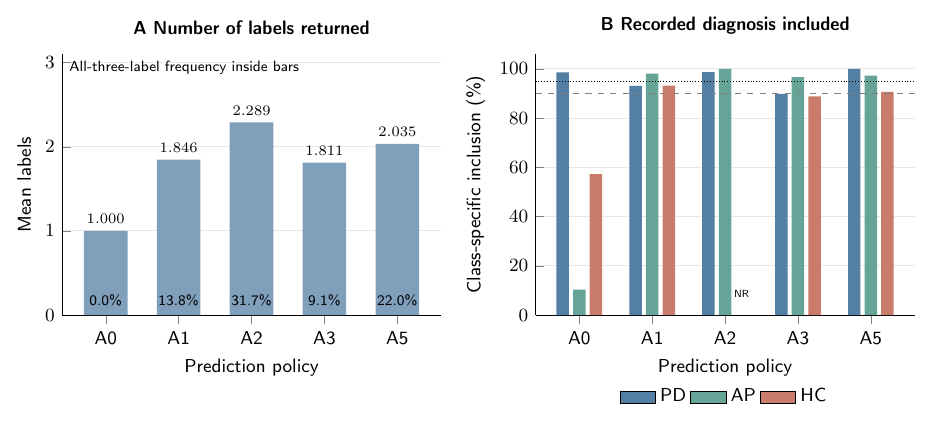}
\caption{\textbf{Diagnostic inclusion and set size at J3.} The policies use the same 10,302 repeated evaluation records from 3,434 participants. Left: mean number of labels, with the frequency of returning all three labels annotated. Right: inclusion within each diagnostic class; A0 shows recall and the set-valued policies show coverage. Dashed lines mark the 90\% PD/HC and 95\% AP targets. Estimates are descriptive; uncertainty intervals are not plotted. The A2 HC bar is omitted (NR) to protect a suppressed count. A3 is the designated challenger to A2; A5 is exploratory.}\label{fig:2}
\end{figure}

At J3, moving from A0, which returns one label, to conformal A1 using the same ensemble increased overall label inclusion from 89.29\% to 93.35\% and AP inclusion from 53/513 (10.33\%) to 503/513 (98.05\%), with mean size increasing from 1.000 to 1.846 labels. A2 increased overall inclusion to approximately 99\% and AP inclusion to 513/513 (100\%), averaging 2.289 labels (\TableRef{tab:4}; \FigureRef{fig:2}). Relative to A1, A2 added 1.95 percentage points of AP inclusion and 0.443 labels, with higher overall inclusion. Because A0 and A1 used the same ensemble probabilities, their comparison directly demonstrates how returning a set can retain a recorded category omitted when only the most probable label is reported. A2 then shows the additional inclusion and list length associated with combining alternative analyses.

\begin{table}[htbp]
\caption{\textbf{Label inclusion and set size across the whole cohort.}}\label{tab:4}
\centering\tablebodyfont
\setlength{\tabcolsep}{3pt}\renewcommand{\arraystretch}{1.15}
\setlength{\TableWidth}{\dimexpr\linewidth-10\tabcolsep\relax}
\begin{tabular}{@{}>{\raggedright\arraybackslash}p{0.14\TableWidth}>{\raggedleft\arraybackslash}p{0.12\TableWidth}>{\raggedright\arraybackslash}p{0.31\TableWidth}>{\raggedleft\arraybackslash}p{0.13\TableWidth}>{\raggedleft\arraybackslash}p{0.16\TableWidth}>{\raggedleft\arraybackslash}p{0.14\TableWidth}@{}}
\toprule
\textbf{Budget / policy} & \textbf{Overall (\%)} & \textbf{PD / AP / HC inclusion (\%)} & \textbf{Mean labels} & \textbf{Full sets (\%)} & \textbf{Robust cost} \\
\midrule
J3 / A0 & 89.29 & 98.63 / 10.33 / 57.24 & 1.000 & 0.00 & 0.706 \\
J3 / A1 & 93.35 & 93.08 / 98.05 / 93.25 & 1.846 & 13.77 & 0.848 \\
J3 / A2 & $\approx99$ & 98.74 / 100.00 / NR & 2.289 & 31.71 & 0.751 \\
J3 / A3 & 90.11 & 89.89 / 96.69 / 88.86 & 1.811 & 9.11 & 0.910 \\
J3 / A5 & 98.75 & 100.00 / 97.27 / 90.65 & 2.035 & 22.02 & 0.587 \\
\midrule
J4 / A0 & 89.75 & 99.02 / 10.72 / 58.21 & 1.000 & 0.00 & 0.726 \\
J4 / A1 & 93.29 & 92.98 / 97.86 / 93.58 & 1.843 & 13.56 & 0.880 \\
J4 / A2 & $\approx99$ & 98.83 / 100.00 / NR & 2.294 & 32.47 & 0.749 \\
J4 / A3 & 89.87 & 89.48 / 97.66 / 89.27 & 1.849 & 9.40 & 0.955 \\
J4 / A5 & 98.73 & 100.00 / 97.66 / 90.33 & 2.054 & 22.72 & 0.598 \\
\bottomrule
\end{tabular}
\par\medskip
\begingroup\singlespacing\tablefootnotefont\raggedright
Each row for a given policy and budget uses the same 10,302 evaluation records from 3,434 participants across three repeats: PD 8,559, AP 513 and HC 1,230. For A0, overall inclusion equals accuracy and class inclusion equals recall. Other rows report coverage of prediction sets. Full sets contain all three labels. Cost uses penalties specified for this study. These are descriptive estimates; paired A3/A5 minus A2 intervals are in Supplement S4. Coverage targets are 90\%/95\%/90\% for PD/AP/HC, assessed separately from set size and cost. A2 had $<5$ uncovered HC records at each budget. The HC percentage is withheld (NR), and overall A2 coverage is rounded to the nearest whole percentage point, to prevent recovery of these counts.
\par\endgroup
\end{table}

A3, the designated main challenger to A2, reduced J3 mean size by 0.478 labels (paired 95\% interval for A3 minus A2: \(-0.492\) to \(-0.464\)), from 2.289 to 1.811. The frequency of returning all three labels fell from 31.71\% to 9.11\%. PD/AP/HC coverage was 89.89\%/96.69\%/88.86\%; PD and HC point estimates were below their 90\% targets. The AP coverage difference versus A2 was \(-3.31\) percentage points (95\% interval \(-5.68\) to \(-1.30\)). The HC coverage difference was approximately $-11$ percentage points, with a 95\% interval rounded outward to $-14$ to $-8$ points. These HC summaries are coarsened to protect the suppressed A2 count. A3 also returned sets containing only HC for 24 records with recorded PD or AP. 

The exploratory A5 enlargement averaged 2.035 labels, 0.254 fewer than A2, with 22.02\% full sets and lower robust cost under the penalties specified for this study (\TableRef{tab:4}). PD/AP/HC coverage was 100\%/97.27\%/90.65\%; the one-sided bootstrap lower bound for HC was about 88.60\%, below the 90\% target. Every saved J3 A5 set contained PD (\TableRef{tab:set-composition}), so 100\% PD coverage coexisted with no PD exclusion. This is an empirical property of these results, not a claim that the algorithm always forces PD inclusion. A5 varied in AP/HC inclusion and set size.

J4 gave the same pattern: A2 remained broad, A3 PD and HC coverage remained below 90\%, and A5 had lower mean size and cost than A2. A3 returned sets containing only HC for 17 disease records. Among A3 outputs containing only AP, 180/298 were incorrect at J3 (60.4\%) and 195/325 at J4 (60.0\%; Supplement S4.2).

\begin{table}[htbp]
\caption{\textbf{J3 set composition and frequencies of returned labels.}}\label{tab:set-composition}
\centering\tablebodyfont\renewcommand{\arraystretch}{1.15}
\begin{tabular*}{0.85\linewidth}{@{\extracolsep{\fill}}lrrr@{}}
\toprule
\textbf{Returned set} & \textbf{A2} & \textbf{A3} & \textbf{A5} \\
\midrule
\(\{PD\}\) & 188 & 2139 & 1905 \\
\(\{AP\}\) & 30 & 298 & 0 \\
\(\{PD, AP\}\) & 6149 & NR & 5697 \\
\(\{HC\}\) & 69 & 445 & 0 \\
\(\{PD, HC\}\) & 0 & $<5$ & 431 \\
\(\{AP, HC\}\) & 599 & 1217 & 0 \\
\(\{PD, AP, HC\}\) & 3267 & 939 & 2269 \\
\midrule
PD present (\%) & 93.22 & 80.97 & 100.00 \\
AP present (\%) & 97.51 & NR & 77.32 \\
HC present (\%) & 38.20 & NR & 26.21 \\
\bottomrule
\end{tabular*}
\par\medskip
\begingroup\singlespacing\tablefootnotefont\raggedright
Counts sum to 10,302 repeated records per policy. ``Present'' percentages use all records as denominator, regardless of recorded diagnosis; they describe how often a label is returned, not class coverage. All observed A5 sets contain PD, whereas AP and HC inclusion vary. $<5$ denotes one to four records. NR withholds companion quantities so that the suppressed count cannot be recovered by subtraction. Supplement S4, Table S4 uses the same reporting convention.
\par\endgroup
\end{table}

A2 most often returned \(\{\mathrm{PD},\mathrm{AP}\}\) (6,149 of 10,302 evaluation records), excluding HC from these lists while retaining both PD and AP. AP appeared in 97.51\% of all A2 lists. Its 100\% AP coverage therefore describes preservation of AP as a diagnostic possibility and should be interpreted alongside how frequently AP was included overall.

\hypertarget{whole-cohort-robustness}{%
\subsection{Whole cohort robustness to variation across scenarios}\label{whole-cohort-robustness}}

All six policies were compared in the same five SSI groups within each assessment budget, using 10,302 test records from 3,434 participants for each combination of policy and budget (Supplement S4, Table S10). SSI measures disagreement across scenarios, separately from the information deficit measured by \(Q\).

\begin{figure}[htbp]
\centering
\includegraphics[width=0.92\linewidth]{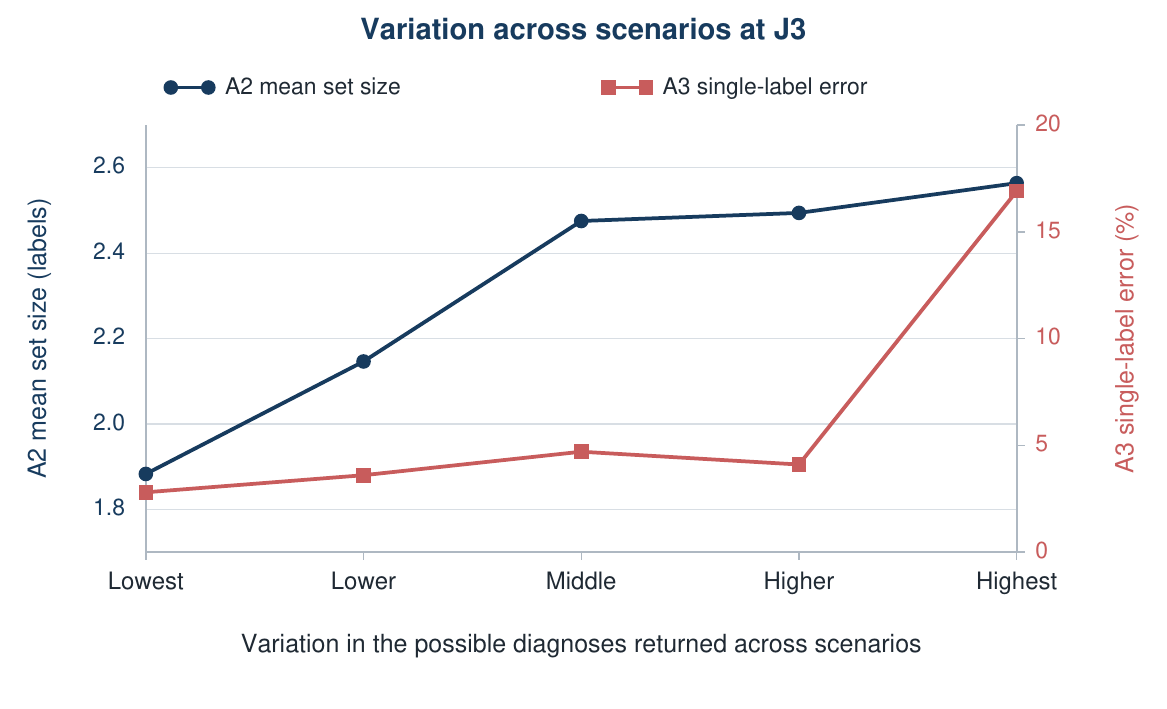}
{\figurecaptionfont Variation across scenarios and the number of labels returned\par}
\caption{\textbf{Variation in the possible diagnoses returned across scenarios.} The analysis uses 10,302 test prediction records from 3,434 participants under the primary assessment budget (J3: levels 1--3). Records are ordered by the Scenario Sensitivity Index (SSI), which measures how much the possible diagnoses vary across the 21 scenarios. The five groups run from lowest to highest SSI, with lower, middle and higher in between. Highest variation means scores above the 80th percentile, roughly the top 20\% of records; equal scores stay together. \SectionRef{endpoints-and-benchmark-interpretation} describes the calculation and grouping. The left axis shows the average number of labels returned by A2, the union of the sets from each scenario. The right axis shows the error percentage among A3 outputs containing only one label (singletons). Group sizes and numbers of outputs containing one label differ. Plotted error rates are rounded to 0.1 percentage points. These variation groups are separate from the four information deficit groups defined using \(Q\). \TableRef{tab:highest-variation} compares all six policies in the highest variation group at both budgets; Supplement S4, Table S10, panels A--B gives the full comparison across all five variation groups.}
\label{fig:3}
\end{figure}

A2 returned more labels as this variation increased. At the primary assessment budget (J3), its mean size increased from 1.883 labels in the lowest variation group to 2.564 in the highest variation group. The proportion of records receiving all three labels increased from zero to 56.39\% (\FigureRef{fig:3}; Supplement S4, Table S10, panel A).

\TableRef{tab:highest-variation} compares all six policies in the highest variation group at each assessment budget. At J3, A0 always returned one label. The average numbers of labels returned were 1.988 for A1, 2.564 for A2, 1.777 for A3, 1.821 for A4 and 2.104 for A5.
Using the ensemble alone, A1 included the recorded diagnosis in 93.20\% of PD records and 93.73\% of HC records. A2 covered all AP records in this group. Selected coverage percentages and companion quantities are withheld to protect sparse events, including repeated-record events whose distinct participant count has not been verified (\TableRef{tab:highest-variation}). HC coverage was 79.70\% under A3, 80.60\% under A4 and 85.97\% under A5. These policies remained below the HC target, so narrower lists did not establish adequate subgroup coverage.
When additional covariates were allowed (J4), A1, A2 and A5 averaged 1.931, 2.558 and 2.154 labels in that budget's highest variation group. A2 returned all three labels for 55.79\% of records, and A5 HC coverage was 84.89\% (\TableRef{tab:highest-variation}).

\begin{table}[htbp]
\caption{\textbf{Prediction policies in the highest variation group.} Highest variation means the roughly 20\% of held-out records with the largest Scenario Sensitivity Index (SSI) scores, above the 80th percentile at each budget. SSI measures variation in the possible diagnoses returned across the 21 scenarios. Cutoffs were calculated separately for J3 and J4, and records with equal scores stayed together.}\label{tab:highest-variation}
\centering
\tablebodyfont
\setlength{\tabcolsep}{4pt}
\renewcommand{\arraystretch}{1.15}
\setlength{\TableWidth}{\dimexpr\linewidth-10\tabcolsep\relax}
\begin{tabular}{@{}>{\raggedright\arraybackslash}p{0.11\TableWidth}>{\raggedleft\arraybackslash}p{0.17\TableWidth}>{\raggedleft\arraybackslash}p{0.18\TableWidth}>{\raggedleft\arraybackslash}p{0.18\TableWidth}>{\raggedleft\arraybackslash}p{0.18\TableWidth}>{\raggedleft\arraybackslash}p{0.18\TableWidth}@{}}
\toprule
\textbf{Policy} & \textbf{Average labels returned} & \textbf{All three labels (\%)} & \textbf{PD coverage} & \textbf{AP coverage} & \textbf{HC coverage} \\
\midrule
\multicolumn{6}{@{}p{\linewidth}@{}}{\textbf{A. Primary assessment budget J3 (levels 1--3)}} \\
\multicolumn{6}{@{}p{\linewidth}@{}}{2,011 records from 1,417 participants; class record counts: PD: 1,574; AP: 102; HC: 335} \\
\addlinespace[3pt]
A0 & 1.000 & 0.00 & NR & $<5$ covered & 39.40\% \\
A1 & 1.988 & 31.18 & 93.20\% & NR & 93.73\% \\
A2 & 2.564 & 56.39 & $<5$ missed & 100.00\% & NR \\
A3 & 1.777 & 17.95 & 82.78\% & NR & 79.70\% \\
A4 & 1.821 & 18.30 & 83.67\% & NR & 80.60\% \\
A5 & 2.104 & 35.06 & 100.00\% & NR & 85.97\% \\
\addlinespace[3pt]
\midrule
\multicolumn{6}{@{}p{\linewidth}@{}}{\textbf{B. Expanded assessment budget J4 (levels 1--4)}} \\
\multicolumn{6}{@{}p{\linewidth}@{}}{2,056 records from 1,453 participants; class record counts: PD: 1,603; AP: 89; HC: 364} \\
\addlinespace[3pt]
A0 & 1.000 & 0.00 & NR & NR & 45.05\% \\
A1 & 1.931 & 26.22 & 91.64\% & NR & 92.31\% \\
A2 & 2.558 & 55.79 & $<5$ missed & 100.00\% & NR \\
A3 & 1.842 & 17.46 & 83.03\% & NR & 82.14\% \\
A4 & 1.846 & 17.46 & 83.09\% & NR & 82.14\% \\
A5 & 2.154 & 35.12 & 100.00\% & NR & 84.89\% \\
\bottomrule
\end{tabular}
\par\medskip
\begin{minipage}{\linewidth}
\singlespacing\tablefootnotefont\raggedright
PD, Parkinson's disease; AP, atypical Parkinsonian category; HC, healthy control. Class percentages count records whose returned set includes their recorded category. For the one-label rule A0, this is the percentage correctly classified within that class. The same records were used for all six policies within each budget; a participant can contribute records in more than one repeat.
\par\smallskip
A0 and A1 use only the primary ensemble: 60\% CatBoost (M2) and 40\% hierarchical model (M4) predictions. A2--A5 also use 20 additional analyses from the adaptive elastic-net model (M5) under different assumptions about missing data. A2 combines scenario-specific sets; A3 combines scores; A4 can add labels to A3 using estimated error risk; A5 can then add labels using study-defined costs. These are descriptive estimates without uncertainty intervals. \SupplementHighlight{Supplement S4, Table S10, panels A--B} reports all five variation groups.
\par\smallskip$<5$ means one to four prediction records (covered or missed, as indicated). NR withholds a value or companion estimate to protect sparse events, including events with unverified numbers of distinct participants. Percentages are withheld with the corresponding counts. Neither marker means zero or missing data. The CSV follows the same rules.
\end{minipage}
\end{table}

\hypertarget{q-distribution-and-model-selection}{%
\subsection{Information deficit: set breadth, subgroup coverage and stability}\label{q-distribution-and-model-selection}}

For the designated primary \(Q\) contrast, A2 mean set size was 1.010 labels higher in Q4 than Q1 (two-sided 95\% interval 0.996--1.023; one-sided 95\% lower bound 0.998). Mean sizes were 1.956 and 2.965 labels, respectively. Records with greater information deficit as weighted by the model therefore received broader A2 sets.

Diagnostic composition differed across \(Q\) strata. HC was absent from Q1, precluding a comparison of HC in Q4 with HC in Q1. Q4 contained 1,889 PD, 132 AP and 495 HC records. These are repeated evaluation records, and participants could change strata across repeats. Detailed counts for the intermediate groups are not reported.

In Q4, A2 returned all three labels for 2,429/2,516 records (96.54\%). For these records, the procedure kept every category open and made the limits of diagnostic narrowing explicit. Exploratory A5 reduced mean size to 2.643 and the frequency of returning all three labels to 64.35\%, but HC inclusion was 401/495 (81.01\%; 95\% interval 75.94--85.72\%), versus 495/495 under A2. Q4 contained 94 of A5's 115 uncovered HC records while accounting for 495 of 1,230 HC records overall. 

A2 covered all 132 AP records in Q4, with counts within each repeat of 45/45, 43/43 and 44/44. The smallest one-sided exact binomial 95\% lower bound was 93.27\%, below the 95\% target. In the separate analysis by participant, all relevant records were covered for 45/45 participants (lower bound 93.56\%). Bootstrap bounds for these results were 100\%, because resampling observed records cannot create failures (Supplements S4.3 and S5.3).

\(Q\) ranked participants similarly across six training divisions. Alternative weighting and penalties for recording status produced similar average scores but changed their spread (Supplement S3, Figure S1 and Tables S2--S3). Detailed selection frequencies are retained in Supplement S3.5. 

\FloatBarrier

\hypertarget{reliability-better-probability-scores-did-not-satisfy-the-ranking-rule}{%
\subsection{Reliability: combining warning signals did not improve error ranking}\label{reliability-better-probability-scores-did-not-satisfy-the-ranking-rule}}

\begin{figure}[htbp]
\centering
\includegraphics[width=\linewidth]{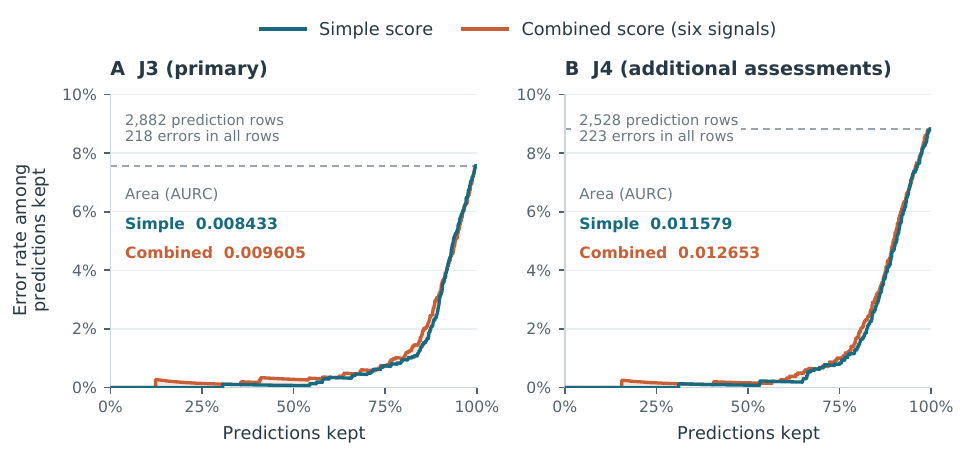}
{\figurecaptionfont Ranking predictions containing one label from most to least reliable\par}
\caption{\textbf{Error ranking among A3 outputs containing one diagnosis.} Panels A and B show risk--coverage curves at J3 (primary) and J4 (additional assessments), respectively, for the simple score and the score combining six signals. Predictions are ranked from most to least reliable. The horizontal axis shows the fraction kept; the vertical axis shows the error rate among those kept. Lower curves and smaller areas under the risk--coverage curve (AURCs) are preferable. Dashed lines show the error rate when all A3 predictions containing one label are kept. Curves combine the 15 testing splits and reproduce the reported AURCs; Supplement S2 gives the calculation. The fraction kept is conditional on A3 returning one label and is distinct from coverage of recorded labels by prediction sets.}
\label{fig:reliability-risk-coverage}
\end{figure}

Among A3 outputs containing one diagnosis, combining the six warning signals was less effective than a simple confidence score at placing correct predictions first (\FigureRef{fig:reliability-risk-coverage}). 
Among the same A3 predictions containing one label at J3, ranking by the combined score using six signals gave an area under the risk--coverage curve (AURC) of 0.009605, compared with 0.008433 when ranking by a single confidence measure selected using training data alone.
The difference, combined minus simple, was +0.001171 (95\% interval +0.000267 to +0.002390). Because smaller areas are better, this positive difference favored the simple score. J4 gave the same direction: +0.001075 (+0.000373 to +0.001923). The combined score had the smaller AURC in only three of 15 testing splits at each budget.

The combined score did improve error Brier score and error log loss. However, these improvements in the numerical estimates of error probability did not translate into a better ordering of predictions by reliability. The predefined comparison therefore did not support adopting the more complex score. Supplement S4, Table S6 gives the detailed differences and intervals.

Interpretation also depends on the error target: the fitted reliability model targets error in the primary ensemble's most probable label, whereas evaluation of A3 singletons uses the returned singleton label. Their 8.3\% J3 and 10.4\% J4 disagreements motivate refitting and threshold validation for the intended target before using reliability to retain A3 singletons.

A separate analysis took the diagnosis with the highest probability for every record. Here the simple score chosen within the training procedure had AURC 0.027100 (95\% interval 0.023202--0.031525). Using the highest diagnostic probability, the gap between the top two probabilities, or the spread of probabilities across diagnoses gave AURCs of 0.027046, 0.027253 and 0.027797, respectively. The difference in AURC between the score selected using training data and the score based on the highest diagnostic probability was 0.000054 (\ensuremath{-}0.000005 to 0.000115). 

\FloatBarrier

\hypertarget{discussion}{%
\section{Conclusion and Discussion}\label{discussion}}

The available clinical records supported prediction sets that retained diagnostic possibilities despite extensive missing information. At J3, the median proportion of unrecorded predictor entries was 75.6\%, yet A1 included the recorded category in 93.35\% of evaluation records while averaging 1.846 categories. Using the same ensemble, A1 retained many AP diagnoses omitted by the single-label rule A0. Combining scenarios through A2 increased inclusion to approximately 99\% while averaging 2.289 categories. These findings demonstrate how allowing several possibilities can preserve recorded diagnoses while making diagnostic ambiguity explicit.

Comparing A2 with A3 and A5 showed how narrowing the list changes diagnostic inclusion. A3, the designated challenger, returned fewer labels than A2, with PD and HC point estimates below the chosen coverage targets. Exploratory A5 regained some coverage, but its 100\% PD inclusion reflected PD being present in every observed J3 set, and its overall HC estimate concealed a large Q4 shortfall. The principal comparison is therefore the joint pattern of class coverage, set composition and size; cost as specified for this study is an additional criterion rather than evidence of clinical utility \citep{Vickers2006}. Additional J4 assessments did not remove these patterns.

The information deficit score \(Q\) provided a way to examine prediction sets across records with different amounts of available information. Records in the highest \(Q\) group received broader A2 lists, and examination of this group revealed an HC inclusion shortfall under exploratory A5 that was less apparent in the whole-cohort average. HC records had more unrecorded predictors than PD records, and HC was absent from Q1 but accounted for 495 Q4 records. Availability patterns and diagnostic composition may therefore contribute to results across \(Q\) strata; the data do not establish why entries were absent \citep{Agniel2018,Molenberghs2008}. The stability of \(Q\) across training divisions supports reproducible assessment within this cohort. \(Q\) was used to examine prediction sets and was not an input to their construction. Further studies should compare it with unweighted missingness and assess its associations within diagnostic groups to determine its additional value.

The reliability comparison identified a simpler option for ranking predictions: simple confidence performed better than the combined score among A3 outputs containing one label. The combined score improved Brier score and log loss, demonstrating that better numerical error estimates and better error ranking are distinct outcomes \citep{Ding2020,Gneiting2007}. The ensemble label used to define training errors also differed from the A3 singleton label for a material fraction of outputs. A4's exploratory enlargement rule therefore requires validation for the intended target before interpreting its threshold as controlling A3 singleton error.

\hypertarget{conclusions}{}\label{conclusions}

Even with extensive missing information, C-OPN records supported prediction sets that retained recorded diagnoses at high rates while leaving relevant alternatives open. Joint assessment of diagnostic inclusion, list length and information deficit showed both the narrowing achieved and the records for which all three categories remained possible. 
This framework provides a practical basis for interpreting incomplete records and refining diagnostic prediction. 
Together, prediction sets and the uncertainty analyses offer a promising way to interpret incomplete clinical datasets by showing where the available information supports diagnostic narrowing and where ambiguity remains.

\hypertarget{limitations}{%
\section{Limitations}\label{limitations}}

This analysis used one C-OPN release, with only 171 participants in a heterogeneous AP/PD-plus category. Repeated evaluation does not increase the number of independent participants or establish performance for each AP condition \citep{Bates2024}. The analyses followed model and policy development in the same cohort; the \(Q\) contrasts were designated before outcome linkage, but the overall chronology remains exploratory. Bootstrap intervals hold predictions fixed, exclude full training, selection and calibration variability, and do not fully resolve dependence from shared model fits \citep{Bates2024}. The binomial sensitivities rely on their stated trial assumptions and do not supply unconditional coverage guarantees.

A typical record lacked about three quarters to four fifths of eligible entries. The sensitivity scenarios represent selected assumptions, not an identified missingness mechanism \citep{Molenberghs2008}. \(Q\)'s coefficient weights, predictor coding and penalties for recording status, and the policy cost matrix, are choices made for this study whose sensitivity matters. Diagnostic composition and record availability constrain interpretation of subgroup comparisons; no causal benefit of collecting a specific item was estimated. Historical documentation of initial ensemble selection is incomplete, so implementation of the fixed blend alone does not establish prespecification of its original selection criteria.

Independent validation across cohorts, sites and time periods \citep{Altman2000} is needed, together with more participants from uncommon diagnostic groups and evaluation of how clinicians use the outputs \citep{Vasey2022}. Future comparisons could include adaptive prediction sets and regularized adaptive prediction sets \citep{Romano2020,Angelopoulos2020}, as well as fitting reliability models for the intended target and simpler missingness summaries. Clinical usefulness and patient benefit remain to be assessed \citep{Vickers2006,Vasey2022}.

\hypertarget{declarations}{%
\section{Declarations}\label{declarations}}

\hypertarget{ethics-approval-and-consent-to-participate}{%
\subsection{Ethics approval and consent to participate}\label{ethics-approval-and-consent-to-participate}}

The original Canadian Open Parkinson Network (C-OPN) study received research ethics approval at participating sites. Participants gave written informed consent, either in person or electronically \citep{Cressatti2024}.
This retrospective cohort study used existing records and involved no new recruitment or data collection. This secondary analysis was approved by the Western University Health Science Research Ethics Board (HSREB; project ID 129896).

\hypertarget{consent-for-publication}{%
\subsection{Consent for publication}\label{consent-for-publication}}

Not applicable. This article reports results for groups of participants and does not include identifiable participant information or images.

\hypertarget{availability-of-data-and-materials}{%
\subsection{Availability of data and materials}\label{availability-of-data-and-materials}}

The C-OPN participant dataset analyzed in this study is not publicly available and cannot be redistributed by the authors under the C-OPN Data Use Agreement. Researchers seeking access should apply directly to C-OPN through its database request process (\url{https://copn-rpco.ca/submit-a-request/}). Applicants should first submit a C-OPN research membership request, then provide a project description, details of the requested data and a signed C-OPN Data Use Agreement. Access is subject to C-OPN review and authorization, the required institutional ethics approval and compliance with the approved project's data-use conditions.

The C-OPN data dictionary used to interpret response codes and questionnaire skip rules cannot be redistributed publicly by the authors. Researchers must request this documentation directly from C-OPN through the access process described above.

The analysis code is maintained at \url{https://github.com/Jiaqi-Bi-Not-PhD/Uncertainty-aware-prediction-conformal-COPN.git}. The repository is currently private and is not publicly accessible. The participant dataset and restricted data dictionary must be obtained separately from C-OPN.

\hypertarget{competing-interests}{%
\subsection{Competing interests}\label{competing-interests}}

None declared.

\hypertarget{funding}{%
\subsection{Funding}\label{funding}}

C-OPN data collection was supported by Parkinson Canada and Brain Canada.

\hypertarget{authors-contributions}{%
\subsection{Authors' contributions}\label{authors-contributions}}

J.B. led the data inspection, cleaning, and wrangling, performed the statistical analyses, contributed to obtaining ethics approval, and prepared the original manuscript draft. 
J.S. and H.D. assisted with data inspection, cleaning, wrangling, analyses, and contributed to obtaining ethics approval. 
J.Y. assisted with data inspection, cleaning, and wrangling. 
C.X. critically reviewed the manuscript and provided scientific feedback. 
O.E.-G. supervised the project, contributed to data inspection, and provided scientific guidance and critical feedback on the manuscript.
\hypertarget{acknowledgements}{%
\subsection{Acknowledgments}\label{acknowledgements}}

We thank the C-OPN participants and the clinical and research teams who collected, managed and shared the data used in this study.

\phantomsection\label{copn-attribution}
\noindent\textsuperscript{*}Data used in this article came from the Canadian Open Parkinson Network (C-OPN) database (\url{https://copn-rpco.ca/}). C-OPN investigators designed and carried out the network study and provided the data. We acknowledge Parkinson Canada's support for C-OPN. A complete list of C-OPN investigators is available at \url{https://copn-rpco.ca/}.

We also thank the Statistical Society of Canada (SSC) for organizing the 2026 Case Studies in Data Analysis Competition, through which we were introduced to this research problem and the C-OPN data.

\end{document}


\renewcommand{\theequation}{S1.\arabic{equation}}
\renewcommand{\thetable}{S\arabic{table}}
\maketitle

This supplement accompanies the manuscript \emph{Uncertainty aware prediction from
incomplete clinical records: a conformal prediction study of Parkinsonian
classification}. The five retained model approaches are designated M1--M5,
consistent with Table~1 and the Methods section of that manuscript.

\paragraph{Notation used throughout.}
Let $x$ denote the predictor information available for one participant at
the specified assessment budget, and let $y$ denote that participant's
recorded diagnostic class. The label set is
$\mathcal{K}=\{\mathrm{PD},\mathrm{AP},\mathrm{HC}\}$: Parkinson's disease,
the atypical Parkinsonian category and healthy control, respectively.
The index $k$ denotes a candidate class in $\mathcal{K}$, and $i$ indexes
records within the sample. For model $m$, where
$m$ is one of M1--M5, $p_{m,k}(x)$ is its estimated class-$k$ probability
before probability calibration, and $\widetilde p_{m,k}(x)$ is the
corresponding probability after calibration. A hat, as in
$\widehat p_k(x)$, denotes the final probability supplied by a fixed
prediction procedure; later subscripts identify the ensemble or scenario.
Each three-class probability vector has nonnegative entries summing to
one. All logarithms below are natural logarithms. The derivations in
S1.1 proceed from a single record to a weighted loss and then to the
coefficient updates. S1.4 supplies the missing-data draws and weights.
S1.5 explains how fitted scores become prediction sets.

\begin{table}[htbp]
\centering
\caption{Retained model prefixes and approaches.}
\label{tab:s1-models}
\renewcommand{\arraystretch}{1.2}
\begin{tabular}{@{}l@{\hspace{1.5em}}l@{}}
\toprule
\textbf{Model prefix} & \textbf{Model approach} \\
\midrule
M1 & Grouped elastic net regression \\
M2 & Boosted decision trees (CatBoost) \\
M3 & Additive model with pairwise interactions \\
M4 & Two-stage hierarchical classification \\
M5 & Adaptive elastic net regression \\
\bottomrule
\end{tabular}
\end{table}

\section*{S1.1\quad Model specifications}

\paragraph{M1: Grouped elastic net regression.}
The class-weighted multinomial regression used an elastic net penalty to
regularize groups of regression coefficients associated with each predictor
variable \cite{Zou2005,Yuan2006}. Predictor eligibility and preprocessing were determined within each
training split. Numeric measurements were screened and scaled, categorical
levels were encoded using training data, and missing numerical entries could
be represented using neutral coordinates and missingness indicators. Its
coefficients provide an interpretable comparison based on regression with the
more flexible prediction approaches.

\paragraph{From encoded measurements to class probabilities.}
A source is one original predictor variable. Encoding can give it several
numerical columns, such as one column for each retained category and a
missingness indicator. Within one training fit, let $n$ be the number of
labelled records, $p$ the total number of encoded columns, $\mathcal J$ the
set of retained sources, $G_j$ the column indices belonging to source $j$,
and $q_j=|G_j|$ their number. The groups partition the design columns.
Write $x_{i\ell}$ for column $\ell$ of record $i$.
Use class indices $1,2,3$ for PD, AP and HC in these derivations.
HC is the reference: its intercept and coefficients are fixed at zero.
The estimated intercept vector $b$ therefore has two entries and the
coefficient matrix $B$ has $p$ rows and two columns. The block
$B_j=B_{G_j,:}$ contains the $q_j$ rows for source $j$ and both
contrasts (PD versus HC and AP versus HC).

The two fitted linear predictors and the three probabilities are
\begin{equation}\label{eq:shared-softmax}
\begin{aligned}
\eta_{ik}&=b_k+\sum_{\ell=1}^{p}x_{i\ell}B_{\ell k},\quad k=1,2,
&\eta_{i3}&=0,\\
D_i^{\rm prob}&=1+\exp(\eta_{i1})+\exp(\eta_{i2}),\\
p_{ik}&=\frac{\exp(\eta_{ik})}{D_i^{\rm prob}},\quad k=1,2,
&p_{i3}&=\frac{1}{D_i^{\rm prob}}.
\end{aligned}
\end{equation}
The coefficient $B_{\ell k}$ changes the
log odds of class $k$ against HC by that amount for a one-unit change in
encoded column $\ell$, holding the other columns fixed.

Let $Y_{ik}=\mathbf 1[y_i=k]$, an indicator equal to one for the recorded
class and zero for the other classes. Exactly one of the three indicators
equals one. The probability assigned to the recorded class can therefore
be written as a product, and taking its negative logarithm gives
\begin{equation}\label{eq:shared-nll}
\begin{aligned}
\Pr(y_i\mid x_i;b,B)&=\prod_{k=1}^3p_{ik}^{Y_{ik}},\\
\ell_i(b,B)&=-\log\Pr(y_i\mid x_i;b,B)
             =-\sum_{k=1}^3Y_{ik}\log p_{ik}\\
&=-\sum_{k=1}^2Y_{ik}\eta_{ik}
  +\left(\sum_{k=1}^3Y_{ik}\right)\log D_i^{\rm prob}\\
&=\log D_i^{\rm prob}-\sum_{k=1}^2Y_{ik}\eta_{ik}.
\end{aligned}
\end{equation}
Thus a small probability for the recorded class produces a large loss.

\paragraph{The grouped M1 objective.}
Let $n_k$ count training records in class $k$, and let $\kappa$ be the
selected multiplier for AP. The record weights are
$w_i=n\kappa^{\mathbf 1[y_i=2]}/(3n_{y_i})$, and $W=\sum_iw_i$.
Before the AP multiplier, each class has total weight $n/3$.
M1 minimizes
\begin{equation}\label{eq:m1-objective}
F_1(b,B)=\frac{1}{W}\sum_{i=1}^nw_i\ell_i(b,B)
+\lambda\left\{\frac{1-\alpha}{2}\|B\|_F^2
+\alpha\sum_{j\in\mathcal J}\sqrt{q_j}\|B_j\|_F\right\}.
\end{equation}
The Frobenius norm $\|B_j\|_F$ is the square root of the sum of squares
of all entries in that block. The tuning parameter $\lambda>0$ controls
overall shrinkage; $\alpha\in[0,1]$ divides the penalty between squared
coefficient shrinkage (ridge) and source selection (the block norm).
The factor $\sqrt{q_j}$ adjusts the selection penalty for block size.
The intercepts are unpenalized. The group penalty shrinks a source's
coefficients jointly across the two diagnostic contrasts and can set
the entire block to zero.
The gradient and block update follow the M5 derivation below by using
one design matrix and setting every adaptive penalty weight to one.

\paragraph{M2: Boosted decision trees (CatBoost).}
The multiclass CatBoost model combined sequentially fitted decision trees \cite{Prokhorenkova2018}.
CatBoost handled missing numeric values directly \cite{CatBoostDocs} and used tokens derived from training data
for missing categorical values and mappings for rare categories.
Its primary clinical predictor representation did not add explicit indicators
of predictor availability or missingness as separate features. Tuning
was restricted to training data. Calibrated M2 probabilities contributed
60\% to the fixed primary probability ensemble.

\paragraph{M3: Additive model with pairwise interactions.}
This interpretable additive approach represented each predictor with a
finite collection of learned contributions. Selected predictor pairs could
also contribute jointly \cite{Lou2013}. Numeric binning,
categorical mappings, explicit tokens for missing values and interaction
selection were based on the applicable training folds. A class-weighted
multinomial fit combined main effects and selected pairwise terms to produce
class probabilities.

\paragraph{M4: Two-stage hierarchical classification.}
For a participant with predictor information $x$, the first stage estimated
$q(x)$, the probability of belonging to the combined PD/AP group rather
than HC. The second stage, trained on PD and AP records, estimated
$r(x)$, the conditional probability of AP given membership in that combined
group. Before probability calibration, the three-class probabilities were
denoted $p_{\mathrm{M4},k}(x)$, where the subscript M4 identifies the
hierarchical model and $k$ is one of the diagnostic labels PD, AP or HC:
\begin{equation*}
\begin{aligned}
p_{\mathrm{M4},\mathrm{PD}}(x)&=q(x)\{1-r(x)\},\\
p_{\mathrm{M4},\mathrm{AP}}(x)&=q(x)r(x),\\
p_{\mathrm{M4},\mathrm{HC}}(x)&=1-q(x).
\end{aligned}
\end{equation*}
The stage learners used additive effects for individual predictor variables
and penalties sensitive to missingness, with separate fitting for each stage
and tuning restricted to training data. Calibrated
M4 probabilities contributed 40\% to the fixed primary probability ensemble.

\paragraph{The two binary fitting problems.}
For stage $h=1$, all labelled training records are used and
$t_i^{(1)}=\mathbf 1[y_i\in\{\mathrm{PD},\mathrm{AP}\}]$.
For stage $h=2$, only PD/AP training records are used and
$t_i^{(2)}=\mathbf 1[y_i=\mathrm{AP}]$.
Each stage learns its own eligible sources and encoding. Let $v_i$ be
its encoded value features and $m_i$ its encoded missingness features.
For the current stage, write
\[
\eta_i=c+v_i^\top\beta+m_i^\top\theta,
\qquad \pi_i=\sigma(\eta_i)=\frac{1}{1+e^{-\eta_i}},
\]
where $c$ is an intercept, $\beta$ contains value coefficients and
$\theta$ contains missingness coefficients. The superscript $\top$
denotes transpose, so $v_i^\top\beta$ means the sum of products of features and
their coefficients. Stage 1 supplies $q(x)=\pi^{(1)}(x)$ and stage 2 supplies
$r(x)=\pi^{(2)}(x)$ in the equations above.

A binary observation has probability $\pi_i^{t_i}(1-\pi_i)^{1-t_i}$.
Since $\log\pi_i=\eta_i-\log(1+e^{\eta_i})$ and
$\log(1-\pi_i)=-\log(1+e^{\eta_i})$, its negative log probability is
\[
-t_i\log\pi_i-(1-t_i)\log(1-\pi_i)
=-t_i\eta_i+[t_i+(1-t_i)]\log(1+e^{\eta_i})
=\log(1+e^{\eta_i})-t_i\eta_i.
\]
Let $n_h$ be the stage's training count, and $n_{h0},n_{h1}$ its two
class counts. Its record weight is
$u_i=n_h\kappa_h^{t_i}/(2n_{h,t_i})$, where $\kappa_1=1$ and
$\kappa_2$ is the selected AP multiplier. Set $U_h=\sum_i u_i$.
Let $\beta_j$ and $\theta_j$ be the value and missingness coefficient
blocks for source $j$, of sizes $d_j$ and $r_j$, respectively; sums
below include only nonempty blocks of the relevant type. The objective is
\begin{equation}\label{eq:m4-objective}
\begin{aligned}
F_h(c,\beta,\theta)
={}&\frac{1}{U_h}\sum_i u_i\{\log(1+e^{\eta_i})-t_i\eta_i\}\\
&+\lambda_h\frac{1-\alpha}{2}
  \left(\|\beta\|_2^2+\gamma_m\|\theta\|_2^2\right)\\
&+\lambda_h\alpha\left(
  \sum_j\sqrt{d_j}\|\beta_j\|_2
  +\gamma_m\sum_j\sqrt{r_j}\|\theta_j\|_2\right).
\end{aligned}
\end{equation}
Here $\|\cdot\|_2$ is the Euclidean norm of a vector.
The stage grid used $\alpha=0.5$,
$\gamma_m=2.5$, separate penalty fractions $0.1$ or $0.01$ for the two
stages, and $\kappa_2\in\{1,1.5,2\}$.

\paragraph{Derivatives and updates.}
Differentiating the binary loss with respect to $\eta_i$ gives
$e^{\eta_i}/(1+e^{\eta_i})-t_i=\pi_i-t_i$.
Multiply by $\partial\eta_i/\partial c=1$,
$\partial\eta_i/\partial\beta=v_i$, or
$\partial\eta_i/\partial\theta=m_i$, and then sum the weighted records:
\[
\begin{aligned}
g_c&=U_h^{-1}\sum_i u_i(\pi_i-t_i),\\
g_\beta&=U_h^{-1}\sum_i u_i v_i(\pi_i-t_i)
                  +\lambda_h(1-\alpha)\beta,\\
g_\theta&=U_h^{-1}\sum_i u_i m_i(\pi_i-t_i)
                  +\lambda_h(1-\alpha)\gamma_m\theta.
\end{aligned}
\]
Take an ordinary gradient step for $c$. For the other coefficients,
first form $V_j=\beta_j-tg_{\beta_j}$ and
$T_j=\theta_j-tg_{\theta_j}$ using step size $t>0$, then use
\[
\begin{aligned}
\beta_j^{\rm new}
 &=\left(1-\frac{t\lambda_h\alpha\sqrt{d_j}}{\|V_j\|_2}\right)_+V_j,\\
\theta_j^{\rm new}
 &=\left(1-\frac{t\lambda_h\alpha\gamma_m\sqrt{r_j}}{\|T_j\|_2}\right)_+T_j.
\end{aligned}
\]
The notation $(a)_+=\max(a,0)$ keeps only a positive value; a zero
input block is assigned a zero output. The derivation of this block
shrinkage is given for M5 below and applies to these vector blocks too.
At the intercept-only solution, set
$\pi_0=\sum_i u_it_i/U_h$ and $c_0=\log\{\pi_0/(1-\pi_0)\}$.
With $g^0_{\beta_j}$ and $g^0_{\theta_j}$ denoting the loss gradients
there, the zero-block conditions give
\[
\lambda_{h,\max}=\max\left\{
\max_j\frac{\|g^0_{\beta_j}\|_2}{\alpha\sqrt{d_j}},\quad
\max_j\frac{\|g^0_{\theta_j}\|_2}{\alpha\gamma_m\sqrt{r_j}}
\right\}.
\]
An absent block type is omitted from the maximum. The selected stage
fraction multiplies this quantity, with a numerical floor of $10^{-8}$.
The solver halves a trial step until the objective does not increase
by more than $10^{-10}$, with at most 30 trials. It stops if no trial is
accepted or if relative objective change is below $10^{-5}$, or when
the supplied iteration cap is reached. The logistic calculation clips
linear predictors to $[-35,35]$ and floors probabilities at $10^{-12}$
before taking logarithms.

\paragraph{Why the class probabilities multiply.}
Let $D$ denote membership in the PD/AP group. The probability rule
$\Pr(A\cap D)=\Pr(D)\Pr(A\mid D)$ gives
$\Pr(\mathrm{AP})=qr$ and $\Pr(\mathrm{PD})=q(1-r)$, while
$\Pr(\mathrm{HC})=1-q$. Adding them gives
$qr+q(1-r)+(1-q)=1$. These are the fitted hierarchical probabilities;
the class weights and penalties affect their estimation. If a stage
has no usable columns or only one observed binary class, its code uses
the unweighted training proportion as a constant probability instead
of solving the regression problem.

\paragraph{M5: Adaptive elastic net regression.}
This regression approach used an initial fit to determine adaptive
penalties, with stronger selection penalties for predictors having smaller
initial contributions \cite{ZouZhang2009,Yuan2006}. It supported separate sensitivity calculations under
assumptions about missing covariates and, where applicable, missing outcome
labels. Its scenario fits were distinct from the fixed M2--M4 primary
probability ensemble.

\paragraph{M5 derivation, step 1: fix the completed-data draws.}
Use the training records, source blocks and HC reference convention
defined above for M1. Let $z_i^{(m)}\in\mathbb R^p$ be record $i$'s
encoded predictor vector in completed-data draw $m$, for
$m=1,\ldots,M$. Here $m$ indexes draws, not model names.
Observed values and missingness indicators are retained; missing value
coordinates are sampled by the procedure fitted on the training data in S1.4.
Each draw supplies one possible completion of each record.
Let $a_{im}\geq0$ be that draw's contribution for record $i$, with
$\sum_{m=1}^M a_{im}=1$. The MAR branch uses $a_{im}=1/M$;
the MNAR branch uses the fixed, normalized weights derived in S1.4.
Record weights $w_i>0$ balance classes, multiply AP contributions and,
in MissCO, incorporate diagnosis-observation weights. Set $W=\sum_iw_i$.
The draws, $a_{im}$ and $w_i$ are fixed during coefficient optimization.

For each draw replace $x_i$ in equation~\eqref{eq:shared-softmax} by
$z_i^{(m)}$ to obtain $\eta_{ik}^{(m)}$ and $p_{ik}^{(m)}$.
All draws use the same $b$ and $B$. The log-probability calculation in
equation~\eqref{eq:shared-nll} then gives
\begin{equation}\label{eq:m5-draw-loss}
\begin{aligned}
\ell_i^{(m)}(b,B)
&=-\log p_{i,y_i}^{(m)}\\
&=\log\{1+e^{\eta_{i1}^{(m)}}+e^{\eta_{i2}^{(m)}}\}
 -\sum_{k=1}^2Y_{ik}\eta_{ik}^{(m)}.
\end{aligned}
\end{equation}
First average the losses across possible completions of each record,
then average across records with their training weights:
\begin{equation}\label{eq:m5-loss}
L(b,B)=\frac{1}{W}\sum_{i=1}^n w_i
                  \sum_{m=1}^M a_{im}\ell_i^{(m)}(b,B).
\end{equation}
This normalization prevents simply increasing the number of draws
from increasing a record's total contribution.

\paragraph{Step 2: distinguish the fitting loss from prediction averaging.}
The loss in equation~\eqref{eq:m5-loss} averages negative logarithms.
For example, if two equally weighted completions give probabilities
0.9 and 0.1 to the recorded class, the fitting contribution is
\[
\frac{-\log(0.9)-\log(0.1)}{2}\simeq1.204,
\qquad
-\log\left\{\frac{0.9+0.1}{2}\right\}\simeq0.693.
\]
The function $-\log u$ has positive second derivative $1/u^2$ for
$u>0$, so it is convex (curves upward). Its value at an average is at
most the average of its values. Consequently,
\[
-\log\left(\sum_m a_{im}p_{i,y_i}^{(m)}\right)
\leq\sum_m a_{im}\{-\log p_{i,y_i}^{(m)}\}.
\]
M5 uses the right-hand side for fitting, while its predictions average
probabilities as described in step 8.

\paragraph{Step 3: calculate the adaptive penalty weights.}
An initial ridge fit uses the same draws and record/draw weights:
\begin{equation}\label{eq:m5-ridge}
(\widetilde b,\widetilde B)
\ \text{approximately minimize}\quad
L(b,B)+\frac{10^{-3}}{2}\|B\|_F^2.
\end{equation}
For each source calculate
\begin{equation}\label{eq:m5-adaptive}
\epsilon_n=n^{-1/2}=\frac{1}{\sqrt n},\qquad
\omega_j=(\|\widetilde B_j\|_F+\epsilon_n)^{-\gamma},\quad\gamma>0.
\end{equation}
Here $n$ is the number of labelled training records. Adding $\epsilon_n$
keeps the weight finite when an initial block is zero. A smaller
initial norm yields a larger $\omega_j$ and hence stronger selection
pressure. The final objective is
\begin{equation}\label{eq:m5-objective}
F(b,B)=L(b,B)+\lambda\left\{
\frac{1-\alpha}{2}\|B\|_F^2
+\alpha\sum_{j\in\mathcal J}\omega_j\sqrt{q_j}\|B_j\|_F
\right\}.
\end{equation}
Neither intercept is penalized. With the draws and adaptive weights
fixed, the objective is convex. At $\alpha=0$ it is ridge
regression, so $\gamma$ has no effect on the exact objective. At
$\alpha=1$ it is an adaptive group lasso. Intermediate values combine
the two penalties.

\paragraph{Step 4: differentiate the smooth part.}
For $k=1,2$, differentiate the logarithm in
equation~\eqref{eq:m5-draw-loss} using
$d\log u/du=1/u$ and $de^v/dv=e^v$:
\[
\frac{\partial\ell_i^{(m)}}{\partial\eta_{ik}^{(m)}}
=\frac{e^{\eta_{ik}^{(m)}}}
 {1+e^{\eta_{i1}^{(m)}}+e^{\eta_{i2}^{(m)}}}-Y_{ik}
=p_{ik}^{(m)}-Y_{ik}.
\]
This residual is predicted probability minus the observed class
indicator. The linear predictor has derivatives
$\partial\eta_{ik}^{(m)}/\partial b_k=1$ and
$\partial\eta_{ik}^{(m)}/\partial B_{\ell k}=z_{i\ell}^{(m)}$.
The chain rule (multiplying these derivatives) and the fixed weights
therefore give
\begin{equation}\label{eq:m5-loss-gradient}
\begin{aligned}
\frac{\partial L}{\partial b_k}
 &=\frac{1}{W}\sum_{i,m}w_i a_{im}(p_{ik}^{(m)}-Y_{ik}),\\
\frac{\partial L}{\partial B_{\ell k}}
 &=\frac{1}{W}\sum_{i,m}w_i a_{im}z_{i\ell}^{(m)}
          (p_{ik}^{(m)}-Y_{ik}).
\end{aligned}
\end{equation}
The shorthand $\sum_{i,m}$ sums over all $n$ records and $M$ draws.
Since $d(B_{\ell k}^2/2)/dB_{\ell k}=B_{\ell k}$, define the smooth
part $f=L+\lambda(1-\alpha)\|B\|_F^2/2$ and obtain
\[
g_{b_k}=\frac{\partial f}{\partial b_k}
       =\frac{\partial L}{\partial b_k},\qquad
g_{B_{\ell k}}=\frac{\partial f}{\partial B_{\ell k}}
       =\frac{\partial L}{\partial B_{\ell k}}
        +\lambda(1-\alpha)B_{\ell k}.
\]
These are the gradients used before block shrinkage. The block-norm
penalty is handled separately because its derivative is not unique
at a zero block.

\paragraph{Step 5: derive the block shrinkage update.}
At the current coefficients take a smooth gradient step with size
$t>0$: $b^{\rm new}=b-tg_b$ and $V_j=B_j-tg_{B_j}$.
For source $j$, write $c_j=\lambda\alpha\omega_j\sqrt{q_j}$.
The remaining update solves a small problem for a trial coefficient
block $A$:
\[
\min_A\left\{\frac{1}{2t}\|A-V_j\|_F^2+c_j\|A\|_F\right\}.
\]
The first term discourages moving far from the gradient proposal;
the second encourages a smaller block. Let $r=\|A\|_F$,
$v=\|V_j\|_F$, and
$\langle A,V_j\rangle=\sum_{\ell,k}A_{\ell k}(V_j)_{\ell k}$.
Expanding the square gives
\[
\|A-V_j\|_F^2=r^2+v^2-2\langle A,V_j\rangle.
\]
For fixed lengths $r$ and $v$, the largest possible inner product is
$rv$ (the Cauchy--Schwarz inequality). It is attained when the two
blocks point in the same direction. For $v>0$, choose
$A=(r/v)V_j$ and reduce the problem to
\[
\min_{r\geq0}\left\{\frac{(r-v)^2}{2t}+c_jr\right\}.
\]
For a positive solution its derivative is $(r-v)/t+c_j=0$, giving
$r=v-tc_j$. Enforcing $r\geq0$ gives $r=(v-tc_j)_+$, hence
\begin{equation}\label{eq:m5-prox}
B_j^{\rm new}=
\begin{cases}
0,&\|V_j\|_F\leq t\lambda\alpha\omega_j\sqrt{q_j},\\[2pt]
\left(1-\dfrac{t\lambda\alpha\omega_j\sqrt{q_j}}{\|V_j\|_F}\right)V_j,
&\|V_j\|_F>t\lambda\alpha\omega_j\sqrt{q_j}.
\end{cases}
\end{equation}
When $V_j=0$, setting $A=0$ minimizes both terms directly. This
piecewise form avoids division by zero and shows exactly when a
source is removed.

\paragraph{Step 6: derive the first penalty on the path.}
A subgradient describes all permissible slopes of a convex function
at a point where the ordinary derivative may not exist. At a zero
block, the subgradients of $\|B_j\|_F$ are exactly the blocks $U$
with $\|U\|_F\leq1$: the defining inequality is
$\|A\|_F\geq\langle U,A\rangle$ for every $A$. Cauchy--Schwarz
proves this when $\|U\|_F\leq1$; choosing $A=U$ shows why a norm
greater than one cannot satisfy it. Thus stationarity at $B_j=0$
requires some such $U$ with
\[
0=\nabla_{B_j}L+\lambda\alpha\omega_j\sqrt{q_j}\,U,
\quad\text{or equivalently}\quad
\|\nabla_{B_j}L\|_F\leq\lambda\alpha\omega_j\sqrt{q_j}.
\]
The ridge derivative is zero at this block. These are the zero-block
Karush--Kuhn--Tucker (KKT) conditions. They must hold together with
the other coefficients' stationarity conditions at an optimum.

To find a penalty for which every block can be zero, set $B=0$.
All draws then have the same probabilities $\pi_k$. The intercept
gradient becomes
\[
\frac{1}{W}\sum_{i,m}w_i a_{im}(\pi_k-Y_{ik})
=\pi_k-\frac{\sum_i w_iY_{ik}}{W},
\]
using $\sum_m a_{im}=1$. Setting it to zero gives
$\pi_k=\sum_iw_iY_{ik}/W$ and $b_k^0=\log(\pi_k/\pi_3)$.
Let $g_j^0=\nabla_{B_j}L(b^0,0)$. For $\alpha>0$, solve each
zero-block inequality for $\lambda$ and take the largest lower bound:
\begin{equation}\label{eq:m5-lambda}
\lambda_{\max}=\max_{j\in\mathcal J}
\frac{\|g_j^0\|_F}{\alpha\omega_j\sqrt{q_j}}.
\end{equation}
For the objective defined above, the intercept-only solution satisfies
the KKT conditions when $\lambda\geq\lambda_{\max}$.
For the MNAR branch, $g_j^0$ uses its shifted designs and its $a_{im}$.
The numerical calculation uses $\max(\omega_j,10^{-12})$ in the denominator.

The study configuration used
$\alpha\in\{0,0.25,0.5,0.75,0.9,1\}$,
$\gamma\in\{0.5,1,2\}$ and AP multipliers
$\kappa\in\{1,1.5,2\}$. Eighteen decreasing penalty fractions ran
from 1 to $10^{-3}$, approximately $10^{-3\ell/17}$ for
$\ell=0,\ldots,17$, using the stored rounded fractions.
For $\alpha=0$, equation~\eqref{eq:m5-lambda} is undefined and
ridge generally has no finite all-zero threshold. The code instead
uses 1 as the reference scale and multiplies it by the same fractions.
The study tuning routine starts the first point from the ridge
initializer and each later point from the preceding fitted point.
Scenario refits use a penalty floor of $10^{-8}$.
Ten draws were used during tuning and twenty for final fits.

\paragraph{Step 7: accept a step and decide when to stop.}
For the scenario-fit solver the initial proposed step is
\[
\begin{aligned}
A_X&=\frac{1}{M}\sum_m\frac{\sum_iw_i\|z_i^{(m)}\|_2^2}{W},\\
t_0&=\min\left\{1,\left[
\max\left\{\tfrac14\max(A_X,10^{-6})
               +\lambda(1-\alpha),10^{-8}\right\}\right]^{-1}\right\}.
\end{aligned}
\]
The MNAR solver uses equal draw weights to propose its initial step and
the MNAR draw weights to evaluate the objective and gradient.

The solver evaluates $F(b^{\rm new},B^{\rm new})$ after the update
in step 5. If this is finite and at most the preceding objective plus
$10^{-10}$, the trial is accepted. Otherwise $t$ is halved and the
trial is repeated, up to 40 trials. If all fail, fitting ends with a
line-search warning and the last accepted coefficients. After an
accepted trial, the next proposed step is $\min(1,1.05t)$.
Let $F_{\rm old},F_{\rm new}$ be consecutive objectives, and let
$\Delta_{\rm par}$ be the largest absolute change in any intercept
or coefficient. Convergence requires both
\[
\frac{|F_{\rm old}-F_{\rm new}|}{\max(1,|F_{\rm old}|)}<10^{-5},
\qquad \Delta_{\rm par}<10^{-4}.
\]
Scenario fits allow at most 500 iterations. Their initial ridge fits
allow at most 250.
A block counts as selected when $\|B_j\|_F>10^{-8}$.
S3.1 describes the ridge initializer used to calculate Q.

For numerical evaluation, the code subtracts the largest linear
predictor before exponentiating. This leaves softmax probabilities
unchanged because numerator and denominator are divided by the same
positive number. The implementation floors the recorded-class probability
at $10^{-15}$ when evaluating the log loss. It uses $p-Y$ for the
gradient even when the floor is active.

\paragraph{Step 8: predict and identify the estimation procedure.}
For a new record, the imputer fitted on the training data generates completed
vectors, and its scenario rule supplies $a_{im}$. The pre-calibration
probability is
\begin{equation}\label{eq:m5-predict}
p_{\mathrm{M5},k}(x_i)=\sum_{m=1}^M
 a_{im}p_k(z_i^{(m)};\widehat b,\widehat B).
\end{equation}
It sums to one over classes because every draw's probabilities sum
to one and $\sum_m a_{im}=1$. The implementation additionally floors
entries at $10^{-15}$ and renormalizes, before any separately fitted
probability-calibration mapping. When there are no eligible
design columns, the scenario routine instead returns the weighted
training class proportions.

This is fixed-draw, imputation-averaged penalized regression.

\section*{S1.2\quad Data preparation and internal evaluation}

The released dataset contained 3,541 unique participants. We excluded 107
participants who could not be assigned an eligible three-class outcome under the diagnostic hierarchy, leaving an
analytic cohort of 3,434 participants (2,853 PD, 171 AP and 410 HC).
Assessment budgets J1--J4 permitted predictors from implementation levels
1 through the budget's number. Thus the primary budget J3 permitted
levels 1--3. J4 also allowed level 4 assessments.
Predictor eligibility, preprocessing, fitting, tuning and probability
calibration were performed using the designated training records.

We linked assessment records using a de-identified participant identifier
and combined comparable measurements. For overlapping entries, we used the
first available value in a fixed priority order. Administrative fields and
variables derived from the recorded diagnosis were excluded from the eligible
predictors. Observed zeros were retained as values, and unexplained blanks
remained missing.

The evaluation used three repeats of five outer folds. Within each outer
split, the approximate allocation was 60\% for fitting, tuning and
probability calibration, 20\% for conformal calibration and 20\% for
held-out testing. Four inner folds supported model tuning and permitted
out-of-fold development predictions \cite{Varma2006}.

\paragraph{Recorded diagnosis and three-class outcome.}
A recognized final diagnosis determined the diagnostic class.
Parkinson's disease mapped to PD. The AP study grouping included progressive
supranuclear palsy, multiple system atrophy, corticobasal syndrome, dementia
with Lewy bodies, frontotemporal dementia, essential tremor and REM sleep
behaviour disorder. When a recognized final diagnosis was unavailable,
the recorded enrolment classification supplied the class. Enrolment
categories recorded as PD, AP/PD-plus and healthy control mapped to PD,
AP and HC, respectively. Records with neither a recognized final diagnosis
nor a usable enrolment classification were excluded from the three-class
analysis. Performance was evaluated for the pooled AP category.

\paragraph{Primary probability ensemble.}
For the fixed clinical ensemble, let $x$ denote one participant's predictor
information at the assessment budget being evaluated. Let
$\mathcal{K}=\{\mathrm{PD},\mathrm{AP},\mathrm{HC}\}$ be the set of three
diagnostic labels, and let $k\in\mathcal{K}$ identify one of those labels:
PD denotes Parkinson's disease, AP the atypical Parkinsonian category,
and HC healthy control. The quantities $\widetilde p_{\mathrm{M2},k}(x)$
and $\widetilde p_{\mathrm{M4},k}(x)$ are the probabilities assigned to
label $k$ by M2 and M4, respectively, after separate probability
calibration. Their fixed weighted average,
$\widehat p_{\mathrm{ens},k}(x)$, is the primary ensemble's probability
for that label:
\begin{equation*}
\widehat p_{\mathrm{ens},k}(x)
=0.60\,\widetilde p_{\mathrm{M2},k}(x)
+0.40\,\widetilde p_{\mathrm{M4},k}(x),
\qquad k\in\mathcal{K}.
\end{equation*}
The subscript ens identifies the ensemble; the same probability is denoted
$\widehat p_{\mathrm{primary},k}(x)$ in the scenario library in S1.5.
The candidate mappings, fitting procedure and selection rule are detailed
in \hyperref[s1:probability-calibration]{Section~S1.3}. Probability calibration parameters were estimated from
development predictions; the separate conformal calibration sample was
reserved for prediction set thresholds.

\section*{S1.3\quad Probability calibration methods}\label{s1:probability-calibration}

\paragraph{Purpose and inputs.}
Probability calibration maps a model's three-class probability vector
to revised probabilities intended to align better with observed outcomes.
For a raw probability vector $p=(p_1,p_2,p_3)$ from one model and record,
the numeric indices 1, 2 and 3 correspond to PD, AP and HC in that order.
Within this section's calibration formulas, $k\in\{1,2,3\}$ indexes
a class component of a vector. Probability and score vectors are treated
as column vectors. The symbols $\log$ and $\exp$ denote the natural
logarithm and exponential functions, respectively.
Each component was clipped to $[\epsilon,1-\epsilon]$ with
$\epsilon=10^{-6}$ and then renormalized to sum to one; denote the
result by $\bar p$. Here $\epsilon$ is a numerical bound that avoids
zero probabilities when taking logarithms. Write $z=(z_1,z_2,z_3)$ with
$z_k=\log\bar p_k$. For any vector $v=(v_1,v_2,v_3)$ with three components, define
$\operatorname{softmax}(v)_k=\exp(v_k)/\sum_{j=1}^{3}\exp(v_j)$;
$j\in\{1,2,3\}$ is a summation index over the three classes. Softmax converts these
three real-valued scores to a probability vector.

\paragraph{Four candidate mappings.}
The four prespecified alternatives \cite{Guo2017,Kull2019} were:
\begin{align*}
\text{No adjustment:}\quad &g_{\mathrm{none}}(p)=\bar p,\\
\text{Temperature scaling:}\quad &g_{\mathrm{temp}}(p)
  =\operatorname{softmax}(z/T),\quad T>0,\\
\text{Vector scaling:}\quad &g_{\mathrm{vector}}(p)
  =\operatorname{softmax}(a\odot z+b),\\
\text{Dirichlet calibration:}\quad &g_{\mathrm{Dir}}(p)
  =\operatorname{softmax}(Wz+b).
\end{align*}
Here $g$ denotes a probability calibration mapping, and its subscript
identifies the candidate family. The fitted scalar $T$ is the temperature;
$a=(a_1,a_2,a_3)$ contains class-specific slopes and
$b=(b_1,b_2,b_3)$ contains intercepts. The symbol $\odot$ denotes
componentwise multiplication, so $(a\odot z)_k=a_kz_k$.
The fitted $3\times3$ matrix $W$ contains the Dirichlet mapping's slopes:
its row identifies the output class and its column the input log
probability. Intercepts are fitted separately for the vector and Dirichlet
families. No adjustment retains
the original probabilities after numerical normalization. Temperature
scaling uses one parameter to soften ($T>1$) or sharpen ($T<1$) the
overall distribution without changing its class ordering. Vector scaling
allows each class its own slope and intercept on log probabilities.
Dirichlet calibration additionally allows one class's original
probability to affect another class's calibrated score through the
off-diagonal entries of $W$.

\paragraph{Fitting and method selection.}
Within each outer split and assessment budget, the M2 and M4 base model
predictions for the development records used for proper fitting were obtained out
of fold using four inner folds. Each mapping was fitted separately for
each ensemble component using these predictions and their observed
diagnostic labels. Let $n$ be the number of development records used to fit
a mapping, and let $i\in\{1,\ldots,n\}$ index those records.
Write $p^{(i)}$ for record $i$'s raw three-class probability vector and
$y_i\in\{1,2,3\}$ for its recorded class code in PD, AP, HC order.
The objective $\mathcal L(g)$ was the
unweighted multiclass log loss of mapping $g$,
\begin{equation*}
\mathcal L(g)=-\frac{1}{n}\sum_{i=1}^{n}
\log\left\{\overline{g(p^{(i)})}_{\,y_i}\right\},
\end{equation*}
where the overbar denotes the same clipping and renormalization applied
to transformed probabilities, and the subscript $y_i$ selects the
probability assigned to record $i$'s recorded class.
Temperature scaling
optimized $\log T\in[-5,5]$. Vector scaling and Dirichlet calibration
used ridge penalties of $10^{-2}\|a-\mathbf 1\|_2^2$ and
$10^{-2}\|W-I\|_F^2$, respectively, favouring an identity mapping;
intercepts were centred to have mean zero. Here $\mathbf 1=(1,1,1)$ is
the vector of ones, $I$ is the $3\times3$ identity matrix,
$\|\cdot\|_2$ is the Euclidean vector norm and $\|\cdot\|_F$ is the
Frobenius matrix norm; the squared norms sum the squared entries of
the indicated vector or matrix.

For each component, candidate mappings were ranked by their unweighted
log loss on the same out-of-fold development predictions used to fit
the calibration parameters. If losses differed by at most $10^{-8}$,
the simpler family was preferred in the order no adjustment,
temperature, vector, then Dirichlet.

Dirichlet calibration was selected for both primary components in
all 15 outer splits under J3 and independently in all 15 under J4.
We applied each fitted map unchanged to its model's probabilities and
combined the results with fixed weights 0.60 for M2 and 0.40
for M4. Blended probabilities were normalized as required.

\paragraph{Difference from conformal calibration.}
Probability calibration modifies numerical class probability estimates.
Conformal calibration instead used the separate conformal calibration
records to determine class-specific cutoffs for the nonconformity score
$1-\widehat p_k(x)$. Here $x$ denotes a participant's predictor
information and $\widehat p_k(x)$ is the final predicted probability
for class $k$, with the class codes defined above.
Those cutoffs determine which PD, AP and HC labels
are included in a prediction set.

\section*{S1.4\quad M5 missing data sensitivity procedures}

\textbf{MissC} denotes missing predictors with observed diagnostic labels;
\textbf{MissCO} denotes missing predictors together with missing diagnostic
labels. For MissCO, inverse probability weighting (IPW) assigned greater
weight to records whose observed outcomes were less likely to have been
recorded, under the fitted observation model. Missing at random (MAR)
scenarios allowed the recording process to depend on observed information.
Missing not at random (MNAR) sensitivity scenarios additionally allowed
prespecified dependence on unobserved information.

M5 screened predictors within the permitted training records. Sources needed
at most 50\% missing entries and observed counts of at least 10 for PD, 1 for
AP and 5 for HC. Preprocessing used a minimum category count of 5, at most
40 categories, a proportion of values parseable as numeric of at least 0.90, winsorization at
the observed training 1st and 99th percentiles, and a quasi-constant threshold
of 0.995. Definitions of missingness states and the
separate deficit penalties used for Q are given in Supplement S3.

For the M5 sensitivity analysis, missing values were drawn from the
observed training distribution for each predictor. The original missingness
indicators were retained.
The final information deficit score $Q$ measures weighted missing predictor
information on a scale from 0 to 1, with larger values indicating more
missing information under the chosen rule. It drew its predictor
weights from the selected MAR MissC fits. Supplement S3 gives its
construction and grouping procedures.

\paragraph{The numerical scenario grid.}
The three MNAR MissC scenarios used a common shift parameter for numerical imputation
$\delta\in\{0,-0.5,+0.5\}$. For measurements treated as numeric by the
imputation procedure, $\delta$ was added to imputed missing entries on the
standardized training data scale. The scale used the mean and standard
deviation of winsorized observed training values. The procedure also
reweighted the completed data draws for each participant: a draw's weight
was proportional to the exponential of $\delta$ times the sum of its
standardized values before the shift at missing numeric entries. These weights
were normalized across draws. At $\delta=0$, the draws had equal weight;
nonzero $\delta$ changed both their values and their relative contributions.
Categorical tilting was disabled. Numerically coded variables could be
treated as numeric, including some variables with discrete values.

The fifteen MNAR MissCO-IPW scenarios crossed these three $\delta$ values
with five settings for diagnosis observation. Here $k$ is a diagnostic label
in $\{\mathrm{PD},\mathrm{AP},\mathrm{HC}\}$. Write $\zeta_k$ for the
class-specific adjustment to the fitted log odds of a diagnosis being
observed; the settings were
\begin{equation*}
(\zeta_{\mathrm{PD}},\zeta_{\mathrm{AP}})
\in\{(0,0),(0,+0.5),(0,-0.5),(+0.5,0),(-0.5,0)\},
\qquad \zeta_{\mathrm{HC}}=0.
\end{equation*}
For a diagnosed training record in class $k$, $\zeta_k$ was added to the
log odds of its fitted probability of diagnosis observation. Log odds means
the logarithm of the probability divided by one minus that probability.
A positive value
increased the assumed observation probability and reduced its inverse
weight; a negative value did the reverse, subject to clipping. The
adjusted observation probability was floored at 0.05, limiting the IPW
component to 20 before multiplication by the other training weights.
The all-zero setting was the reference within each MNAR family. MissC
used zero adjustments for diagnosis observation throughout.

\paragraph{Completed predictors, step 1: learn the sampling distributions.}
All quantities in this paragraph are estimated within the permitted
training records. For a retained numeric source $j$, let $O_j$ be its
records with usable numeric values. Clip each observed value to the
training 1st and 99th percentiles, obtaining $v_{hj}$ for $h\in O_j$.
With $N_j=|O_j|$, calculate
\[
\mu_j=\frac{1}{N_j}\sum_{h\in O_j}v_{hj},\qquad
s_j=\sqrt{\frac{1}{N_j-1}\sum_{h\in O_j}(v_{hj}-\mu_j)^2}.
\]
The implementation substitutes $s_j=1$ if this scale is nonfinite
or at most $10^{-12}$. An observed numeric entry is clipped to the
same limits and standardized as $(v_{ij}-\mu_j)/s_j$.
For a missing numeric entry in draw $m$, select one index $H_{ijm}$
uniformly, with replacement, from $O_j$, and use
\[
z_{ij,\mathrm{value}}^{(m)}
=\frac{v_{H_{ijm},j}-\mu_j}{s_j}.
\]
Thus every observed training value, including repeated values, is
equally likely to be sampled. The accompanying indicator remains
$R_{ij}=1$ for a missing entry and $0$ for an observed entry.
For numeric sources, failure to parse a value as numeric also produces
a missing value coordinate and indicator in this encoding.

For a categorical source, let $n_{jc}$ be the training count in mapped
category $c$, after rare levels are pooled into the other category.
The sampled category $C_{ij}^{(m)}$ satisfies
$\Pr(C_{ij}^{(m)}=c)=n_{jc}/\sum_d n_{jd}$.
The output contains one indicator column per encoded category, plus
the original missingness indicator. Retained levels normally have
at least five observations, up to forty retained levels; if none
meet the count rule, the most frequent levels up to that limit are
used. The other-category column is retained even if its training
probability is zero. An unseen observed category maps to other.

\paragraph{Step 2: shift and reweight the MNAR draws.}
Let $\mathcal N_i$ be the retained numeric sources with missing value
coordinates for record $i$. Write $z_{ij}^{(m,0)}$ for the standardized
sampled value before a sensitivity shift. The shifted value and the
draw's tilt score are
\begin{equation}\label{eq:m5-tilt}
z_{ij}^{(m,\delta)}=z_{ij}^{(m,0)}+\delta
\quad (j\in\mathcal N_i),\qquad
h_{im}=\delta\sum_{j\in\mathcal N_i}z_{ij}^{(m,0)}.
\end{equation}
Observed value coordinates, category coordinates and missingness
indicators are unchanged by this tilt. The sum uses the values
\emph{before} the shift. To convert scores to stable weights, let
$h_i^{\max}=\max_m h_{im}$, subtract it, and compute
\begin{equation}\label{eq:m5-tilt-weights}
\bar h_{im}=\max\{-40,h_{im}-h_i^{\max}\},\qquad
a_{im}=\frac{\exp(\bar h_{im})}{\sum_{r=1}^M\exp(\bar h_{ir})}.
\end{equation}
The centred tilt scores are floored at $-40$ before exponentiation.
The denominator makes the weights sum to one for each record.
At $|\delta|\leq10^{-15}$ the code explicitly
assigns $a_{im}=1/M$. A record without missing numeric sources also
has equal draw weights, even when $\delta\ne0$.

\Needspace{7\baselineskip}
For illustration, take two draws with one missing standardized value,
$-1$ and $+1$, and $\delta=0.5$. The shifted values are $-0.5$ and
$1.5$, while the uncentred tilt scores are $-0.5$ and $+0.5$.
Their weights are
$1/(1+e)\simeq0.269$ and $e/(1+e)\simeq0.731$.
The weighted shifted mean is therefore
$0.269(-0.5)+0.731(1.5)\simeq0.962$, although the original mean
was zero and the direct shift was only 0.5.
Both operations enter the fitting loss and prediction average in S1.1.

\paragraph{Diagnosis observation, step 3: fit the observation probability.}
For MissCO let $C_i=1$ indicate an available modelled diagnosis and
$C_i=0$ an unavailable one. The observation model uses the labelled
training records and the supplied records with missing outcomes. Its
predictors are the original recording indicators for the retained
sources. Standardize each indicator using
its mean and sample standard deviation in this observation-model
sample; replace a nonfinite or near-zero standard deviation by one.
Write the resulting vector as $o_i$ and the sample size as $N$.
The fitted observation probability has the form
\[
\rho_i=\sigma(a+o_i^\top d),
\]
where $a$ is an unpenalized intercept and $d$ a vector of slopes.
The observation model minimizes the binary logistic criterion
\begin{equation}\label{eq:observation-objective}
J(a,d)=-\frac{1}{N}\sum_{i=1}^N
\{C_i\log\rho_i+(1-C_i)\log(1-\rho_i)\}
+\frac{\lambda_o}{2N}\|d\|_2^2,\qquad\lambda_o=1.
\end{equation}
Multiplication by $N$ gives an equivalent sum-loss criterion with
ridge term $\lambda_o\|d\|_2^2/2$. The binary derivative already
derived for M4 gives
\[
\frac{\partial J}{\partial a}=N^{-1}\sum_i(\rho_i-C_i),\qquad
\nabla_dJ=N^{-1}\sum_i o_i(\rho_i-C_i)+\lambda_o d/N.
\]
The implementation takes gradient steps using
$[\tfrac14\max(N^{-1}\sum_i\|o_i\|_2^2,10^{-6})+\lambda_o]^{-1}$,
for at most 300 iterations, stopping when relative objective change
is below $10^{-6}$. Its sigmoid clips linear predictors to
$[-35,35]$, and its logged probabilities are clipped to
$[10^{-12},1-10^{-12}]$. If there are no observation features or
only one value of $C$, it returns the constant sample observation
rate clipped to $[10^{-4},1-10^{-4}]$.

\paragraph{Step 4: add diagnosis-specific sensitivity adjustments.}
For a labelled training record, clip its fitted observation probability
to $\rho_i^c\in[10^{-12},1-10^{-12}]$. Its log odds, adjusted log
odds and adjusted probability are
\begin{equation}\label{eq:observation-tilt}
\begin{aligned}
l_i&=\log\{\rho_i^c/(1-\rho_i^c)\},\\
l_i^\zeta&=\max\{-35,\min(35,l_i+\zeta_{y_i})\},\\
\rho_i^\zeta&=\sigma(l_i^\zeta),\qquad
\rho_i^*=\max(0.05,\rho_i^\zeta).
\end{aligned}
\end{equation}
MAR MissCO applies the floor directly to the unadjusted fitted
$\rho_i$. In the MNAR branch, adding $\zeta_{y_i}$ multiplies the
unclipped observation odds by $e^{\zeta_{y_i}}$, because
$e^{l_i+\zeta_{y_i}}=e^{l_i}e^{\zeta_{y_i}}$.
The class-specific adjustment uses the training diagnosis to calculate that record's weight.

\paragraph{Step 5: combine all record weights.}
Let $n$ and $n_k$ again count labelled training records and records
in class $k$, and let $\kappa$ be the AP multiplier. The weights
supplied to the initial ridge fit and the adaptive fit are
\begin{equation}\label{eq:m5-record-weights}
w_i=
\underbrace{\frac{n}{3n_{y_i}}}_{\text{class balance}}
\underbrace{\kappa^{\mathbf 1[y_i=\mathrm{AP}]}}_{\text{AP multiplier}}
\times
\begin{cases}
1,&\text{MissC},\\
1/\rho_i^*,&\text{MissCO-IPW}.
\end{cases}
\end{equation}
The observation-probability floor caps the IPW factor at 20.
For example, $n=30$, $n_{\mathrm{AP}}=3$, $\kappa=1.5$ and
$\rho_i^*=0.05$ give an AP weight
$(30/9)\times1.5\times20=100$. A participant's draw-$m$ contribution
to equation~\eqref{eq:m5-loss} is $w_i a_{im}/W$. Summing these
contributions over draws gives $w_i/W$, and then over records gives one.

\paragraph{Expected loss under inverse probability weighting.}
Suppose $\rho(V)=\Pr(C=1\mid V)$
is the true, positive observation probability given available
information $V$, and suppose $C$ is conditionally independent of the
full-data outcome $Y$ and loss $H$ given $V$ \cite{Robins1994}. For fixed fitted parameters,
\[
\begin{aligned}
\mathbb E\{CH/\rho(V)\mid V,H\}
 &=\frac{H}{\rho(V)}\mathbb E(C\mid V,H)
 =\frac{H}{\rho(V)}\rho(V)=H,\\
\mathbb E\{CH/\rho(V)\}&=\mathbb E(H).
\end{aligned}
\]
Thus ideal inverse weighting compensates for selective outcome
observation under these assumptions. The implemented criterion also
uses estimated, floored probabilities and a class/AP factor
$c(Y)=n\kappa^{\mathbf 1[Y=\mathrm{AP}]}/(3n_Y)$.
Treating that factor as fixed, ideal inverse weighting targets the
normalized weighted loss $\mathbb E\{c(Y)H\}/\mathbb E\{c(Y)\}$.

\section*{S1.5\quad Conformal prediction sets and scenario aggregation}

\paragraph{Class-specific score thresholds.}
In this section, $x$ denotes one participant's predictor information,
$\mathcal{K}=\{\mathrm{PD},\mathrm{AP},\mathrm{HC}\}$ is the set of
diagnostic labels, and $k\in\mathcal{K}$ identifies a candidate label.
For a fixed prediction procedure, let $\widehat p_k(x)$ be its final
probability assigned to that label. The least ambiguous classification
(LAC) score $s_k(x)$ was $s_k(x)=1-\widehat p_k(x)$; a smaller score
means greater probability assigned to the label \cite{Sadinle2016,Angelopoulos2021}.
Conformal calibration records were kept separate from model fitting,
tuning and probability calibration. Let $n_k$ be the number of calibration
records with recorded diagnosis $k$, and order their scores as
$s_{k,(1)}\leq\cdots\leq s_{k,(n_k)}$; $s_{k,(j)}$ denotes the
$j$th smallest score among those class-$k$ records, with
$j\in\{1,\ldots,n_k\}$.
Let $\alpha_k$ be the target miscoverage rate for class $k$, so the
target coverage is $1-\alpha_k$. The cutoff $q_k$ was
$s_{k,(r_k)}$, the score with rank $r_k$, whenever that rank was available
and the minimum class count was met, where
\begin{equation*}
r_k=\left\lceil(n_k+1)(1-\alpha_k)\right\rceil.
\end{equation*}
The ceiling symbol $\lceil\cdot\rceil$ rounds up to the smallest integer
at least as large as its argument.
The cutoff was $+\infty$ when there were no calibration records in the
class, the requested rank exceeded $n_k$, or the minimum class count was
not met. The minimum count was 19 for AP, with no additional minimum for
PD or HC. An infinite cutoff automatically retained the affected label.
The targets were $\alpha_{\mathrm{PD}}=\alpha_{\mathrm{HC}}=0.10$ and
$\alpha_{\mathrm{AP}}=0.05$. This separate calibration within each recorded
class is also called label-conditional or Mondrian calibration. The same
rank rule was used for the combined scores in A3, with newly calculated
cutoffs for those scores.

\paragraph{Scenario library and A0--A2.}
The 21 scenarios were grouped into five families: the primary M2/M4
ensemble (one scenario), M5 MAR MissC (one), M5 MAR MissCO-IPW (one),
M5 MNAR MissC (three), and M5 MNAR MissCO-IPW (fifteen). The numerical
MNAR grid is specified in S1.4. The index $s\in\{1,\ldots,21\}$
identifies a scenario.
The quantity $\widehat p_{sk}(x)$ denotes scenario $s$'s final
probability of class $k$ for record $x$. The primary scenario uses
$\widehat p_{\mathrm{primary},k}(x)$, the calibrated M2/M4 ensemble
probability: 0.60 times M2's separately calibrated probability for label $k$
plus 0.40 times M4's. This is the quantity denoted
$\widehat p_{\mathrm{ens},k}(x)$ in S1.2. The symbols A0--A5 name the
six reporting policies; $A0(x),\ldots,A5(x)$ denote the sets of diagnostic
labels they return for record $x$.
A0 returned the singleton containing the highest-probability class
from this primary vector.
A1 applied the class-specific LAC cutoffs calibrated for this same primary
ensemble.

For scenario $s$, let $q_{sk}$ be its own conformal calibration cutoff
for class $k$, obtained by the rank rule above using that scenario's
scores. Its initial prediction set was
$C_s(x)=\{k\in\mathcal{K}:1-\widehat p_{sk}(x)\leq q_{sk}\}$:
the labels whose scores did not exceed their cutoffs.
If empty, it was enlarged with that scenario's highest-probability label;
write the resulting nonempty set as $\widetilde C_s(x)$.
The subscript primary in a set identifies the primary
ensemble scenario, one of the 21 scenarios.
A1 returned $\widetilde C_{\mathrm{primary}}(x)$ and
$A2(x)=\bigcup_{s=1}^{21}\widetilde C_s(x)$.
Thus A2 combined already calibrated sets, with each scenario's own cutoffs.
The union symbol $\bigcup$ retains every label appearing in at least one
of these 21 sets.

\paragraph{A3: aggregate scores, then calibrate.}
Let $f\in\{1,\ldots,5\}$ index the scenario families, $n_f$ be the
number of scenarios in family $f$, and $f(s)$ identify the family
containing scenario $s$. Each family received total weight $1/5$.
Scenario $s$ had weight $w_s=1/(5n_{f(s)})$, giving weights $1/5$ for each
of the three single-scenario families, $1/15$ for each MNAR MissC scenario,
and $1/75$ for each MNAR MissCO-IPW scenario. For each record and candidate
label, the combined score was the weighted empirical upper quantile
\begin{equation*}
a_k(x)=\inf\left\{t:\sum_{s=1}^{21}w_s
\mathbf{1}\bigl[1-\widehat p_{sk}(x)\leq t\bigr]\geq\rho\right\},
\qquad \rho=0.90.
\end{equation*}
Here $a_k(x)$ is the combined score for class $k$.
The symbol $t$ is a candidate
score threshold, and $\mathbf 1[\cdot]$ equals 1 when its condition holds
and 0 otherwise. The scenario weights $w_s$ sum to one. The infimum
$\inf$ selects the smallest threshold at which the cumulative weight of
scores at or below that threshold reaches the quantile level $\rho$.
Scores were sorted in ascending order and the first score
whose cumulative weight reached $\rho$ was selected, using a numerical
tolerance of $10^{-15}$. Each family contributes one fifth of the total
weight regardless of its size. The quantile $\rho$ controls aggregation
across scenarios; the class targets determine the conformal cutoffs.

The same combined scores were calculated for conformal calibration
records. Applying the rank rule within each recorded class gave new
cutoffs $q_k^{A3}$ for the combined scores $a_k(x)$. The initial A3 set was
$\{k\in\mathcal{K}:a_k(x)\leq q_k^{A3}\}$. If empty, it was enlarged with
\begin{equation*}
k^*(x)=\operatorname*{arg\,min}_{k\in\mathcal{K}}
\frac{a_k(x)}{q_k^{A3}+10^{-12}}.
\end{equation*}
Here $\operatorname*{arg\,min}$ returns the label $k^*(x)$ attaining the
smallest combined score relative to its own cutoff; $10^{-12}$ is a
numerical stabilizer in the denominator. This fallback can differ from
the primary ensemble's highest-probability label. Ties in this
fallback, or in the fallback to a scenario's most probable label,
were resolved in PD, AP, HC order.

\Needspace{10\baselineskip}
\paragraph{A4--A5: costs and enlargement.}
Let $y\in\mathcal{K}$ denote a possible recorded diagnostic class, and
let $H$ be the matrix of study-defined misclassification penalties.
Its entry $H_{yk}$ is the penalty for reporting label $k$ when the
recorded class is $y$. Rows and columns follow PD, AP, HC order:
\begin{equation*}
H=\begin{pmatrix}0&3&5\\4&0&6\\1&1&0\end{pmatrix}.
\end{equation*}
For a nonempty candidate set $S\subseteq\mathcal{K}$, let $|S|$ be its
number of diagnostic labels. The cost $L(y,S)$ if the recorded class
were $y$ was
$L(y,S)=b(|S|)+\min_{k\in S}H_{yk}$, with penalties for set size
$b(1)=0$, $b(2)=0.5$ and $b(3)=1$.
The minimum selects the smallest penalty among labels in $S$.
The maximum expected cost across scenarios $R(S\mid x)$ was
\begin{equation*}
R(S\mid x)=\max_{s=1,\ldots,21}
\sum_{y\in\{\mathrm{PD},\mathrm{AP},\mathrm{HC}\}}
\widehat p_{sy}(x)L(y,S).
\end{equation*}
Thus candidate sets were compared using the same scenario probabilities
that supplied the LAC scores. The summation averages $L(y,S)$ over
possible recorded classes $y$, weighted by $\widehat p_{sy}(x)$,
the probability that scenario $s$ assigns to class $y$ for record $x$;
the maximum then selects the largest of these 21 expected costs.

A singleton contains one label, a pair contains two, and the full set is
$\mathcal{K}$. A superset retains every label in the starting set; a
strict superset also contains at least one additional label.
A4 retained A3 pairs and three-label sets. An A3 singleton was retained
only if its fitted Support Reliability Score (SRS) was finite, a valid
fitted general threshold was available, and SRS met that threshold.
For a finite score below the threshold, or an invalid threshold, A4
selected the strict superset with the lowest $R(S\mid x)$. A nonfinite
singleton score led to the full three-label set. SRS is one minus the
fitted probability that the primary ensemble's highest-probability label
is incorrect, multiplied by 100 to give a score on a scale from 0 to 100;
larger values indicate lower estimated error risk. A nonfinite score
means a missing or numerically invalid value. Supplement S2.2 describes
the error model. The thresholds were fitted
using out-of-fold predictions from the proper fitting records for general
errors in the primary ensemble's most probable labels. Development criteria used an overall error fraction at
most 0.10, AP precision at least 0.40 and an HC error fraction at most 0.025.

A5 compared A4 with all its supersets and selected the set minimizing
$R(S\mid x)$. HC singletons additionally required valid HC thresholds,
sufficient SRS, sufficiently low maximum combined PD and AP probability across
scenarios, and validation that the record was not out of distribution (non-OOD).
Validated OOD status was unavailable in this study, so an HC singleton
was always restricted to strict supersets. Any nonfinite SRS in A5
led to the full set. An AP singleton could persist only after meeting
the A4 reliability criteria. Cost ties favored smaller sets, then
the fixed set order PD, AP, PD/AP, HC, PD/HC, AP/HC, PD/AP/HC.
A4 retains or enlarges the A3 set, and A5 retains or enlarges the A4
set: $A3(x)\subseteq A4(x)\subseteq A5(x)$.

\paragraph{Validity scope.}
For a scoring rule fixed before conformal calibration, exchangeability of
calibration and test observations within a diagnostic class gives marginal
coverage of at least $1-\alpha_k$ for that class in the same population
\cite{Sadinle2016,Angelopoulos2021}. S4.3 reports observed coverage in
the highest information deficit group.

\paragraph{Why the rank rule gives coverage.}
Fix one class $k$ and a scoring rule trained without conformal
calibration outcomes \cite{Sadinle2016,Angelopoulos2021}. Under the within-class exchangeability
assumption stated above, the $n_k$ calibration scores and the next
class-$k$ test score are interchangeable. With no ties, the test
score's rank $R$ among these $n_k+1$ scores is equally likely to be
any integer from 1 to $n_k+1$. For an available calibration rank
$r_k\leq n_k$, the test score is at most the calibration cutoff
$s_{k,(r_k)}$ exactly when $R\leq r_k$. Therefore
\[
\Pr\{k\in C(X)\mid Y=k\}
=\frac{r_k}{n_k+1}
\geq\frac{(n_k+1)(1-\alpha_k)}{n_k+1}
=1-\alpha_k.
\]
Here $X,Y$ denote a new record and its class, and $C(X)$ the initial
prediction set. The inequality follows from rounding the requested
rank upward. With tied scores, a hypothetical random ordering of
ties still gives the rank argument, while including every score equal
to the cutoff can only add labels. Coverage is therefore at least
this level. If the rank is unavailable or the minimum count rule
sets $q_k=+\infty$, that class is always included and its coverage
is one. For AP, $n_k=18$ requests
$\lceil19(0.95)\rceil=19>18$; $n_k=19$ requests
$\lceil20(0.95)\rceil=19$, a finite cutoff at the largest score.

The same reasoning applies to A3's fixed aggregate scoring function,
using its own calibration scores and newly calculated cutoffs.
For any sets $C(X)\subseteq D(X)$, the record-wise indicator satisfies
\[
\mathbf 1\{Y\in C(X)\}\leq\mathbf 1\{Y\in D(X)\}.
\]
Taking conditional probabilities preserves the inequality. This
proves that empty-set fallback, the A2 union and the enlargements from A3 to A4
and from A4 to A5 cannot reduce inclusion relative to the sets they
contain.


\renewcommand{\theequation}{S2.\arabic{equation}}
\allowdisplaybreaks[1]
\maketitle

\section*{S2.1\quad Prediction set summaries and scenario variation}

The returned prediction set is the core set. Class coverage is the number
of prediction records whose set contains the recorded class divided by the
number of records in that class. Mean size is the average number of labels
returned. The frequency of full sets is the fraction receiving all three labels.
The mention set is the union of the core set and A2; labels in that union
but outside the core are included only for sensitivity analysis.
Supplement S1.5 defines the study-specific loss matrix, penalties for set size
and maximum expected cost across scenarios. Realized set loss applies the same loss
function to the recorded diagnosis. These costs have study-defined units.

For a nonempty set $S_i$ returned for prediction record $i$, Set Information
Gain (SIG) measures narrowing:
\[
SIG_i=\log_2\frac{3}{|S_i|}
\in\{\log_2 3,\log_2(3/2),0\}.
\]
Here $|S_i|$ counts the returned labels. The values are approximately 1.58496,
0.58496 and 0 bits for one, two and three labels, respectively.

For each record, the Scenario Sensitivity Index (SSI) averages pairwise
Jaccard distance among the $L=21$ nonempty sets returned by individual scenarios \cite{Jaccard1912}:
\[
SSI_i=\frac{100}{\binom{L}{2}}\sum_{s<t}
\left(1-\frac{|S_{is}\cap S_{it}|}{|S_{is}\cup S_{it}|}\right).
\]
The indices $s$ and $t$ identify different scenarios, and $S_{is}$ is the set
returned by scenario $s$ for record $i$, including the empty-set fallback
in Supplement S1.5. Intersection ($\cap$) selects shared labels and union
($\cup$) combines labels; vertical bars count them. The denominator is the
210 distinct scenario pairs. SSI ranges from 0 to 100. SSI is zero when
all 21 scenarios return the same set.

Within each assessment budget, held-out records pooled across three
cross-validation repeats were ordered by SSI. Empirical 20th, 40th, 60th
and 80th percentiles defined the lowest, lower, middle, higher and highest
variation groups. Equal scores stayed together. All six policies were
compared within the same groups.

\Needspace{16\baselineskip}
\section*{S2.2\quad Reliability inputs and fitting}

The multicomponent error model used six inputs: normalized predictive
entropy, ambiguity between the two most probable classes, disagreement between the primary models,
weighted missing information, similarity of missingness patterns and sensitivity
to missing data scenarios. For the primary three-class probability vector
$p=(p_1,p_2,p_3)$, normalized entropy \cite{Shannon1948} and ambiguity were
\[
u_H=-\frac{\sum_{k=1}^3p_k\log p_k}{\log 3},
\qquad u_M=1-(p_{(1)}-p_{(2)}),
\]
where $p_{(1)}$ and $p_{(2)}$ are the largest and second-largest probabilities.
Let $H(p)$ denote the normalized entropy above. For calibrated M2 and M4
vectors $p$ and $q$, model disagreement was the weighted Jensen--Shannon
divergence \cite{Lin1991}
\[
u_J=H(0.60p+0.40q)-0.60H(p)-0.40H(q).
\]
Similarity of missingness patterns was one minus total variation distance
between primary clinical predictions and predictions from the missingness control, where
$\mathrm{TV}(p,q)=\tfrac12\sum_{k=1}^3|p_k-q_k|$.
Mechanism sensitivity was the largest total variation distance between
an M5 scenario and the M5 MAR MissC reference, across the 20 M5 scenarios.

The input for missing information used M1 coefficient blocks. If $C_j$ contains
the coefficients for predictor $j$ across its encoded columns and
nonreference diagnostic classes, its weight and the input for each record were
\[
w_j=\frac{\lVert C_j\rVert_F}{\sum_h\lVert C_h\rVert_F},
\qquad D_i=\sum_jw_jd(M_{ij}).
\]
The Frobenius norm is the square root of the sum of squared coefficients.
Sums cover eligible predictors within the budget; if every norm is zero,
weights are uniform. The recording state penalty $d$ is 0 for observed
values, 0.5 for entries that were verified as not applicable or were structurally absent, and
1 for other missing entries. Development records use their matching
out-of-fold M1 fit. Calibration and test records use the final development
fit. The reliability model uses the M1-weighted deficit $D_i$. The
subgroup analysis uses the M5-weighted score Q (S3).

Development minima and ranges rescaled each input to $[0,1]$, with
values outside the development range clipped to that interval; a constant
input used a denominator of one. A ridge-penalized logistic regression
with nonnegative coefficients combined these inputs into an estimated
error probability $\widehat r_i$.
The Support Reliability Score (SRS) was $100(1-\widehat r_i)$; higher values
indicate lower fitted error risk. These scores supplied the A4/A5 rules
specified in Supplement S1.5.

\paragraph{Step 1: put the six inputs on a common scale.}
Let $u_{ij}$ be uncertainty input $j\in\{1,\ldots,6\}$ for development
record $i\in\{1,\ldots,n\}$. For each input calculate its development
minimum $m_j$, maximum $M_j$ and denominator
$s_j=M_j-m_j$ if $M_j>m_j$, otherwise $s_j=1$. The scaled input is
\[
x_{ij}=\min\{1,\max[0,(u_{ij}-m_j)/s_j]\}.
\]
All records use the same development minima and ranges, including later
calibration and test records. A constant input is zero for all
development records. Clipping prevents a later value from falling
outside the fitted scale.

\paragraph{Step 2: define the error outcome and objective.}
Let $\widehat y_i$ be the primary ensemble's highest-probability label
and let $e_i=\mathbf 1[\widehat y_i\ne y_i]$. An error has $e_i=1$,
and a correct label has $e_i=0$. With intercept $c$ and slopes
$\beta=(\beta_1,\ldots,\beta_6)$, the error probability is
\[
\eta_i=c+\sum_{j=1}^6x_{ij}\beta_j,\qquad
r_i=\sigma(\eta_i)=\frac{1}{1+e^{-\eta_i}}.
\]
The error outcome has Bernoulli probability
$r_i^{e_i}(1-r_i)^{1-e_i}$. Taking negative logs, substituting the
logistic expressions and averaging gives
\[
\begin{aligned}
-\frac1n\sum_i\log\{r_i^{e_i}(1-r_i)^{1-e_i}\}
&=-\frac1n\sum_i\{e_i\log r_i+(1-e_i)\log(1-r_i)\}\\
&=\frac1n\sum_i\{\log(1+e^{\eta_i})-e_i\eta_i\}.
\end{aligned}
\]
The constrained fitting criterion adds ridge shrinkage:
\begin{equation}\label{eq:reliability-objective}
\min_{c\in\mathbb R,\ \beta_j\geq0}
\left[\frac1n\sum_i\{\log(1+e^{\eta_i})-e_i\eta_i\}
+0.1\sum_{j=1}^6\beta_j^2\right].
\end{equation}
Each development record contributes equally to the loss. The intercept is
unrestricted and unpenalized.

\paragraph{Step 3: obtain the gradient and explain the constraint.}
Differentiating $\log(1+e^{\eta_i})-e_i\eta_i$ gives $r_i-e_i$.
The chain rule and $d(0.1\beta_j^2)/d\beta_j=0.2\beta_j$ yield
\[
g_c=\frac1n\sum_i(r_i-e_i),\qquad
g_{\beta_j}=\frac1n\sum_i x_{ij}(r_i-e_i)+0.2\beta_j.
\]
For a positive fitted slope, stationarity requires $g_{\beta_j}=0$.
At the boundary $\beta_j=0$, it requires $g_{\beta_j}\geq0$:
the only permitted movement is to increase the slope, and that
movement must not lower the objective. These are the bound-constrained
optimality conditions. The code uses L-BFGS-B, an optimizer that
respects the nonnegative bounds, and checks its success status.
It starts with slopes zero and intercept
$\log\{\bar e/(1-\bar e)\}$, with the empirical error fraction
$\bar e$ clipped to $[10^{-6},1-10^{-6}]$ for this initialization.
Probabilities inside logarithms are floored at $10^{-12}$.

To see the effect of nonnegative slopes, differentiate the prediction
with respect to a scaled input:
\[
\frac{\partial r_i}{\partial x_{ij}}
=r_i(1-r_i)\beta_j\geq0,\qquad
\frac{\partial\{100(1-r_i)\}}{\partial x_{ij}}
=-100r_i(1-r_i)\beta_j\leq0.
\]
Increasing one uncertainty input while holding the others fixed
therefore cannot reduce the fitted error risk or increase SRS.
Rescaling with clipping also preserves this monotonic direction, becoming
flat outside its range.

\Needspace{14\baselineskip}
\paragraph{Step 4: select the general A4 reliability cutoff.}
Write $S_i=100(1-\widehat r_i)$ for the development SRS.
The evaluated A4/A5 rules used the original threshold procedure for
primary argmax predictions. It enumerates
$\mathcal T=\{0,100,S_1,\ldots,S_n\}$ after removing duplicates.
For each candidate $t$, define
\[
\begin{aligned}
A(t)&=\{i:S_i\geq t\},\qquad N(t)=|A(t)|,\\
E(t)&=\frac{\sum_{i\in A(t)}\mathbf 1[\widehat y_i\ne y_i]}{N(t)},\\
P_{\mathrm{AP}}(t)&=
\frac{\sum_{i\in A(t)}\mathbf 1[\widehat y_i=\mathrm{AP},\ y_i=\mathrm{AP}]}
 {\sum_{i\in A(t)}\mathbf 1[\widehat y_i=\mathrm{AP}]}.
\end{aligned}
\]
A candidate is feasible only if $N(t)>0$, at least one retained
prediction is AP, $E(t)\leq0.10$, and $P_{\mathrm{AP}}(t)\geq0.40$.
An empty AP denominator makes precision undefined and the candidate
infeasible. Among feasible candidates, choose the cutoff with the largest
retained count $N(t)$; break a tie by the smaller SRS cutoff $t$.
If no cutoff is feasible, the routine returns 100 and marks the cutoff
invalid. A4 then enlarges the A3 singleton.

\paragraph{Step 5: select the two HC cutoffs.}
First restrict to records whose primary argmax is HC. Among candidate
SRS cutoffs $\{100\}\cup\{S_i:\widehat y_i=\mathrm{HC}\}$, define
\[
H(t)=\{i:\widehat y_i=\mathrm{HC},\ S_i\geq t\},\qquad
E_H(t)=\frac{\sum_{i\in H(t)}\mathbf 1[y_i\ne\mathrm{HC}]}{|H(t)|}.
\]
Require a nonempty set and $E_H(t)\leq0.025$. Maximize $|H(t)|$,
breaking ties by the smaller $t$, and call the selected cutoff $t_H$.
If none is feasible, the HC thresholds are marked invalid.

Next calculate each record's largest PD/AP probability across the
21 scenarios,
\[
d_i^{\max}=\max_s\{\widehat p_{s,\mathrm{PD}}(x_i)
                         +\widehat p_{s,\mathrm{AP}}(x_i)\}.
\]
Enumerate $v\in\{0\}\cup\{d_i^{\max}:i\in H(t_H)\}$ and retain
$H(t_H,v)=\{i\in H(t_H):d_i^{\max}\leq v\}$.
Among nonempty sets whose observed non-HC fraction is at most 0.025,
choose the largest retained count, breaking ties by the larger $v$.
Call it $v_H$. With general cutoff $t_*$ from step 4, the
stored HC SRS cutoff is $\max(t_*,t_H)$, and the disease-probability
cutoff is $v_H$. If either search finds no feasible cutoff, the HC cutoffs are invalid;
the code uses numerical defaults 100 and 0 when the first step fails,
or $t_H$ and 0 if the second fails.

The base-model predictions were obtained out of fold. The reliability
model was fitted and its cutoffs selected on the same development records.
The selected cutoffs remained unchanged during conformal calibration and
testing.

Simple candidates were one minus the maximum primary probability,
one minus the margin between the two largest probabilities, and normalized entropy. Nonnegative
univariate logistic maps supplied error probabilities, while AURC used
the raw uncertainty scores. Using cross-fitted development predictions, we selected
the comparator by lowest AURC, then error Brier score, then the candidate
order above.

\Needspace{14\baselineskip}
\section*{S2.3\quad Error targets, risk ranking and comparison criterion}

Let $e_i$ equal 1 when the evaluated predicted label disagrees with the
recorded diagnosis, and 0 otherwise. For $n$ eligible records, binary error
Brier score and error log loss \cite{Gneiting2007} were
\[
\frac1n\sum_{i=1}^n(\widehat r_i-e_i)^2,
\qquad
-\frac1n\sum_{i=1}^n
\{e_i\log\widehat r_i+(1-e_i)\log(1-\widehat r_i)\}.
\]
Lower scores indicate more accurate error probabilities.

For errors $e_{(j)}$ ordered from lowest to highest estimated risk,
the area under the risk--coverage curve \cite{Ding2020,ElYaniv2010} was
\[
AURC=\frac1n\sum_{m=1}^n\frac1m\sum_{j=1}^me_{(j)}.
\]
Here $m$ is the number of predictions retained. Coverage on this curve is
the fraction of predictions retained. AURC averages the error fractions over
all retained prefixes. Lower values are better. The calculation used deterministic
tie ordering and expanded bootstrap multiplicities consistently. Predictions
were ranked within each fold, and fold estimates were weighted by
the number of eligible records in each fold.

The evaluation of A3 singletons used the returned singleton label as its
correctness target. This differed from the primary highest-probability
label in 238 of 2,882 J3 singleton records and 262 of 2,528 J4 records.
The separate analysis of all records used the primary highest-probability
label throughout.
Bootstrap evaluation held the fitted supporting models fixed
(Supplement S5).

The combined score could be retained only if the one-sided 95\% upper
bound for its AURC minus the AURC of the simple score selected during training was
below zero at J3. The combined score did not meet the J3 AURC criterion
(S4.2).


\maketitle

This supplement accompanies the manuscript \emph{Uncertainty aware prediction from
incomplete clinical records: a conformal prediction study of Parkinsonian
classification}. It details the training, data partition and percentile rules for
the information deficit score, \(Q\), whose predictor weights were derived
from the selected adaptive elastic net model (M5).
For record $i$, the score is
\[
Q_i=\sum_{j\in\mathcal J}\nu_jd(M_{ij}),\qquad
\nu_j=\frac{\lVert B_j\rVert_F/\sqrt{q_j}}
{\sum_{h\in\mathcal J}\lVert B_h\rVert_F/\sqrt{q_h}}.
\]
Here $\mathcal J$ contains eligible predictors represented in the selected
MAR MissC M5 fit, $B_j$ is predictor $j$'s coefficient block across its
encoded columns and the two estimated diagnostic-class contrasts
(PD versus HC and AP versus HC), $q_j$ counts those columns and
$\lVert\cdot\rVert_F$ is the Frobenius norm. The nonnegative weights sum
to one; excluded predictors and zero coefficient blocks carry no weight.
The recording state $M_{ij}$ refers to the prepared record before imputation.
Its penalty is 0 for an observed entry, including valid zero; 0.5 only for
entries that were not applicable or structurally absent, as confirmed by response codes
or verified skip rules; and 1 for other missing entries. Unexplained blanks
and ambiguous `n/a' entries received a penalty of 1.
Q ranges from 0 to 1 under this weighting scheme.

\paragraph{From coefficients to a weighted deficit.}
The HC coefficients are fixed at zero. Source blocks contain only slope coefficients.
For source $j$, sum the squared entries in its two estimated columns,
take the square root and then adjust for its encoded size:
\[
N_j=\sqrt{\sum_{\ell\in G_j}
 (B_{\ell,\mathrm{PD}}^2+B_{\ell,\mathrm{AP}}^2)},\qquad
A_j=\frac{N_j}{\sqrt{q_j}},\qquad A=\sum_{h\in\mathcal J}A_h.
\]
Here $G_j$ is the set of column indices for source $j$.
Every encoded column in that source block counts in $q_j$, including
its missingness indicator. Thus a numeric source has its standardized
value column and missingness column ($q_j=2$); a categorical source
with $c_j$ encoded category columns, including other, has $q_j=c_j+1$.
Moreover,
\[
A_j=\sqrt{\frac{1}{q_j}\sum_{\ell\in G_j}
 (B_{\ell,\mathrm{PD}}^2+B_{\ell,\mathrm{AP}}^2)}
\]
shows that this is the square root of the mean squared contrast
length per encoded column.

Provided $A>0$, divide each adjusted norm by $A$ to obtain
$\nu_j=A_j/A$. Then $\nu_j\geq0$ and
$\sum_j\nu_j=(\sum_j A_j)/A=1$. Since every recording penalty lies
between zero and one,
\[
0=\sum_j\nu_j\,0\ \leq\ \sum_j\nu_jd(M_{ij})=Q_i
\ \leq\ \sum_j\nu_j\,1=1.
\]
For example, adjusted norms 2 and 1 give weights $2/3$ and $1/3$.
If the first source is missing with penalty 1 and the second is
verified structurally absent with penalty $1/2$, the score is
$Q_i=(2/3)1+(1/3)(1/2)=5/6$.

Q is undefined when all coefficient blocks are zero. The calculation stops
if the denominator is zero or if a norm or the denominator is nonfinite.
Every positive adjusted block norm contributes to the Q weights.

Q uses predictors eligible at the assessment budget. For those predictors,
response codes and verified skip rules identify structural absences. The
C-OPN data dictionary is available by request from C-OPN. See the
manuscript's Availability of data and materials.

\section*{S3.1\quad Exclusion from model selection and fitting}

For each division of the training records and each outer validation split, the records for model development
(proper fitting) were divided into four inner folds. For each target fold,
candidate models were trained on two of the remaining folds and evaluated
on the third. The selected model was refitted on all three and used to
score the target fold. Repeating this rotation yielded one out-of-fold
Q score per development record.

Calibration and outer test records were scored using a final model selected
from the out-of-fold evidence for model development and fitted on all
records used for model development. The weights were specific to the selected model
for the relevant division of training records, outer split and training subset.

\paragraph{Numerical fitting used to obtain Q's coefficients.}
The Q fits use the MAR loss, $10^{-3}$ initial ridge penalty, adaptive
weights and group penalty derived in Supplement S1.1. Their numerical
ridge initializer uses L-BFGS-B on that smooth ridge objective, with
at most 1,000 iterations, 4,000 function evaluations, function tolerance
$10^{-12}$ and gradient tolerance $10^{-6}$. A successful optimizer
status is required. The code also checks for finite, nonincreasing
objective values, maximum absolute gradient at most $10^{-5}$, and one
accepted original gradient step satisfying both original stopping
conditions; that verified step supplies the initializer coefficients.
Adaptive weights are then computed from those coefficients using
$(\|\widetilde B_j\|_F+n^{-1/2})^{-\gamma}$.

The subsequent grouped fit uses the same gradient, block update,
line search and 500-iteration limit as S1.1, with cached arrays to
avoid repeated calculations. Q candidate selection requires convergence
in every required training/validation
pair. Eligible candidates are ordered by higher unweighted balanced
accuracy, higher AP average precision, lower unweighted multiclass
log loss and finally the prespecified candidate order. The selected
penalty is its fraction times the reference penalty calculated
from that training set. The final penalized coefficients determine the source weights.

\section*{S3.2\quad Dividing the training records and counting model fits}

We checked Q scores and participants' ordering by Q under six ways of dividing
the training records between model development and prediction set calibration.
Prediction set calibration uses separate records with known diagnoses to
calculate cutoffs for including diagnostic labels in a prediction set. Three
divisions reserved 25\%, 30\% or 35\% of the outer training records for this
step; 25\% was the reference. The other three used alternative selections of
calibration records while keeping the outer test records fixed. These choices
changed the records available to fit M5 and derive Q's weights.

Combining the six divisions of the training records with 15 outer validation
splits gave 90 combinations. Each combination required four fits of selected models for the
target inner folds and one final fit on all records used for model development, giving
450 fits of selected models in total. Each of the 3,434 participants
was scored in all 90 combinations, yielding 309,060 Q evaluations.

\section*{S3.3\quad Percentile cutoffs and group assignment}

For each division of the training records and each outer split, \(c_{25}\), \(c_{50}\) and \(c_{75}\) were
calculated from the primary out-of-fold \(Q\) scores from model development only.
Under the `higher' rule, each cutoff is the observed score at the rank
given below. For sorted scores
\(Q_{(1)}\leq\cdots\leq Q_{(n)}\), the rule was
\[
  c_p = Q_{(1+\lceil (n-1)p\rceil)},
  \qquad p\in\{0.25,0.50,0.75\}.
\]
Here, \(c_{0.25}=c_{25}\), \(c_{0.50}=c_{50}\) and \(c_{0.75}=c_{75}\).
The four ordered groups were assigned by
\[
\begin{aligned}
  Q1\;\text{(lowest)} &: Q\leq c_{25}, \\
  Q2\;\text{(lower middle)} &: c_{25}<Q\leq c_{50}, \\
  Q3\;\text{(upper middle)} &: c_{50}<Q\leq c_{75}, \\
  Q4\;\text{(highest)} &: Q>c_{75}.
\end{aligned}
\]

Tied scores stayed together, and all four group labels were retained even
when a middle group was empty. Within each division and split, the
development cutoffs were applied unchanged to calibration and test records.
Primary Q defined the groups. The three sensitivity scores were analysed
as continuous variables.

\section*{S3.4\quad Spread of Q scores under different calculations and divisions of records}

The main Q calculation adjusts predictor weights for the number of numerical
columns used to represent each variable and assigns entries verified as not applicable or
structurally absent a half penalty (0.5). Three alternative rules remove
the column count adjustment, assign these entries no penalty (0), or assign them
a full penalty (1), changing one choice at a time. Tables S2--S3
describe the resulting score distributions.

\Needspace{16\baselineskip}
\begingroup
\small
\setlength{\tabcolsep}{3pt}
\renewcommand{\arraystretch}{1.16}
\setlength{\TableWidth}{\dimexpr\linewidth-8\tabcolsep\relax}
\begin{longtable}{@{}>{\raggedright\arraybackslash}p{0.28\TableWidth}>{\raggedright\arraybackslash}p{0.22\TableWidth}>{\raggedright\arraybackslash}p{0.19\TableWidth}>{\raggedright\arraybackslash}p{0.1\TableWidth}>{\raggedright\arraybackslash}p{0.21\TableWidth}@{}}
\caption*{\textbf{Table S2. Q scores under four calculation rules}}\\
\toprule
\textbf{Definition} & \textbf{Mean (SD)} & \textbf{5th / 25th percentiles} & \textbf{Median} & \textbf{75th / 95th percentiles} \\
\midrule
\endfirsthead
\multicolumn{5}{l}{\footnotesize Table S2. Q scores under four calculation rules\ (continued)}\\
\toprule
\textbf{Definition} & \textbf{Mean (SD)} & \textbf{5th / 25th percentiles} & \textbf{Median} & \textbf{75th / 95th percentiles} \\
\midrule
\endhead
\midrule
\multicolumn{5}{r}{\footnotesize Continued on next page}\\
\endfoot
\bottomrule
\endlastfoot
Not applicable/structural penalty 1 & 0.370389\newline (0.365383) & 0.000000 / 0.029507 & 0.245872 & 0.788696 / 0.941405 \\
Not applicable/structural penalty 0 & 0.361658\newline (0.368168) & 0.000000 / 0.018477 & 0.230495 & 0.784331 / 0.941379 \\
Main calculation & 0.366023\newline (0.366648) & 0.000000 / 0.024037 & 0.239504 & 0.785156 / 0.941379 \\
Without column count adjustment & 0.366637\newline (0.376595) & 0.000000 / 0.025624 & 0.193571 & 0.821786 / 0.958087 \\
\end{longtable}
\endgroup

\textbf{Table S2. Q scores under four calculation rules.} Every row summarizes 309,060 repeated Q evaluations from the same 3,434 people. Units are score points on {[}0,1{]}; smaller means less weighted deficit under that rule. SD (standard deviation) describes spread around the mean; the 25th and 75th percentiles bound the middle half of scores. Removing the column count adjustment changes the median from 0.239504 to 0.193571 despite similar averages.

\begingroup
\small
\setlength{\tabcolsep}{3pt}
\renewcommand{\arraystretch}{1.16}
\setlength{\TableWidth}{\dimexpr\linewidth-10\tabcolsep\relax}
\begin{longtable}{@{}>{\raggedright\arraybackslash}p{0.26\TableWidth}>{\raggedright\arraybackslash}p{0.148\TableWidth}>{\raggedright\arraybackslash}p{0.148\TableWidth}>{\raggedright\arraybackslash}p{0.148\TableWidth}>{\raggedright\arraybackslash}p{0.148\TableWidth}>{\raggedright\arraybackslash}p{0.148\TableWidth}@{}}
\caption*{\textbf{Table S3. Main Q under six ways of dividing the training records}}\\
\toprule
\textbf{Division of training records} & \textbf{Mean} & \textbf{25th percentile} & \textbf{Median} & \textbf{75th percentile} & \textbf{95th percentile} \\
\midrule
\endfirsthead
\multicolumn{6}{l}{\footnotesize Table S3. Main Q under six ways of dividing the training records\ (continued)}\\
\toprule
\textbf{Division of training records} & \textbf{Mean} & \textbf{25th percentile} & \textbf{Median} & \textbf{75th percentile} & \textbf{95th percentile} \\
\midrule
\endhead
\midrule
\multicolumn{6}{r}{\footnotesize Continued on next page}\\
\endfoot
\bottomrule
\endlastfoot
25\% reserved (reference) & 0.365753 & 0.023440 & 0.243662 & 0.777551 & 0.939620 \\
30\% reserved & 0.364982 & 0.023439 & 0.231349 & 0.786530 & 0.944014 \\
35\% reserved & 0.367321 & 0.024028 & 0.228696 & 0.783094 & 0.944128 \\
Alternative selection 1 & 0.365231 & 0.023537 & 0.246585 & 0.783894 & 0.937817 \\
Alternative selection 2 & 0.363950 & 0.025717 & 0.240610 & 0.792459 & 0.935983 \\
Alternative selection 3 & 0.368902 & 0.023978 & 0.247829 & 0.791255 & 0.943809 \\
\end{longtable}
\endgroup

\textbf{Table S3. Main Q under six ways of dividing the training records.} Each row contributes 15 × 3,434 repeated Q evaluations. The first three rows reserve 25\%, 30\% or 35\% of the outer training records for prediction set calibration, which determines the cutoffs for including diagnostic labels. The remaining rows use alternative selections of calibration records while keeping the test records fixed (S3.2).

\begin{figure}[htbp]
\centering
\includegraphics[width=\linewidth]{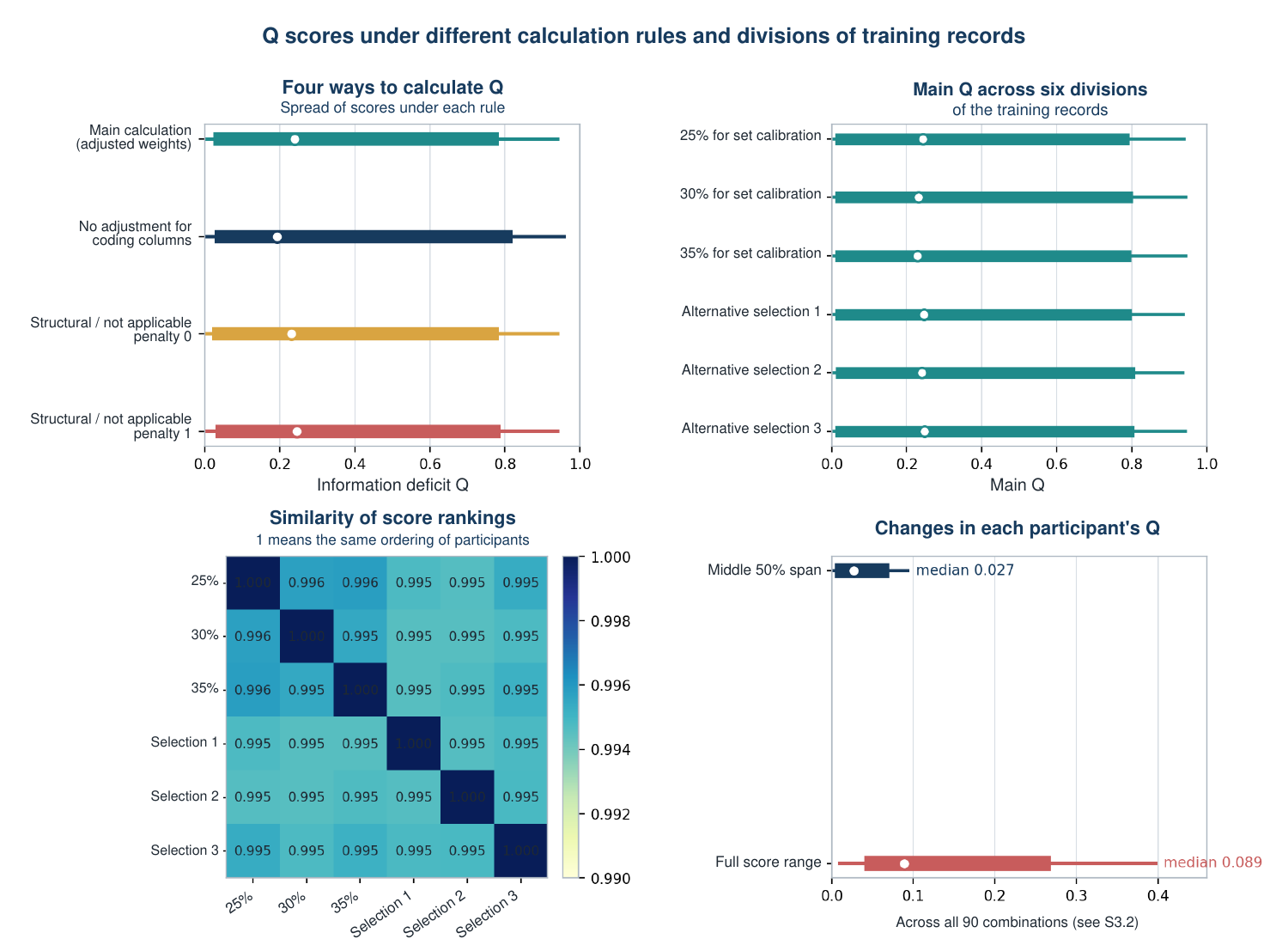}
\captionsetup{font=small}
\caption{\textbf{Q scores under four calculation rules and six ways of dividing the training records.} The main rule adjusts predictor weights for the number of numerical columns and gives entries verified as not applicable or structurally absent a half penalty (0.5). The alternatives remove this column count adjustment or change only that penalty to 0 or 1. Upper panels show how scores are spread across records on the 0--1 scale: white points mark the middle score (median), thick bars span the middle 50\% of scores, and thin lines span the middle 90\% (5th--95th percentiles). The left panel compares the four rules. The right uses the main rule with training records divided between model development and prediction set calibration, which determines the cutoffs for including diagnostic labels. The percentage labels show the share of training records reserved for calibration; alternative selections 1--3 change which records are reserved while keeping the test records fixed. The lower-left heatmap compares participants' ordering by Q between the six divisions (Spearman rank correlation; 1 means the same ordering). The lower-right panel shows each participant's span of the middle 50\% and full range (highest minus lowest) across the 90 combinations of a division of the training records and an outer validation split; the medians of these spans across participants are 0.027 and 0.089, respectively. The figure summarizes 309,060 Q evaluations from 3,434 participants. Tables S2--S3 give the numerical summaries.}
\label{fig:s3-q-consistency}
\end{figure}

Across all Q score evaluations, the main calculation gave an average of 0.3660,
a median of 0.2395, and a middle half of scores between 0.0240 and 0.7852.
Assigning no penalty (0) to entries verified as not applicable or structurally absent
gave an average of 0.3617; assigning a full penalty (1) gave 0.3704. Removing
the adjustment for the number of numerical columns gave an average of 0.3666,
but changed the median to 0.1936. Average Q under the main rule ranged from
approximately 0.3640 to 0.3689 across the six divisions of the training records. Tables S2--S3
give the numerical summaries, and Figure~\ref{fig:s3-q-consistency} shows the
spread of scores and the consistency of participants' ordering by Q.

\clearpage

\section*{S3.5\quad Selection-frequency details}
Across the 450 selected fits described in S3.2, 172 distinct configurations were used. The most frequent pairing was adaptive exponent \(\gamma=0.5\) with an AP loss multiplier of 1.5 (159 fits). Elastic-net mixing parameters 0.05 and 0.25 were selected in 175 and 151 fits. Among the 90 combinations of training division and outer split, the five fits used three distinct configurations in four cases, four in 39 cases and five in 47 cases.
\begin{figure}[htbp]
\centering
\includegraphics[width=\linewidth]{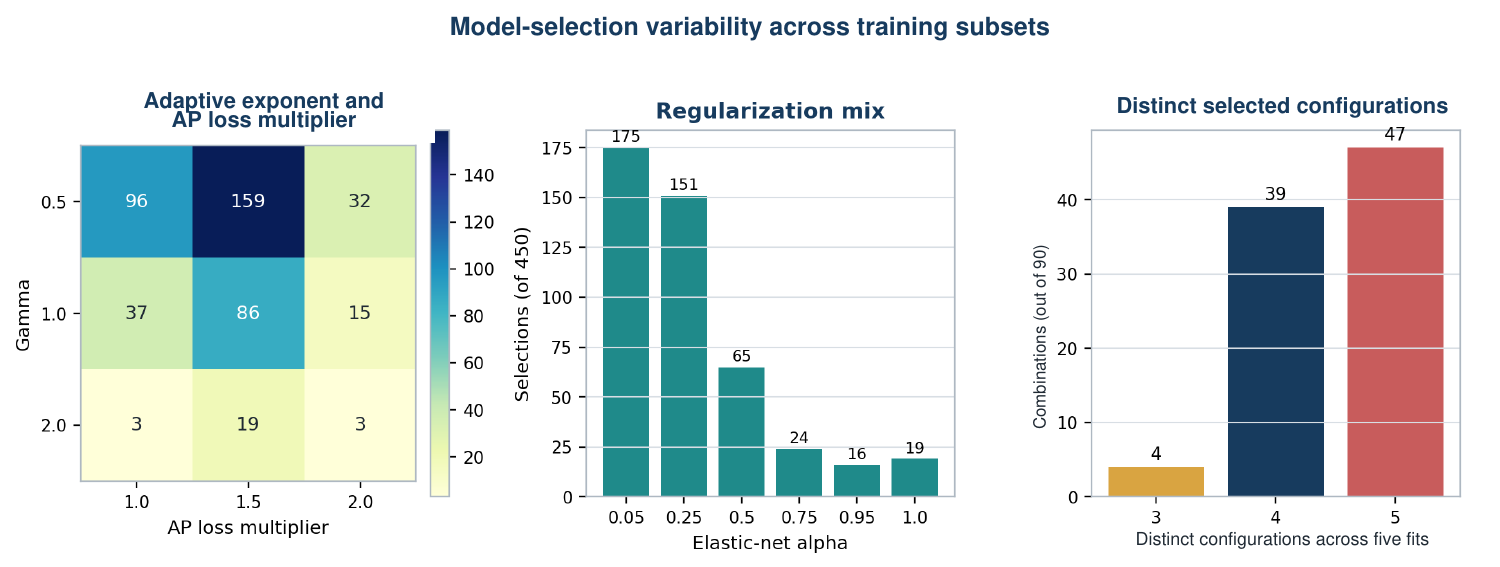}
\caption{Selection frequencies across training subsets. The heatmap and middle panel summarize 450 selected fits. The right panel summarizes the 90 combinations of training division and outer split. We compared 972 candidates using balanced accuracy, AP average precision and log loss, with deterministic tie-breaking.}
\end{figure}


\maketitle

This supplement accompanies the manuscript \emph{Uncertainty aware prediction from incomplete clinical records: a conformal prediction study of Parkinsonian classification}.

These analyses use held-out predictions from 3,434 participants, each
contributing one test prediction in each of three cross-validation repeats
(10,302 records per budget). Tables S4--S7 report set composition, policy
contrasts and reliability. Tables S8--S9 report Q4 coverage and overall
efficiency. Table S10 compares the six policies across SSI groups.
Supplement S5 specifies resampling and inference limits.

\section*{S4.1 Set composition and global comparisons}

\begingroup
\small
\setlength{\tabcolsep}{3pt}
\renewcommand{\arraystretch}{1.16}
\setlength{\TableWidth}{\dimexpr\linewidth-14\tabcolsep\relax}
\begin{longtable}{@{}>{\raggedright\arraybackslash}p{0.09\TableWidth}>{\raggedright\arraybackslash}p{0.13\TableWidth}>{\raggedright\arraybackslash}p{0.13\TableWidth}>{\raggedright\arraybackslash}p{0.13\TableWidth}>{\raggedright\arraybackslash}p{0.13\TableWidth}>{\raggedright\arraybackslash}p{0.13\TableWidth}>{\raggedright\arraybackslash}p{0.13\TableWidth}>{\raggedright\arraybackslash}p{0.13\TableWidth}@{}}
\caption*{\textbf{Table S4. J3 counts of nonempty label sets}}\\
\toprule
\textbf{Policy} & \textbf{\{PD\}} & \textbf{\{AP\}} & \textbf{\{PD,AP\}} & \textbf{\{HC\}} & \textbf{\{PD,HC\}} & \textbf{\{AP,HC\}} & \textbf{Full} \\
\midrule
\endfirsthead
\multicolumn{8}{l}{\footnotesize Table S4. J3 counts of nonempty label sets\ (continued)}\\
\toprule
\textbf{Policy} & \textbf{\{PD\}} & \textbf{\{AP\}} & \textbf{\{PD,AP\}} & \textbf{\{HC\}} & \textbf{\{PD,HC\}} & \textbf{\{AP,HC\}} & \textbf{Full} \\
\midrule
\endhead
\midrule
\multicolumn{8}{r}{\footnotesize Continued on next page}\\
\endfoot
\bottomrule
\endlastfoot
A2 & 188 & 30 & 6149 & 69 & 0 & 599 & 3267 \\
A3 & 2139 & 298 & NR & 445 & $<5$ & 1217 & 939 \\
A5 & 1905 & 0 & 5697 & 0 & 431 & 0 & 2269 \\
\end{longtable}
\endgroup

\textbf{Table S4. J3 counts of nonempty label sets.} Each method totals 10,302 held-out rows. $<5$ denotes one to four records. NR denotes additional withholding to prevent recovery from totals or related outputs. A5 required verified in-distribution status to retain HC alone. That status was unavailable.

\begingroup
\small
\setlength{\tabcolsep}{3pt}
\renewcommand{\arraystretch}{1.16}
\setlength{\TableWidth}{\dimexpr\linewidth-10\tabcolsep\relax}
\begin{longtable}{@{}>{\raggedright\arraybackslash}p{0.13\TableWidth}>{\raggedright\arraybackslash}p{0.06\TableWidth}>{\raggedright\arraybackslash}p{0.12\TableWidth}>{\raggedright\arraybackslash}p{0.23\TableWidth}>{\raggedright\arraybackslash}p{0.16\TableWidth}>{\raggedright\arraybackslash}p{0.3\TableWidth}@{}}
\caption*{\textbf{Table S5. Paired A3/A5 minus A2 policy contrasts}}\\
\toprule
\textbf{Source} & \textbf{J} & \textbf{Contrast} & \textbf{Metric} & \textbf{Difference} & \textbf{95\% interval} \\
\midrule
\endfirsthead
\multicolumn{6}{l}{\footnotesize Table S5. Paired A3/A5 minus A2 policy contrasts\ (continued)}\\
\toprule
\textbf{Source} & \textbf{J} & \textbf{Contrast} & \textbf{Metric} & \textbf{Difference} & \textbf{95\% interval} \\
\midrule
\endhead
\midrule
\multicolumn{6}{r}{\footnotesize Continued on next page}\\
\endfoot
\bottomrule
\endlastfoot
Scenario aggregation & J3 & A3 minus A2 & Mean labels & -0.477868 & -0.492431--\allowbreak{}-0.464373 \\
Scenario aggregation & J3 & A3 minus A2 & Mean SIG & 0.384080 & 0.372452--\allowbreak{}0.396007 \\
Scenario aggregation & J3 & A3 minus A2 & Robust expected cost & 0.159338 & 0.138980--\allowbreak{}0.178981 \\
Scenario aggregation & J3 & A3 minus A2 & AP core coverage & -0.033138 & -0.056819--\allowbreak{}-0.012983 \\
Scenario aggregation & J3 & A3 minus A2 & HC core coverage & $\approx -0.11$ & -0.14--\allowbreak{}-0.08 \\
Scenario aggregation & J4 & A3 minus A2 & Mean labels & -0.445545 & -0.460010--\allowbreak{}-0.431855 \\
Scenario aggregation & J4 & A3 minus A2 & Mean SIG & 0.349782 & 0.338542--\allowbreak{}0.361134 \\
Scenario aggregation & J4 & A3 minus A2 & Robust expected cost & 0.206016 & 0.185399--\allowbreak{}0.226980 \\
Scenario aggregation & J4 & A3 minus A2 & AP core coverage & -0.023392 & -0.040565--\allowbreak{}-0.008546 \\
Scenario aggregation & J4 & A3 minus A2 & HC core coverage & $\approx -0.10$ & -0.14--\allowbreak{}-0.08 \\
Cost enlargement & J3 & A5 minus A2 & Mean labels & -0.253931 & -0.268880--\allowbreak{}-0.239953 \\
Cost enlargement & J3 & A5 minus A2 & Mean SIG & 0.213725 & 0.202514--\allowbreak{}0.225745 \\
Cost enlargement & J3 & A5 minus A2 & Robust expected cost & -0.164173 & -0.173164--\allowbreak{}-0.155975 \\
Cost enlargement & J3 & A5 minus A2 & AP core coverage & -0.027290 & -0.048898--\allowbreak{}-0.008769 \\
Cost enlargement & J3 & A5 minus A2 & HC core coverage & $\approx -0.09$ & -0.12--\allowbreak{}-0.06 \\
Cost enlargement & J4 & A5 minus A2 & Mean labels & -0.239662 & -0.253354--\allowbreak{}-0.226170 \\
Cost enlargement & J4 & A5 minus A2 & Mean SIG & 0.199214 & 0.188748--\allowbreak{}0.210064 \\
Cost enlargement & J4 & A5 minus A2 & Robust expected cost & -0.150692 & -0.159085--\allowbreak{}-0.142889 \\
Cost enlargement & J4 & A5 minus A2 & AP core coverage & -0.023392 & -0.040565--\allowbreak{}-0.008546 \\
Cost enlargement & J4 & A5 minus A2 & HC core coverage & $\approx -0.09$ & -0.12--\allowbreak{}-0.07 \\
\end{longtable}
\endgroup

\textbf{Table S5. Selected paired policy contrasts from the A3 and A5 comparisons.} Each estimate is the named policy minus A2. Confidence limits use paired
resampling of participant clusters with predictions held fixed. These policy
contrasts and the Q subgroup comparisons used their respective bootstrap
procedures (Supplement S5). Set size is in labels, SIG in bits and cost in
study-defined units. To protect suppressed counts, HC coverage differences
are rounded to the nearest percentage point and their interval limits are
rounded outward.

None of the 3,024 participants with PD or AP received an HC-only output under A5. The one-sided exact upper bound for this event was approximately 0.099\%. A5 core sets omitted AP for eight of 171 unique AP participants at J3 and nine at J4 under the participant-level event rule. The broader mention set had zero AP exclusions, with a one-sided upper bound about 1.74\%.

The A5 singleton bootstrap yielded 1,290 valid replicates at J3 and none
at J4, out of 2,000 per budget. Both counts fell below the required 1,900,
so no conditional interval is reported.

\Needspace{10\baselineskip}
\section*{S4.2 Reliability of single-label predictions}

\begingroup
\small
\setlength{\tabcolsep}{3pt}
\renewcommand{\arraystretch}{1.16}
\setlength{\TableWidth}{\dimexpr\linewidth-8\tabcolsep\relax}
\begin{longtable}{@{}>{\raggedright\arraybackslash}p{0.07\TableWidth}>{\raggedright\arraybackslash}p{0.2\TableWidth}>{\raggedright\arraybackslash}p{0.42\TableWidth}>{\raggedright\arraybackslash}p{0.18\TableWidth}>{\raggedright\arraybackslash}p{0.13\TableWidth}@{}}
\caption*{\textbf{Table S6. Reliability contrasts for A3 singletons}}\\
\toprule
\textbf{J} & \textbf{Metric} & \textbf{Difference (95\% interval)} & \textbf{One-sided upper} & \textbf{Valid replicates} \\
\midrule
\endfirsthead
\multicolumn{5}{l}{\footnotesize Table S6. Reliability contrasts for A3 singletons\ (continued)}\\
\toprule
\textbf{J} & \textbf{Metric} & \textbf{Difference (95\% interval)} & \textbf{One-sided upper} & \textbf{Valid replicates} \\
\midrule
\endhead
\midrule
\multicolumn{5}{r}{\footnotesize Continued on next page}\\
\endfoot
\bottomrule
\endlastfoot
J3 & AURC & 0.001171\newline (0.000267--\allowbreak{}0.002390) & 0.002116 & 2000 \\
J3 & Error Brier score & -0.001266\newline (-0.001518--\allowbreak{}-0.001019) & -0.001059 & 2000 \\
J3 & Error log loss & -0.007954\newline (-0.009460--\allowbreak{}-0.006513) & -0.006735 & 2000 \\
J4 & AURC & 0.001075\newline (0.000373--\allowbreak{}0.001923) & 0.001747 & 2000 \\
J4 & Error Brier score & -0.001816\newline (-0.002172--\allowbreak{}-0.001484) & -0.001521 & 2000 \\
J4 & Error log loss & -0.009063\newline (-0.010682--\allowbreak{}-0.007511) & -0.007697 & 2000 \\
\end{longtable}
\endgroup

\textbf{Table S6. Reliability contrasts for A3 singleton correctness.} Differences are each metric for the multicomponent score minus the same metric for the simple score selected during training. Negative differences favor the multicomponent score. Retention required the J3 one-sided upper bound for the AURC difference to be below zero. The observed bound was 0.002116.

Among AP-only A3 outputs, there were 180 errors among 298 outputs at J3 (60.4\%) and 195/325 at J4 (60.0\%). Within these outputs, error discrimination was limited and calibration imperfect. The uncertainty estimates are conditional on the supporting quantities held fixed across training and test splits (S2, S5).

\begingroup
\small
\setlength{\tabcolsep}{3pt}
\renewcommand{\arraystretch}{1.16}
\setlength{\TableWidth}{\dimexpr\linewidth-6\tabcolsep\relax}
\begin{longtable}{@{}>{\raggedright\arraybackslash}p{0.46\TableWidth}>{\raggedright\arraybackslash}p{0.14\TableWidth}>{\raggedright\arraybackslash}p{0.27\TableWidth}>{\raggedright\arraybackslash}p{0.13\TableWidth}@{}}
\caption*{\textbf{Table S7. Error ranking for highest-probability predictions}}\\
\toprule
\textbf{Estimand} & \textbf{Estimate} & \textbf{95\% interval} & \textbf{Valid replicates} \\
\midrule
\endfirsthead
\multicolumn{4}{l}{\footnotesize Table S7. Error ranking for highest-probability predictions\ (continued)}\\
\toprule
\textbf{Estimand} & \textbf{Estimate} & \textbf{95\% interval} & \textbf{Valid replicates} \\
\midrule
\endhead
\midrule
\multicolumn{4}{r}{\footnotesize Continued on next page}\\
\endfoot
\bottomrule
\endlastfoot
AURC nested selected & 0.027100 & 0.023202--\allowbreak{}0.031525 & 2000 \\
AURC nested selected minus one minus max primary probability & 0.000054 & -0.000005--\allowbreak{}0.000115 & 2000 \\
AURC nested selected minus one minus primary top two margin & -0.000153 & -0.000374--\allowbreak{}0.000059 & 2000 \\
AURC nested selected minus primary normalized entropy & -0.000697 & -0.001273--\allowbreak{}-0.000146 & 2000 \\
AURC one minus max primary probability & 0.027046 & 0.023163--\allowbreak{}0.031458 & 2000 \\
AURC one minus primary top two margin & 0.027253 & 0.023339--\allowbreak{}0.031640 & 2000 \\
AURC primary normalized entropy & 0.027797 & 0.023895--\allowbreak{}0.032257 & 2000 \\
\end{longtable}
\endgroup

\textbf{Table S7. Error ranking for highest-probability predictions.} Each score is evaluated against the same outcome: an error in the primary ensemble's most probable label. AURC lies in {[}0,1{]} and lower is better. Estimates aggregate fold-specific risk--coverage curves with weights based on row counts and use intervals from resampling participant clusters. The simple score was selected using development predictions.

For prediction error, the selected simple score had a Brier score of
0.082908, log loss of 0.272215, and AUROC of 0.833046.

\Needspace{8\baselineskip}
\section*{S4.3 Targeted information deficit subgroup checks}

These exploratory checks use the fixed J3 predictions with 25\% of each outer training sample reserved for conformal calibration. Q groups use quartiles of development scores, applied unchanged to test records with equal scores kept together (S3.3). Q4 contains 2,516 prediction rows from 875 unique participants. Participants can change groups across repeats.

\begin{table}[htbp]
\centering
\small
\setlength{\tabcolsep}{4pt}
\renewcommand{\arraystretch}{1.2}
\setlength{\TableWidth}{\dimexpr\linewidth-4\tabcolsep\relax}
\caption*{\textbf{Table S8. Output breadth and class coverage in the highest information deficit group (Q4)}}
\begin{tabular}{@{}>{\raggedright\arraybackslash}p{0.34\TableWidth}>{\raggedright\arraybackslash}p{0.33\TableWidth}>{\raggedright\arraybackslash}p{0.33\TableWidth}@{}}
\toprule
\textbf{Metric} & \textbf{A2} & \textbf{A5} \\
\midrule
Mean labels & 2.965 & 2.643 \\
All three labels & 2429/2516 (96.54\%) & 1619/2516 (64.35\%) \\
PD coverage & 1845/1889 (97.67\%) & 1889/1889 (100\%) \\
AP coverage & 132/132 (100\%) & $<5$ not covered \\
HC coverage & 495/495 (100\%) & 401/495 (81.01\%) \\
\bottomrule
\end{tabular}
\end{table}

\noindent $<5$ denotes one to four uncovered records; the corresponding numerator and percentage are withheld.

The prespecified A2 Q4 minus Q1 contrast in mean size was 1.009735 labels
(two-sided 95\% interval 0.996083--1.023213; one-sided 95\% lower bound
0.998252).

Coverage is the share of prediction rows whose set includes the recorded class. The two-sided 95\% interval from resampling participant clusters for A5 HC coverage in Q4 is 75.94--85.72\%. Q1 contained no HC records. HC counts and estimates for Q2 and Q3 are withheld to protect sparse counts. Q4 contained 94 of the 115 uncovered HC rows, while accounting for 495 of the 1,230 HC prediction rows. Overall HC coverage was 1,115/1,230 (90.65\%).

For A2, the 132 covered AP rows in Q4 comprised 45/45, 43/43 and 44/44 across the three repeats. Their one-sided exact 95\% lower bounds were 93.56\%, 93.27\% and 93.42\%, respectively. None reached the 95\% target. All Q4 records were covered for each of the 45 AP participants; the one-sided exact lower bound for this participant-level event was 93.56\%. The participant bootstrap lower bound was 100\%. S5.3 gives the calculation of the exact bound and its assumptions.

\Needspace{24\baselineskip}
\begingroup
\small
\setlength{\tabcolsep}{3pt}
\renewcommand{\arraystretch}{1.16}
\setlength{\TableWidth}{\dimexpr\linewidth-6\tabcolsep\relax}
\begin{longtable}{@{}>{\raggedright\arraybackslash}p{0.3\TableWidth}>{\raggedright\arraybackslash}p{0.4\TableWidth}>{\raggedright\arraybackslash}p{0.17\TableWidth}>{\raggedright\arraybackslash}p{0.13\TableWidth}@{}}
\caption*{\textbf{Table S9. Global A5 versus A2 efficiency checks}}\\
\toprule
\textbf{Improvement metric} & \textbf{Estimate (95\% interval)} & \textbf{One-sided lower} & \textbf{Threshold} \\
\midrule
\endfirsthead
\multicolumn{4}{l}{\footnotesize Table S9. Global A5 versus A2 efficiency checks\ (continued)}\\
\toprule
\textbf{Improvement metric} & \textbf{Estimate (95\% interval)} & \textbf{One-sided lower} & \textbf{Threshold} \\
\midrule
\endhead
\midrule
\multicolumn{4}{r}{\footnotesize Continued on next page}\\
\endfoot
\bottomrule
\endlastfoot
absolute mean size reduction & 0.253931\newline (0.240048--\allowbreak{}0.268103) & 0.241507 & 0.20 \\
relative mean size reduction & 0.110923\newline (0.104811--\allowbreak{}0.117033) & 0.105751 & 0.10 \\
mean SIG increase & 0.213725\newline (0.202565--\allowbreak{}0.225221) & 0.204004 & 0.15 \\
relative full set reduction & 0.305479\newline (0.275021--\allowbreak{}0.334364) & 0.280777 & 0.20 \\
\end{longtable}
\endgroup

\textbf{Table S9. Global A5 versus A2 efficiency checks.} Positive values denote improvement. The size requirement was an absolute reduction at least 0.20 labels or a relative reduction at least 10\%; SIG increase at least 0.15 bits and relative full set reduction at least 20\% were additional requirements. A5 met the overall efficiency criteria at J3. Its Q4 HC coverage was 81.01\%, below the 90\% target (Table S8).

\clearpage
\section*{S4.4 Variation across scenarios}

Table 6 in the main manuscript compares all six policies in the highest variation group at both assessment budgets. Table S10 shows the full five groups, ordered by the Scenario Sensitivity Index (SSI), which measures variation in the possible diagnoses returned across scenarios: lowest, lower, middle, higher and highest variation.

\begin{center}\small\textbf{Table S10. Robustness summaries for the whole cohort.}\end{center}
\begingroup\small\setlength{\tabcolsep}{4pt}\renewcommand{\arraystretch}{1.08}
\setlength{\TableWidth}{\dimexpr\linewidth-10\tabcolsep\relax}
\noindent\textbf{A. Primary assessment budget J3 (levels 1--3)}\par\smallskip
\noindent\begin{tabular}{@{}>{\raggedright\arraybackslash}p{0.08\TableWidth}>{\raggedleft\arraybackslash}p{0.14\TableWidth}>{\raggedleft\arraybackslash}p{0.21\TableWidth}*{3}{>{\raggedleft\arraybackslash}p{0.19\TableWidth}}@{}}
\toprule\textbf{Policy} & \textbf{Average labels} & \textbf{All three labels, count (\%)} & \textbf{PD coverage} & \textbf{AP coverage} & \textbf{HC coverage} \\ \midrule
\multicolumn{6}{@{}l@{}}{\textbf{Lowest variation: 2,451 records (PD: 2,191; AP: 135; HC: 125)}} \\
A0 & 1.000 & 0 (0.00) & 98.63\% & 15.56\% & NR \\
A1 & 1.883 & 0 (0.00) & 97.81\% & 100.00\% & 100.00\% \\
A2 & 1.883 & 0 (0.00) & 97.81\% & 100.00\% & NR \\
A3 & 1.883 & 0 (0.00) & 97.81\% & 100.00\% & 100.00\% \\
A4 & 1.905 & 0 (0.00) & 97.81\% & 100.00\% & 100.00\% \\
A5 & 1.982 & 99 (4.04) & 100.00\% & 100.00\% & 100.00\% \\
\midrule
\multicolumn{6}{@{}l@{}}{\textbf{Lower variation: 2,071 records (PD: 1,756; AP: 98; HC: 217)}} \\
A0 & 1.000 & 0 (0.00) & NR & 14.29\% & 84.33\% \\
A1 & 1.635 & NR & 90.72\% & $<5$ missed & $<5$ missed \\
A2 & 2.146 & 303 (14.63) & 97.67\% & 100.00\% & NR \\
A3 & 1.718 & $<5$ & 90.89\% & $<5$ missed & $<5$ missed \\
A4 & 1.752 & NR & 91.17\% & $<5$ missed & $<5$ missed \\
A5 & 1.943 & 304 (14.68) & 100.00\% & $<5$ missed & $<5$ missed \\
\midrule
\multicolumn{6}{@{}l@{}}{\textbf{Middle variation: 1,715 records (PD: 1,387; AP: 85; HC: 243)}} \\
A0 & 1.000 & 0 (0.00) & 98.27\% & NR & 51.44\% \\
A1 & 1.827 & 274 (15.98) & 88.75\% & $<5$ missed & 89.30\% \\
A2 & 2.475 & 815 (47.52) & NR & 100.00\% & NR \\
A3 & 1.828 & 194 (11.31) & 85.58\% & $<5$ missed & 88.89\% \\
A4 & 1.863 & 196 (11.43) & 85.94\% & $<5$ missed & 88.89\% \\
A5 & 2.094 & 499 (29.10) & 100.00\% & $<5$ missed & 89.30\% \\
\midrule
\multicolumn{6}{@{}l@{}}{\textbf{Higher variation: 2,054 records (PD: 1,651; AP: 93; HC: 310)}} \\
A0 & 1.000 & 0 (0.00) & 99.21\% & NR & NR \\
A1 & 1.889 & NR & 92.85\% & $<5$ missed & NR \\
A2 & 2.494 & 1,015 (49.42) & NR & 100.00\% & NR \\
A3 & 1.840 & NR & 88.73\% & $<5$ missed & NR \\
A4 & 1.869 & 381 (18.55) & 88.92\% & $<5$ missed & NR \\
A5 & 2.076 & 662 (32.23) & 100.00\% & $<5$ missed & NR \\
\midrule
\multicolumn{6}{@{}l@{}}{\textbf{Highest variation: 2,011 records (PD: 1,574; AP: 102; HC: 335)}} \\
A0 & 1.000 & 0 (0.00) & NR & $<5$ covered & 39.40\% \\
A1 & 1.988 & 627 (31.18) & 93.20\% & NR & 93.73\% \\
A2 & 2.564 & 1,134 (56.39) & $<5$ missed & 100.00\% & NR \\
A3 & 1.777 & 361 (17.95) & 82.78\% & NR & 79.70\% \\
A4 & 1.821 & 368 (18.30) & 83.67\% & NR & 80.60\% \\
A5 & 2.104 & 705 (35.06) & 100.00\% & NR & 85.97\% \\
\bottomrule\end{tabular}\par\endgroup\medskip
\noindent $<5$ denotes one to four prediction records, covered or missed as indicated. NR marks a value or related estimate withheld to protect sparse counts. Percentages are withheld with their corresponding counts.
\clearpage
\begin{center}\small\textbf{Table S10. Robustness summaries for the whole cohort (continued).}\end{center}
\begingroup\small\setlength{\tabcolsep}{4pt}\renewcommand{\arraystretch}{1.08}
\setlength{\TableWidth}{\dimexpr\linewidth-10\tabcolsep\relax}
\noindent\textbf{B. Expanded assessment budget J4 (levels 1--4)}\par\smallskip
\noindent\begin{tabular}{@{}>{\raggedright\arraybackslash}p{0.08\TableWidth}>{\raggedleft\arraybackslash}p{0.14\TableWidth}>{\raggedleft\arraybackslash}p{0.21\TableWidth}*{3}{>{\raggedleft\arraybackslash}p{0.19\TableWidth}}@{}}
\toprule\textbf{Policy} & \textbf{Average labels} & \textbf{All three labels, count (\%)} & \textbf{PD coverage} & \textbf{AP coverage} & \textbf{HC coverage} \\ \midrule
\multicolumn{6}{@{}l@{}}{\textbf{Lowest variation: 2,313 records (PD: 2,098; AP: 123; HC: 92)}} \\
A0 & 1.000 & 0 (0.00) & 99.09\% & NR & NR \\
A1 & 1.864 & 0 (0.00) & 97.95\% & 100.00\% & 100.00\% \\
A2 & 1.864 & 0 (0.00) & 97.95\% & 100.00\% & NR \\
A3 & 1.864 & 0 (0.00) & 97.95\% & 100.00\% & 100.00\% \\
A4 & 1.864 & 0 (0.00) & 97.95\% & 100.00\% & 100.00\% \\
A5 & 1.929 & 78 (3.37) & 100.00\% & 100.00\% & 100.00\% \\
\midrule
\multicolumn{6}{@{}l@{}}{\textbf{Lower variation: 2,002 records (PD: 1,692; AP: 102; HC: 208)}} \\
A0 & 1.000 & 0 (0.00) & 98.17\% & 19.61\% & NR \\
A1 & 1.654 & 19 (0.95) & 91.67\% & $<5$ missed & $<5$ missed \\
A2 & 2.137 & 274 (13.69) & 97.87\% & 100.00\% & NR \\
A3 & 1.769 & $<5$ & 91.37\% & $<5$ missed & $<5$ missed \\
A4 & 1.770 & $<5$ & 91.43\% & $<5$ missed & $<5$ missed \\
A5 & 1.965 & 283 (14.14) & 100.00\% & $<5$ missed & $<5$ missed \\
\midrule
\multicolumn{6}{@{}l@{}}{\textbf{Middle variation: 1,907 records (PD: 1,525; AP: 98; HC: 284)}} \\
A0 & 1.000 & 0 (0.00) & 98.56\% & NR & 57.04\% \\
A1 & 1.865 & 345 (18.09) & 88.79\% & 100.00\% & NR \\
A2 & 2.507 & 966 (50.66) & 99.08\% & 100.00\% & NR \\
A3 & 1.888 & 244 (12.79) & 84.07\% & 100.00\% & NR \\
A4 & 1.888 & 244 (12.79) & 84.07\% & 100.00\% & NR \\
A5 & 2.147 & 613 (32.14) & 100.00\% & 100.00\% & NR \\
\midrule
\multicolumn{6}{@{}l@{}}{\textbf{Higher variation: 2,024 records (PD: 1,641; AP: 101; HC: 282)}} \\
A0 & 1.000 & 0 (0.00) & $<5$ missed & NR & 50.35\% \\
A1 & 1.896 & 494 (24.41) & 93.17\% & $<5$ missed & 92.20\% \\
A2 & 2.473 & 958 (47.33) & NR & 100.00\% & NR \\
A3 & 1.879 & NR & 88.06\% & NR & 88.30\% \\
A4 & 1.880 & 362 (17.89) & 88.06\% & NR & 88.30\% \\
A5 & 2.099 & 645 (31.87) & 100.00\% & NR & 89.36\% \\
\midrule
\multicolumn{6}{@{}l@{}}{\textbf{Highest variation: 2,056 records (PD: 1,603; AP: 89; HC: 364)}} \\
A0 & 1.000 & 0 (0.00) & NR & NR & 45.05\% \\
A1 & 1.931 & 539 (26.22) & 91.64\% & NR & 92.31\% \\
A2 & 2.558 & 1,147 (55.79) & $<5$ missed & 100.00\% & NR \\
A3 & 1.842 & 359 (17.46) & 83.03\% & NR & 82.14\% \\
A4 & 1.846 & 359 (17.46) & 83.09\% & NR & 82.14\% \\
A5 & 2.154 & 722 (35.12) & 100.00\% & NR & 84.89\% \\
\bottomrule\end{tabular}\par\endgroup\medskip
\noindent $<5$ denotes one to four prediction records, covered or missed as indicated. NR marks a value or related estimate withheld to protect sparse counts. Percentages are withheld with their corresponding counts.
\par\smallskip At each budget, we evaluated every policy on 10,302 held-out prediction records from 3,434 participants across three repeats. Equal SSI scores stay together, so group sizes differ. Coverage percentages are observed proportions within each SSI group.


\maketitle

We selected A3 as the primary alternative to A2.
The Q analysis used A2 at J3 as its primary policy and A5 as an exploratory
comparison. We specified the exploratory Q comparisons after developing the
models and policies in this cohort and before linking Q to outcomes.

\section*{S5.1\quad Repeated records and resampling of participant clusters}

A participant contributes one held-out prediction in each of three
cross-validation repeats. Metrics for repeated records average over those
assessments, while resampling must preserve dependence within participants.
For bootstrap replicate $b$, participant identifiers were sampled with
replacement and their multiplicities $m_i^{(b)}$ applied to all eligible
records for participant $i$ \cite{Field2007}. For numerator $u_{ir}$ and denominator
indicator $v_{ir}$ at repeat $r$, the ratio estimate was
\[
\widehat\theta^{(b)}=
\frac{\sum_i m_i^{(b)}\sum_r u_{ir}}
{\sum_i m_i^{(b)}\sum_r v_{ir}}.
\]
Each eligible prediction record contributes to the estimate. Participants
are the resampling units. The 10,302 held-out records and 309,060 Q
evaluations came from 3,434 participants.

For paired policy or group contrasts, we used the same multiplicities on both
sides and calculated the difference within each replicate. Participants
could contribute to different Q groups across repeats.

\section*{S5.2\quad Intervals, analysis criteria and their limits}

The analysis linking Q to outcomes used 2,000 replicates, random seed 20260819 and
at least 1,900 valid replicates per interval. Linear percentile interpolation
gave the 2.5th and 97.5th percentiles for two-sided 95\% intervals and the
relevant 5th or 95th percentile for one-sided 95\% bounds \cite{Efron1986}. The A3 singleton reliability analysis used seed 20260802 and the ``higher'' quantile convention for its
retention comparison. The analysis of error ranking across all records used
resampling of participant clusters with linear percentile interpolation.

An undefined denominator makes that replicate unusable for the relevant
conditional metric. Intervals required at least 1,900 of 2,000 usable
replicates. Q1 contained no HC records.

The Q analysis specified a positive difference in mean A2 set size (Q4 minus Q1)
with a positive one-sided lower bound, and Q4 class coverage estimates
and one-sided bootstrap lower bounds at or above their targets.
The AURC criterion for retaining the combined reliability score is
specified in Supplement S2.3. Table S9 in S4 gives the A5 efficiency
criteria and estimates.

The bootstrap resampled participants while holding the fitted prediction
procedures fixed. The reported intervals and one-sided bounds are
pointwise.
When every eligible record is covered, every bootstrap replicate with a
nonzero denominator also has 100\% coverage.

\section*{S5.3\quad Exact AP coverage checks}

Within each repeat, a participant contributes at most one outer test
record to Q4. For $x>0$ covered records among $n$ independent Bernoulli
trials, the one-sided exact lower bound at confidence level $1-\eta$ is
the $\eta$ quantile of a Beta$(x,n-x+1)$ distribution \cite{Clopper1934}. When $x=n$,
\[
p_L=\eta^{1/n}.
\]
At $\eta=0.05$, 43 observations with no failures give 0.932703 and 45 give
0.935596. The minimum summarizes the three repeat-specific bounds.
Each bound assumes independent Bernoulli trials within its repeat. Shared training data can induce dependence among prediction errors.